# Biomedical systems biology workflow orchestration and execution with PoSyMed

Simon Süwer[1], Zoe Chervontseva[1], Kester Bagemihl[1], Jan Baumbach[1,3], Olga Tsoy[1,2], Andreas Maier[1]

[1] Institute for Computational Systems Biomedicine, University of Hamburg, Hamburg, Germany
[2] Faculty of Science, Computer Science, Vrije Universiteit Amsterdam, Amsterdam, Netherlands
[3] Department for Mathematics and Computer Science, University of Southern Denmark, Odense, Denmark

## Abstract

The rapid growth of scientific software has created practical barriers for bioinformatics research. Although powerful statistical, artificial intelligence (AI)-based methods are now widely available, their effective use is often hindered by fragmented distribution, inconsistent documentation, complex dependencies, and difficult-to-reproduce execution environments. As a result, reusing published tools and workflow adaptation to own date remains technically demanding and time-intensive, even for experienced users.

Here, we present PoSyMed, an open and modular platform for the controlled integration, composition, and execution of bioinformatics tools and workflows. PoSyMed combines a backend-centered platform architecture with formal tool descriptions, controlled container-based build and execution processes, persistent workflow state, and a dialogue-based user interface. Large language models (LLM) are integrated not as autonomous decision-makers, but as human-computer interface with bounded semantic assistants that help identify tools, propose workflow steps, and support parameterization within a typed, validated, and human-supervised execution environment.

PoSyMed is designed to improve reproducibility, traceability, and transparency in practical biomedical analysis within one platform. We describe the system architecture and evaluate its behavior across representative biological software scenarios with respect to workflow support, interaction design, and platform extensibility.

PoSyMed is publicly available at https://apps.cosy.bio/posymed.

## 1. Introduction

The exponential growth of biological data challenges our ability to convert it into biomedical knowledge [1]. High-throughput technologies such as next-generation sequencing generate petabytes of data, but the software infrastructure required to analyze this data stays behind [2,3]. Additionally, despite remarkable advances in AI for data analysis, many modern bioinformatics methods remain difficult for most of the research community to use [4].

The bioinformatics tool landscape also suffers from being highly fragmented: numerous programs perform equivalent functions and are distributed across heterogeneous repositories [5,6]. These tools are often inadequately documented and offer only limited user-friendly interfaces [7,8]. Thus, researchers spend substantial time installing, configuring, and troubleshooting bioinformatics tools before starting the actual analysis [7–9]. This is not only a productivity problem but also a threat to the FAIR principles for scientific data stewardship [10,11].

Workflow management systems, such as Galaxy, Nextflow, and Snakemake, address this fragmentation issue through standardization, automation, and scalability of bioinformatic pipelines [12–14]. These systems use containerization technologies, such as Docker and Apptainer, and enable reproducible and portable analyses [15]. However, developing and debugging complex workflows still remains a major hurdle for many users, especially in experimental and clinically oriented research groups [16]. Consequently, access to modern bioinformatics methods remains limited despite the availability of powerful tools [17,18]. Meanwhile, large language models (LLMs) are reshaping how humans interact with computational systems. LLMs can generate code and perform increasingly complex planning and decision-making tasks in scientific contexts [19–22]. In some cases, LLMs can already assist in the generation of analysis pipelines from natural-language prompts, for example by proposing preprocessing, model selection, and evaluation steps within existing workflow environments [19–22]. This development has the potential to lower access barriers and accelerate parts of the analysis process [23]. This remains problematic not only because workflows are difficult to construct, but also because hidden assumptions in tools and parameter choices can affect the interpretation of scientific results [23]. Moreover, existing LLM solutions typically address only parts of the problem, such as tool recommendation without robust workflow generation or automation without adequate interaction and control mechanisms for researchers [23–25].

In this paper, we introduce PoSyMed (Population Systems Medicine) (Figure 1), an AI-supported and human-in-the-loop controlled framework for biomedicine-friendly bioinformatics analysis, which includes: (1) the integration of LLMs for tool identification and parameterization, (2) the explicit preservation of human control through transparent, iterative decision-making, (3) the execution of tools in centrally built, validated, and platform-controlled docker containers, and (4) the generation of structured execution reports that document hyperparameters, inputs, outputs, and runtime behavior.

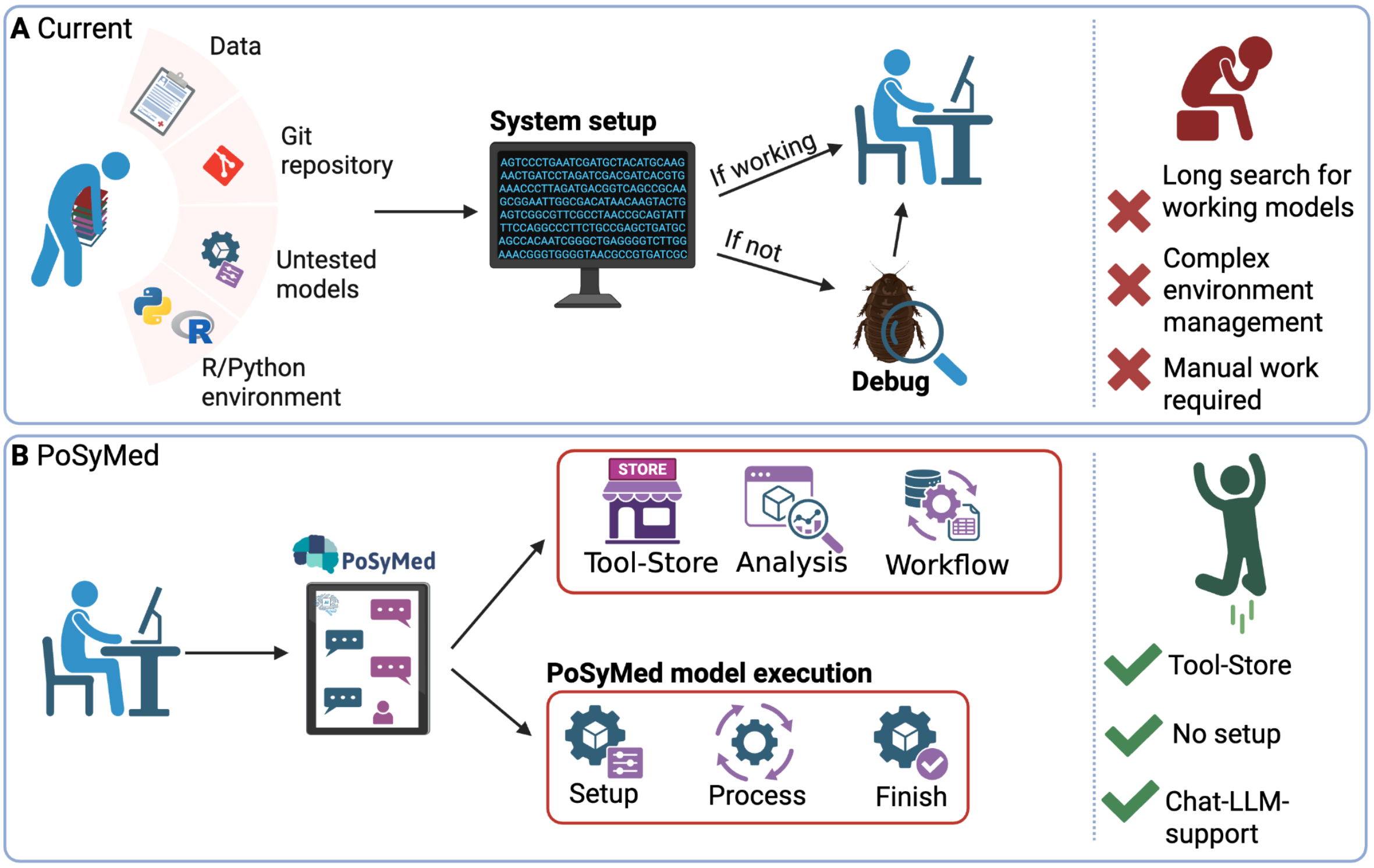

**Figure 1**: **PoSyMed Overview:**The figure compares the current reality of highly manual workflow of biomedical model analysis (A) with the PoSyMed approach (B). In the conventional scenario, researchers are responsible for configuring and testing data, repositories, software environments, and models themselves. This process involves a high level of setup effort, complex environment management, and repeated debugging cycles. PoSyMed integrates these steps into a central platform. A unified user interface gives users access to a controlled *Tool-Store* as well as analysis and workflow functions. Model execution takes place in clearly defined phases. This eliminates the need for local configuration and creates reproducible execution contexts.

The framework derives analysis pipelines from natural language inputs, executes them in containerized environments, and documents each step in a traceable manner. At the same time, researchers retain control over critical decisions, from the refinement of the resulting pipeline to the interpretation of intermediate results. PoSyMed is designed as an open, extensible architecture based on modern software engineering and cloud-native design principles [26]. This architecture supports the reproducibility and traceability of FAIR-oriented scientific workflows [27]. We evaluated the PoSyMed framework on practical bioinformatics analysis scenarios. Our results suggest that PoSyMed reduces technical barriers to complex bioinformatics analysis and supports scientific productivity while preserving user oversight and data analysis transparency.

## 2. Related Work

The broader bioinformatics and clinical analytics landscape is still strongly shaped by classical workflow management systems, in which analyses are typically represented as directed acyclic graphs (DAGs) of tasks and data dependencies [12,28]. Although languages such as Nextflow and Snakemake enable scalable and reproducible execution through containerization and declarative rules [29,30]. They primarily describe workflows as technical flows of tasks and files rather than as explicitly typed biomedical analysis steps [29,30]. Standardized workflow languages such as Common Workflow Language (CWL) and Workflow Description Language (WDL) improve portability across computational environments through machine-readable workflow and tool specifications, yet they primarily encode execution structure and interface compatibility rather than semantically informed workflow composition [12,31,32].

Whereas traditional workflow management systems primarily represent analyses as directed task and data flows, scientific software platforms extend this model by treating execution itself as an explicit, persistent, and reconstructible part of the scientific process [33]. Workflow managers such as AiiDA support reproducible and auditable analyses by recording each calculation step and input-output relationship [34]. Similarly, REANA represents analyses as sequences of versioned program runs to support reproducibility across changing computational infrastructures. As a result, scientific outcomes are treated not merely as isolated computing events, but as reconstructible processes linked to explicit execution histories [35,36]. In biomedical practice, Galaxy facilitates the formal description of analysis tools through pre-defined wrappers [14]. This enables the automatic persistence of parameters, tool versions, and dependencies as provenance information, thereby improving transparency and reusability across studies [14]. Galaxy further improves accessibility and reproducibility by exposing tool versions and execution parameters within an interactive platform [37]. However, it uses limited file formats, interface conventions, and wrapper-level metadata.

OpenSAFELY, a secure analytics platform for NHS electronic health records, demonstrates particularly clearly that reproducibility in clinical settings hinges on secure, governed, and tightly controlled execution environments [38]. In OpenSAFELY, analysis code is executed exclusively within controlled data environments. This design supports both scientific integrity and auditability [38]. Platforms such as DataSHIELD further show that policy-based restrictions on analysis functions and server-side controlled execution can support reproducible analyses without requiring the transfer of sensitive data. In such systems, execution logic is explicitly constrained by institutionally defined rules [39]. Platforms from bioimaging and multi-omics research, such as BIOMERO and qPortal, further emphasize that reproducible analysis remains scientifically interpretable only when execution is tightly linked to domain-specific metadata management and FAIR-compliant data organization [40–42]. These projects illustrate that computations are recorded as contextualized components within a broader scientific lifecycle [40,41].

Across this literature, reproducibility problems are repeatedly linked to implicit assumptions about data, software environments, and parameter dependencies that often become apparent only at runtime [43–45]. In response, workflow management systems have evolved toward more controlled scientific execution platforms that separate workflow specification, orchestration, persistence, and formalize execution through stable application programming interfaces (APIs) or explicit process engines. In such platforms, the execution engine becomes a reproducible and auditable core element of scientific infrastructure [33–35]. An overview can be found in table 1.

## 3. PoSyMed Platform

PoSyMed treats scientific analysis as a controlled, system-managed process beyond sole workflow management. Classical workflow management systems primarily focus on modeling and executing processing steps. By contrast, PoSyMed shapes analysis as a scientific research lifecycle that begins with controlled workflow provisioning and ends with the persistent archival of execution records and results (see table 1, figure 2). In this lifecycle, the platform acts not merely as a technical tool but as the central control layer for execution, validation, and long-term record keeping. It supports the consistent enforcement of scientific and security-related rules throughout this life cycle [14,38].

**Table 1: Comparison of representative workflow and execution platforms in relation to PoSyMed:** *Persistent provenance* denotes the explicit support of retaining and exposing execution lineage, as opposed to general reproducibility claims. *Controlled server-side execution* denotes platform-managed remote or protected execution environments, as opposed to local or user-managed workflow execution. *Workflow planning* denotes native support for defining, composing or explicitly specifying multi-step analytical workflows or pipelines. A *platform-controlled build and validation pipeline* denotes the central management of the build and validation of executable artifacts before use. *LLM support* refers to the native, platform-level integration of large language model functionality rather than external or user-added tools.

| Platform | Persistent provenance | Controlled server-side execution | Workflow planning | Build pipeline | LLM support |
|---|---|---|---|---|---|
| AiiDA | ✓ | X | ✓ | X | X |
| REANA | X | ✓ | ✓ | X | X |
| Galaxy | ✓ | ✓ | ✓ | X | ✓ |
| Open SAFELY | X | ✓ | ✓ | X | X |
| DataSHIELD | X | ✓ | ✓ | X | X |
| BIOMERO | ✓ | ✓ | X | X | X |
| qPortal | X | X | ✓ | X | X |
| Nextflow | ✓ | X | ✓ | X | X |
| Snakemake | X | X | ✓ | X | X |
| CWL | X | X | ✓ | X | X |
| WDL | X | X | ✓ | X | X |
| **PoSyMed** | ✓ | ✓ | ✓ | ✓ | ✓ |

This is ensured by a separation between specification and execution. At the specification level, each tool defines its required input formats, permissible hyperparameters, and declared outputs (see section 3.1). Additionally, only tool images produced through the platform-controlled build pipeline are eligible for execution. The pipeline builds reproducible container images with explicitly linked repositories and automatically checks them for vulnerabilities and malware. Although each tool may have multiple published versions, a version only becomes executable after a linked container image has been successfully built and published through the platform pipeline. Once published, both the tool version and its associated image are immutable (see section 3.1.3). These published tool versions can be accessed via an internal public *Tool-Store*. Each analysis context is defined by a combination of a validated tool version, concrete inputs, and a hyperparameter configuration. At the execution level, the platform preserves a reproducible reference to the input data used for the run, together with the associated configuration.

This separation allows scientific results to be linked directly to specific tool versions, inputs, and parameter configurations, thereby strengthening reproducibility and auditability. Comparable decoupling principles can be found in platforms such as REANA [36]. In PoSyMed, however, this decoupling is extended to the full analysis lifecycle, including tool provisioning, validated containerization, execution, and persistent archiving of analysis steps. This distinction is central to the PoSyMed architecture: specification-level objects describe what may be executed, whereas execution-level objects record what was actually executed.

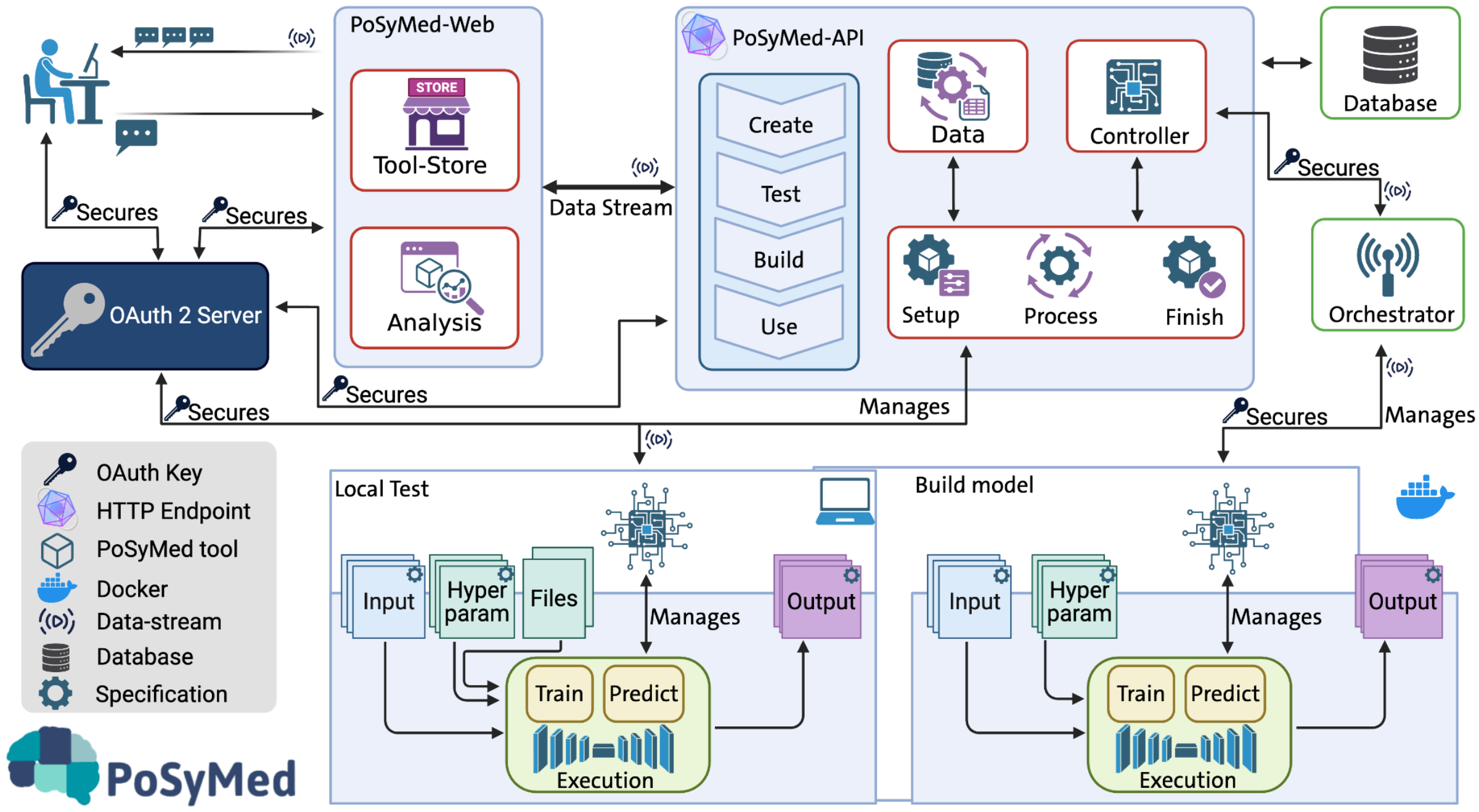

**Figure 2**: **PoSyMed Overview:** PoSyMed is an API-centered platform for controlled and reproducibly analysis execution. User interaction through the PoSyMed web interface, including the *Tool-Store* and analysis modules, is protected by centrally managed authentication and authorization via an OAuth2 server. The PoSyMed backend API acts as a central control instance and a RESTful API for tool creation through to execution. It persistently manages tool definitions, metadata, and execution states in synchronization with the platform database. Actual runtime execution is delegated to a separate orchestration layer that coordinates containerized analyses and encapsulates access to infrastructure resources. Depending on their type, tools may be used for local testing, analysis, or model training. Each individual run is controlled via the linked data (input/output, structure, and hyperparameters). In local testing mode, tools may access local data sources.

The PoSyMed platform adopts key zero-trust principles to allow the analysis of sensitive data. Each analysis job runs in an isolated Docker container. Role-based access control, authentication, and authorization are implemented through OAuth2/OpenID Connect. Further details of the platform architecture are described in the supplementary section 4. The architecture can be deployed in controlled infrastructure environments in which analyses are executed close to sensitive data rather than requiring unrestricted data export [38,46]. Supplementary 6 provides further user-facing examples of the PoSyMed interface, including tool-store views, workflow composition, development and pipeline inspection, experiment execution, LLM interaction, and orchestration administration.

## 3.1 Tools

PoSyMed is a container-based tool framework that enables the standardized, reproducible, and verifiable execution of data-driven analyses (see S4). It allows for the integration of methods ranging from statistical evaluation procedures to complex machine learning (ML) models within a formally specified and technically controlled environment. In this framework, tools are encapsulated as containerized units with explicitly defined input and output interfaces, as well as formalized hyperparameter schemas. Each tool's behavior is described through formal metadata. This metadata specifies expected inputs (e.g., clinical variables, type of data, type of measurement), declared outputs (e.g., predictions, clusters), and configurable hyperparameters that control execution behavior. It also specifies the tool type (see section 3.1.2). In this context, tool typing is more than just organizational metadata. It determines which roles a tool can play within workflows and how its outputs can be interpreted later on.

A central design principle of PoSyMed is community-driven extension through new tool contributions. Researchers and developers can create and version tools independently, while publication on the platform remains subject to centralized validation and controlled build procedures (see 3.1.3). Overall, PoSyMed's container-based tool framework enforces formal, machine-readable descriptions of inputs, outputs, and hyperparameters. This structured approach is central to reproducibility, traceability, and secure orchestration in analysis pipelines, especially in clinical environments [24, 25].

### 3.1.1 Platform analysis lifecycle

When the user starts the analysis, the container is downloaded and executed. The container startup triggers the initialization of the PoSyMed runtime engine. During initialization, the tool establishes a persistent bidirectional WebSocket connection to the platform backend. In development mode, this channel supports interactive transmission of configuration updates, test executions, and runtime outputs. In production, the same channel carries runtime telemetry, including logs, metrics, status transitions, and execution control signals. For local tool development, PoSyMed also supports a non-containerized mode that can connect to the API. By contrast, production execution remains strictly containerized in order to preserve a homogeneous, verifiable, and security-controlled runtime environment.

Figure 3 illustrates the PoSyMed analysis lifecycle. Each tool execution is governed by that explicit lifecycle with clearly defined phases. The lifecycle is represented both in the platform control plane and within the container runtime. Each phase corresponds to a stable system condition that is observable through status messages and runtime telemetry.

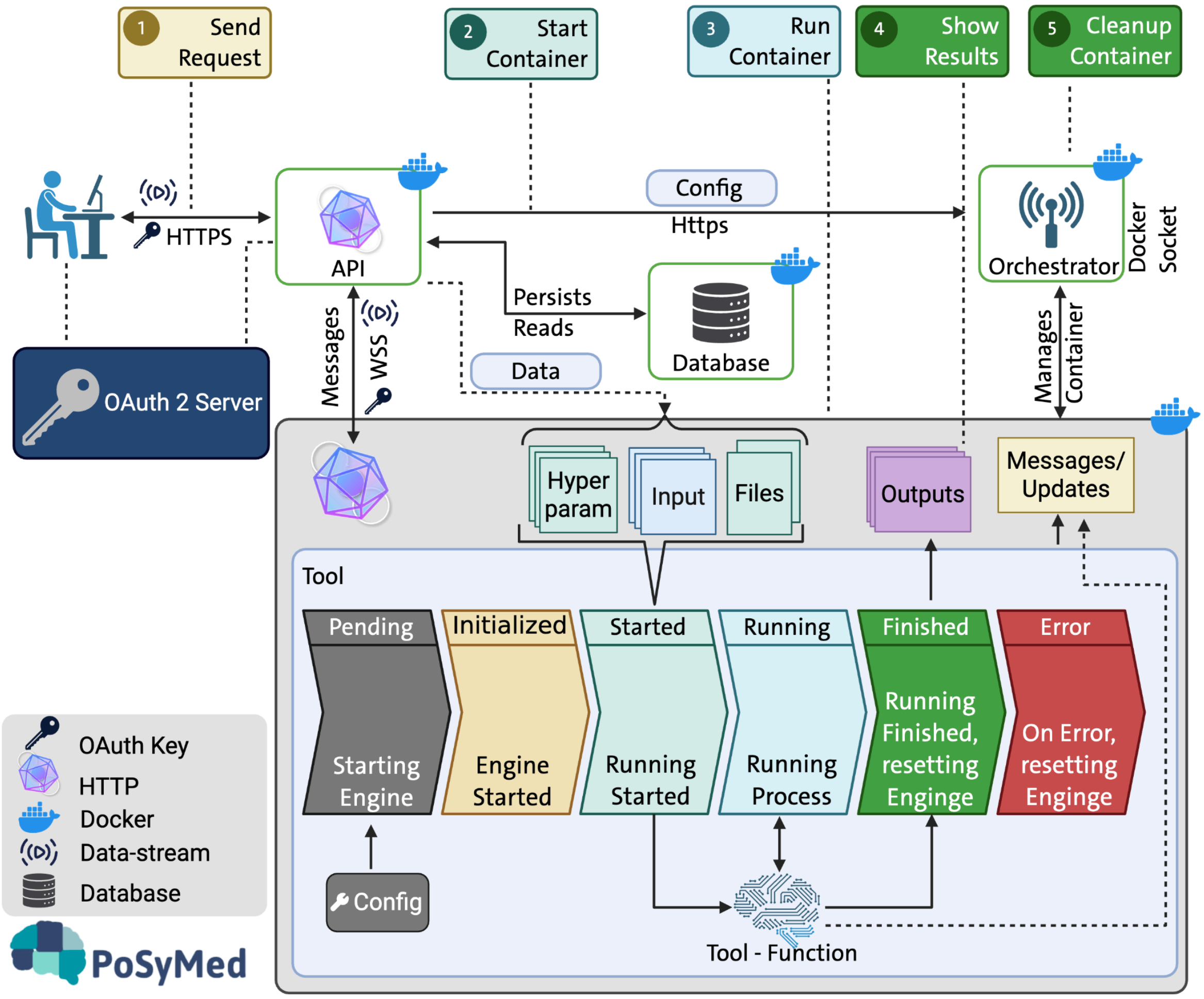

**Figure 3**: **PoSyMed Tool Lifecycle Overview:** A user request is transmitted to the platform via HTTPS, and it is subsequently authenticated via an OAuth2 server. The PoSyMed API is responsible for executing processes, reading, and validating configuration and input data, persisting metadata and provenance in the database, and initiating an isolated container run via the orchestrator (Orch-API/Docker Socket). Within the container, the tool undergoes a sequence of explicit state transitions (Pending → Initialized → Started → Running → Finished/Error). During this process, the tool transfers hyperparameters, inputs, and files, executes its function, and generates outputs. These outputs are then transferred back to the platform. Status changes, logs, and runtime updates are continuously transmitted to the frontend as messages/streams. Subsequent to the completion or error, a controlled reset/cleanup of the container is initiated.

### Pending

On container start, the engine initializes automatically and the tool enters the *Pending* state. In this phase, the tool establishes an authenticated connection to the platform API. Authentication is based on OAuth2, with runtime credentials provisioned by the orchestration layer. Initialization completes only after successful authentication and connection establishment. During this phase, adjustable platform-defined timeout constraints are enforced (default 5 min). On connection failure, the system records the error and produces a structured report. In parallel, configuration (input, output, and hyperparameter schema) is loaded and validated to ensure that received parameters conform to the expected schema.

**Initialized**

In the Initialized state, the tool is operational and can process incoming WebSocket messages. Relevant message types include commands to start execution of the tool, termination requests, and configuration updates during development. Until a start command is received, the tool remains idle.

**Started**

When a run is started, the corresponding WebSocket message contains the information required for execution, including hyperparameters and input data. These elements are validated against the currently loaded configuration. Inputs may be provided through several controlled mechanisms (see S5). Alternatively, the message may contain a download reference that allows the tool to retrieve required data from the platform in a controlled manner. If validation fails, execution is not started.

Once initialization and input validation have completed, the tool's scientific function is invoked. Aspects such as orchestration, status management, and error handling are not delegated to the scientific tool logic, but are handled in a standardized manner by the platform. If invocation succeeds, the tool transitions to the *Running* state.

**Running**

In the *Running* state, the tool communicates continuously and bidirectionally with the platform. Status changes, logs, and metrics are transmitted in real time, enabling live monitoring of execution progress. This runtime telemetry is an integral part of the PoSyMed execution model. It supports debugging during tool development as well as systematic observation of performance, resource usage, and reproducibility across tool versions. In addition, users receive continuous progress updates derived from this live telemetry.

**Finished**

After successful completion, the tool transitions to *Finished* state. In this phase, declared output artifacts are transferred back to the platform (Fig. 3). An artifact is any generated, stored, and reusable result of a computational process. These output artifacts may include result files, model weights, and visualizations. The engine is then reset in a controlled manner, leaving the container in a clean and well-defined final state. In PoSyMed, successful completion is a clearly defined and semantically meaningful state linked to verifiable outputs and documented configuration.

**Error Handling and Recoverability**

In PoSyMed, errors are modeled not as exceptional events outside the system but as explicit states within the lifecycle. When an error occurs, the run transitions into the *Error state*. This state is associated with controlled engine reset and structured error reporting. Error messages, stack traces, and runtime context are returned for user inspection. Explicit error states ensure that failed executions remain traceable and reproducible whenever inputs and configuration are preserved. This design supports robust production pipelines while preserving scientific traceability and methodological transparency.

### 3.1.2 Types of Tools in PoSyMed

The lifecycle of containerized tools defines how analyses are executed operationally in PoSyMed. To model complex pipelines in a methodologically robust and maintainable way, PoSyMed additionally distinguishes tools by their scientific role. Different tools fulfill different scientific functions, such as data preparation, analysis, and result evaluation. Accordingly, PoSyMed assigns each tool an explicit type. By constraining each tool to a well-defined purpose, typing improves interpretability and reuse and thereby strengthens transparency and reproducibility. We define an initial set of types that can be extended by the community.

**Preprocessing**
This type includes data preprocessing operations such as normalization, encoding, and imputation. These operations do not generate persistent model weights. This distinction separates data preparation from downstream analytical procedures.

**Analysis**
This tool type fulfills supervised machine learning tasks. The PoSyMed framework creates versionable analysis components that support both training and prediction. Each training run generates a distinct model instance. Accordingly, a single tool version may produce multiple model versions depending on the training data and hyperparameter configuration. The framework creates a formal and traceable link between configuration, training run, and resulting model. After training is complete, the trained model, including its weights, is stored as a containerized artifact using the PoSyMed pipeline (see section 3.1.3). All generated model weights are versioned and can later be reloaded for selection, comparison, and reproducible execution.

**Evaluation**
Tools of this type calculate performance metrics and evaluation outputs without altering the data or models. The explicit separation of analysis and evaluation supports systematic comparisons between models and pipelines.

**Algorithmic-Analysis**
This type of analytical tool covers algorithmic procedures that do not produce persistent model artifacts, such as clustering or heuristic methods. In these cases, results are derived directly from data and parameterization rather than from reusable model versions. This distinction prevents such outputs from being misinterpreted as trainable model artifacts.

**Data Transformation**
This type includes structural data operations, such as splitting, aggregating, or transforming datasets. The emphasis is on structural adjustment rather than statistical preprocessing. Explicit modeling makes data-changing steps transparent and reproducible.

**Exporter**
These tools act as workflow entry points and load data from external sources. They may lack a classical upstream input interface, but they define the provenance of incoming data.

### 3.1.3 Tool-Building

Studies on software reproducibility show that even slight variations in dependency versions can produce dissimilar results [47]. In clinical, data-intensive contexts, these deviations jeopardize scientific validity [48,49]. At the same time, public container registries may expose vulnerable base layers, insecure dependencies, and embedded secrets [50–54]. This creates substantial security risks, especially when sensitive health data are involved [50–52]. Build and distribution processes in public registries can themselves be compromised, meaning that tool integrity cannot be inferred from popularity or mere availability [55].

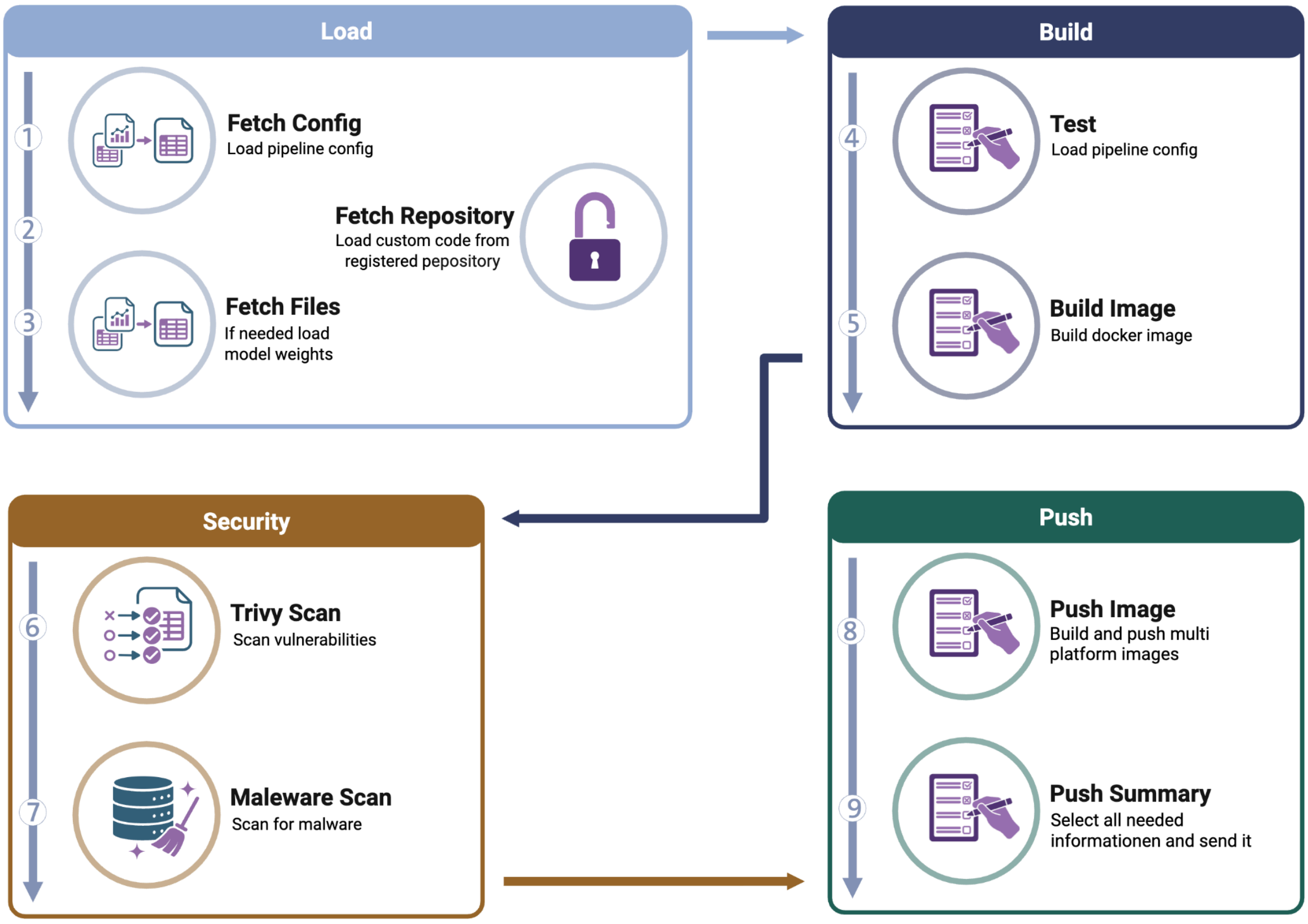

**Figure 4: The PoSyMed tool building pipeline:** This pipeline is used for secure app deployment. The necessary files are loaded, including the model weights and the tool's source code. Based on these files, a reproducible Docker image is created. Then, the image is automatically checked for vulnerabilities and malware. Only successfully tested images are pushed to the central registry.

To support these guarantees, the controlled containerization build pipeline is launched in a separate container for each run. The pipeline follows a server-controlled process chain designed for reproducibility and auditability (see Fig. 4 and S5). Initially, the pipeline obtains a complete build configuration from the backend. This configuration contains the repository source, the target image name, and the required security credentials. Based on this information, the source repository is cloned into a clean, newly created working environment. Additionally, centrally managed artifacts, such as configuration files or runtime components, can be injected in a controlled manner. The Docker image is then generated with a unique metadata-derived name, excluding local overrides, and subsequently validated by automated runtime tests. In addition, a structured vulnerability analysis of the image is performed by trivy[1] and its results are documented. Before the final image construction, for multiple platforms, the entire workspace is automatically scanned for malware using ClamAV[2]. Only after all mandatory checks have been completed successfully is the multiple platform image published to the central registry. It is archived together with cryptographically referenceable metadata, scan results, and source commit information.

PoSyMed containerization process shifts the primary trust anchor from decentralized developer environments to a centrally monitored build pipeline as part of the framework.
In open container ecosystems, artifacts often originate from heterogeneous and partially controlled sources. PoSyMed mitigates this risk by restricting builds to validated, explicitly linked repositories. This helps prevent documented attack patterns, such as dependency confusion and targeted source diversion [56,57].

## 3.2 LLM-Agents

LLMs have recently emerged as a powerful class of AI systems capable of processing and generating natural language, thereby enabling question answering, information synthesis and, to a limited extent, technical reasoning [58–60]. Its capability is rooted in large-scale training on heterogeneous text corpora. This enables LLMs to generate fluent, contextually coherent language and serve as a universal interface in many domains [61]. In biomedical research, they thus open up new possibilities for translating complex analyses into language-guided workflows and facilitating access to data-driven methods [62–64]. However, this generality is a structural limitation. Without domain-specific context or a controlled evidence base, their precision and reliability decline, particularly in specialized biomedical tasks [65–68]. A key risk is their tendency to hallucinate, meaning they generate plausible-sounding statements that are factually incorrect or misleading [65–67]. In critical contexts, such as analyses of clinical data or the derivation of therapeutically actionable hypotheses, this necessitates technical countermeasures and a strict validation logic before outputs from LLMs can be considered usable [69–71].
To mitigate this risk, PoSyMed incorporates Retrieval-Augmented Generation (RAG) and In-Context Learning (ICL) into its multi-agent system. These technologies allow relevant workflow contexts, tools, and data artifacts to be included in the respective model interactions. This helps ensure that responses are based on verifiable system states and computed results rather than on the model's internal assumptions [61,72–74].
In PoSyMed, each agent is explicitly typed, constrained to a narrow action space, supported by ICL, and bound to tool interfaces whose outputs are written back into a persistent planning state. This creates a traceable and auditable reasoning loop with explicit intermediate representations rather than an unverified end-to-end reasoning chain generated by a single model [75–77].

[1] https://trivy.dev/
[2] https://www.clamav.net/

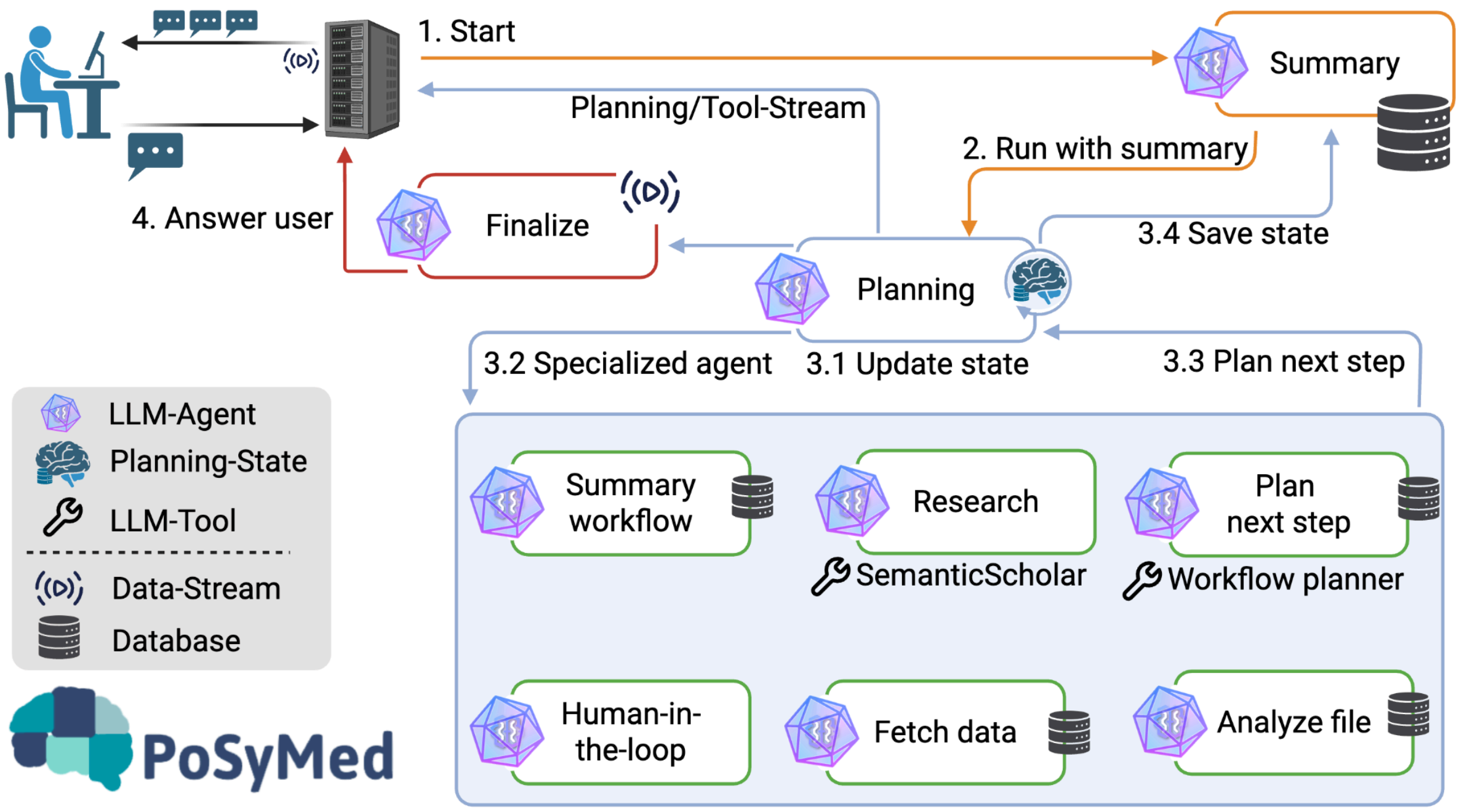

**Figure 5: Agent-based human-in-the-loop support architecture in PoSyMed:** PoSyMed implements a multi-agent system to generate a response: (1) A summary component condenses the relevant conversation context and provides a stable working basis. (2) A planning agent breaks down the user query into subtasks, selects specialized agents, and controls their execution. Results from tool and agent steps are written back to a persistent planning state (PlanState) (3.1–3.4) so that intermediate results remain traceable, reusable, and auditable. (3) Specialized agents cover central subtasks, e.g., Analyze file, Fetch data, Research (literature search), Human-in-the-loop (targeted queries in case of underdetermination), and a workflow planner for selecting the next workflow step. (4) A separate Finalize agent generates the final answer exclusively from the user prompt and consolidated, evidence-based state [77].

Each response is generated through the following sequence of stages (Fig. 5). First, a summary component condenses the relevant conversation history into a stable working context. Because LLMs are vulnerable to jailbreak and instruction-overriding attacks, prompt-injection safeguards are applied at this stage [78–80]. The *Planning agent* then decomposes the user request into simpler subtasks and selects appropriate specialized agents. If a subtask requires analysis of uploaded files, it is routed through a dedicated analysis toolchain. The resulting outputs are written to the *PlanState*, where they can be reused by downstream agents or incorporated into the final answer. A retrieval-supported agent accesses the *Tool-Store* in order to identify the next workflow action. Files referenced by the user can be resolved and analyzed by the LLM. Rather than selecting tools freely, this agent identifies candidates through retrieval, validates tool compatibility, checks whether user-referenced data match the tool requirements, and fills in or requests hyperparameters as needed. If mandatory entries are missing, the agent asks a precise question and halts the process via a *Human-in-the-loop* (HITL) mechanism. This HITL path is, therefore, not an exception, but a deliberate safety mechanism. It prevents underdetermined requirements from being resolved through speculative assumptions and transfers the decision back to the user when uncertainty cannot be resolved from available tool evidence. Literature search can be invoked selectively when a user request requires external scientific evidence that is not available from the current workflow state, for example to compare candidate methods or contextualize analysis results, without bypassing the structured management of tools and persistent state. The final answer is generated by a separate *Finalize agent*. This agent receives only the user prompt and the consolidated, evidence-based state and uses these inputs to generate a concise answer.

The separation of summarization, planning, and response generation increases architectural robustness. If intermediate steps fail or remain incomplete, the final response reflects that uncertainty explicitly. To mitigate residual hallucination risk, PoSyMed includes a dedicated output-safety mechanism via an output guardrail [81,82].

## 3.3 PoSyMed workflows

In PoSyMed, workflows are designed as platform-managed, typed DAGs. Therefore, a workflow describes a user-defined sequence of processing steps and a persisted execution specification. Within this specification, each node refers to a specific, version-controlled, published tool definition. This binding includes the declared input and output interfaces, as well as the associated hyperparameter schema. Workflow composition therefore takes place based on formally described tools that are already known to the platform at the time of graph construction.

Each workflow node represents an executable analysis step. Edges not only encode a topological sequence but also a named mapping between the outputs of an upstream node and the inputs of a downstream node. Therefore, an edge is only valid if the output configuration referenced by the source node is compatible with the input configuration referenced by the target node.

Before execution, graphs are checked for structural consistency, interface compatibility, and executability. This results in an ordered, executable representation. At runtime, the individual workflow nodes are converted into execution-specific objects with concrete inputs and hyperparameters and are then executed by the orchestration layer in isolated environments.

As workflow nodes refer to previously described tool interfaces, the platform can allocate results, logs, and provenance information to the relevant graph objects. This coupling allows tracing back which tool version was executed at which position in the graph, with which inputs and parameterizations.

# 4. Evaluation

We evaluated the PoSyMed framework based on three complementary aspects: (i) the ability to integrate and execute tools, (ii) workflow assembly capability, and (iii) the quality and reliability of LLM-supported interaction.

To evaluate tool integration and execution, PoSyMed was tested as an end-to-end platform by implementing and executing a total of 20 bioinformatics tools selected as practically useful examples, although not intended to be fully representative of the broader bioinformatics tool landscape (see section 3.1 and table 2). For this evaluation, we defined tool success as the ability of a tool to complete the full platform-managed lifecycle: (a) successful implementation, (b) reproducible publication of the tool by the building pipeline,

(c) execution of the tool, and (d) returned its expected status information, logs, and outputs to the platform in the intended form.

The workflow evaluation checked PoSyMed's compositional performance, i.e., its ability to orchestrate tools as repeatable and traceable DAG workflows (see section 3.3). To this end, 10 workflows covering different common structures were created and executed. The evaluation assessed the following: (1) whether workflows could be specified as syntactically and semantically valid, (2) whether the orchestration would deterministically implement the expected execution order and parameter binding, and (3) whether identical workflow specifications would produce consistent execution artifacts in repeated runs.

Evaluating LLM-based agent systems is challenging because LLM responses involve interpretation and decision-making processes [83]. Unlike deterministic evaluations, there is often no single objective truth that goes beyond simple accuracy metrics [83]. Therefore, a robust evaluation requires a combined approach of automated metrics and manual validation [83].
We evaluate our LLM agent on two levels: (1) whether all tools required for a question are invoked correctly, and (2) whether the resulting user communication is complete and comprehensible (see section 3.2). For this purpose, we designed an evaluation set of 120 questions across six thematic topics, with 20 questions per topic (S3). Additionally, 30 questions, with five in each topic, evaluate the HITL capability by providing predefined follow-up answers when the agent requests clarification. In this way, the benchmark covers both single-turn and clarification-driven multi-turn interactions. Responses were first evaluated using LLM-as-a-Judge, where a separate LLM simulates human judgment for tasks such as evaluation, ranking, and criticism [84]. Since such judges can show bias and instability, the PoSyMed evaluation was additionally supervised by us, and incorrect judgments or obvious execution artifacts were corrected where necessary.

## 5. Results

### 5.1 Tool development and execution

We assessed a diverse set of tools, covering the core categories of bioinformatics operations that can extract biomedical knowledge from omics cohort-based data (see table 2). The evaluated tools therefore encompass data acquisition and transformation, supervised analysis, and various data-driven methods, including clustering, network-based analysis, factor-based modeling, and single-cell analysis. This selection was intended to test PoSyMed's ability to operate across heterogeneous tool interfaces, programming languages (R/Python), input requirements, and hyperparameter configurations while maintaining the framework's general applicability to various biomedical scenarios. Detailed tool development specifications and implementation overviews are provided in supplementary 1 and all tool definition files, build artifacts, execution reports, and related materials are available in a public repository[3].

**Table 2: Tool development and execution evaluation**: Overview of the bioinformatics tools evaluated in PoSyMed, including the tool category, number of execution runs, and

[3] https://github.com/SimonSuewerUHH/posymed-experiments

report. The table summarizes the heterogeneity of the integrated tools and demonstrates the platform's capacity to facilitate the publication and execution of reproducible tools. Run details are explained in supplementary 1.

| Name | Tool-ID | App-Type | Runs | Report |
|---|---|---|---|---|
| UciFetch | 1 | Extractor | 1,2 | R.1 |
| SpectralClustering | 2 | Algorithmic-Analysis | 3,4 | R.2 |
| SqlSelectExtractor | 3 | Extractor | 5 | R.3 |
| OneHotEncoding | 4 | Pre-Processing | 6,7 | R.4 |
| Agglomerative Clustering | 5 | Algorithmic-Analysis | 8,9 | R.5 |
| UnPast [85] | 6 | Algorithmic-Analysis | 10 | R.6 |
| BIRCH [86] | 7 | Algorithmic-Analysis | 11,12 | R.7 |
| GaussianMixture Model | 8 | Algorithmic-Analysis | 13,14 | R.8 |
| DysRegNet [87] | 9 | Algorithmic-Analysis | 15 | R.9 |
| Spycone [88] | 10 | Algorithmic-Analysis | 16 | R.10 |
| MoSBi [89] | 11 | Algorithmic-Analysis | 17 | R.11 |
| Clusteval [90] | 12 | Evaluation | 18,19,20 | R.12 |
| LiniearRegression | 13 | Analysis (Trained + Model executed) | 21 | R.13 |
| train-test-split | 14 | Pre-Processing | 22,23 | R.14 |
| BranchMergeAggregator | 15 | Post-Processing | 24,25,26 | R.15 |
| ClusterComparisonEvaluator | 16 | Evaluation | 27,28 | R.16 |
| MOFA2 [91] | 17 | Algorithmic-Analysis | 29,30 | R.17 |
| BiCoN [92] | 18 | Algorithmic-Analysis | 31 | R.18 |
| SCANet [93] | 19 | Algorithmic-Analysis | 32 | R.19 |
| SpliceDrift | 20 | Algorithmic-Analysis | 33 | R.20 |

A total of 20 tools with 165 documented runs were evaluated. Of these, 20 tools were successfully implemented, published, and executed, corresponding to an end-to-end success rate of 100%. The documented runs prove that inputs, hyperparameters, and output objects were managed in a reproducible manner on the platform side.

## 5.2 Workflow capability

The PoSyMed system was evaluated using five workflow patterns: single-step, linear, branched, comparative multi-model, and split-and-merge. For example, in a comparative multi-tool workflow, a dataset was first retrieved using *UciFetch*, then preprocessed using *OneHotEncoding*, and then the encoded feature matrix was passed in parallel to alternative clustering tools, such as *AgglomerativeClustering* and *GaussianMixtureModel*. The resulting cluster assignments were then compared in a downstream evaluation step using the *Comparative-Clustering-Benchmark* tool. In a split-and-merge pattern example, an input CSV file was divided into parallel outputs using *train-test-split*. The resulting branches were then recombined using the *Branch-Merge-Aggregator* tool for downstream processing. While branched comparison and split-and-merge workflows may have similar graph shapes, they differ in meaning: in the former, a common intermediate result is used to compare alternative analysis methods, whereas in the latter, data is explicitly partitioned and then combined again. Overall, ten workflows were defined and executed, with two sample workflows per pattern (see table 3).

Across these workflow classes, PoSyMed supported the construction of valid workflows and their execution in the expected deterministic order. Linear workflows allowed extraction, preprocessing, and downstream analysis to be combined in a consistent manner, supporting stable and reproducible analyses. Branching and comparison workflows additionally demonstrated that a common intermediate output can be reused across multiple downstream tools. This enabled a systematic comparison of different analysis methods within a unified execution context. Split-and-merge workflows further expanded this insight by demonstrating that PoSyMed can also represent more complex DAG topologies. In this process, data is initially divided into parallel branches and subsequently recombined for downstream evaluation. These results are consistent with the platform's intended role as a system for repeatable and traceable workflow orchestration, rather than merely isolated tool execution. Detailed workflow definitions and graphs are provided in Supplementary 2, and all workflow files, execution reports, and related artifacts are available in a public repository[4].

**Table 3: Workflow capability evaluation of representative PoSyMed workflow patterns**: The evaluation comprises ten workflows spanning five structural classes: single-stage, linear, branching, comparative multi-model, and split-and-merge. For each workflow, the table lists the pattern type, the tools involved, the executed runs, and the corresponding report. The IDs in Tools match the implemented tools from section 5.1.

| Name | Pattern type | Tools | Runs | Report |
|---|---|---|---|---|

[4] https://github.com/SimonSuewerUHH/posymed-experiments

| | | | | |
|---|---|---|---|---|
| UCI dataset fetch | Single-step |   | 1,2 | R.1 |
| CSV to Train-Test-Split | Single-step | 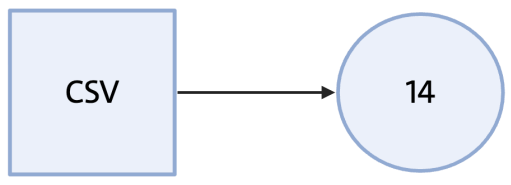  | 3,4 | R.2 |
| Fetch–encode workflow | Linear | 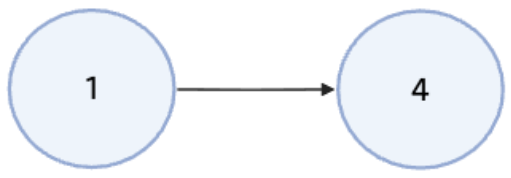  | 5 | R.3 |
| CSV encode to cluster | Linear | 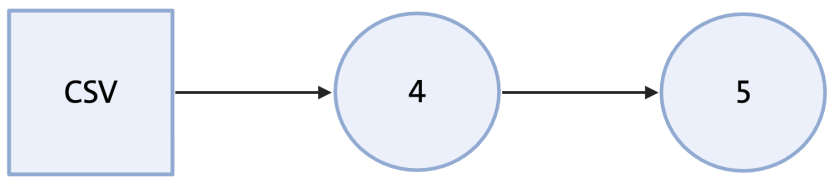  | 6 | R.4 |
| Shared preprocessing with clustering branches | Branching | 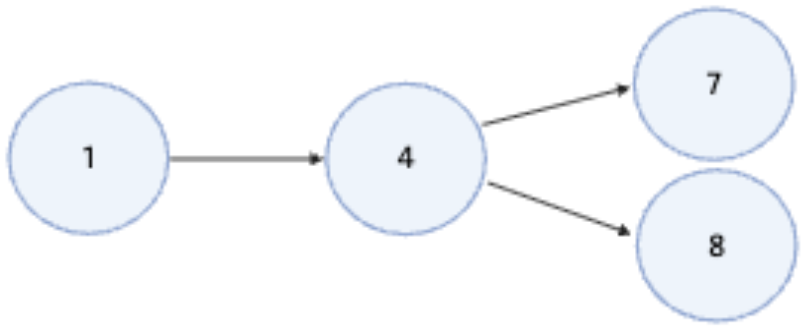  | 7 | R.5 |
| Shared input with analysis branches | Branching | 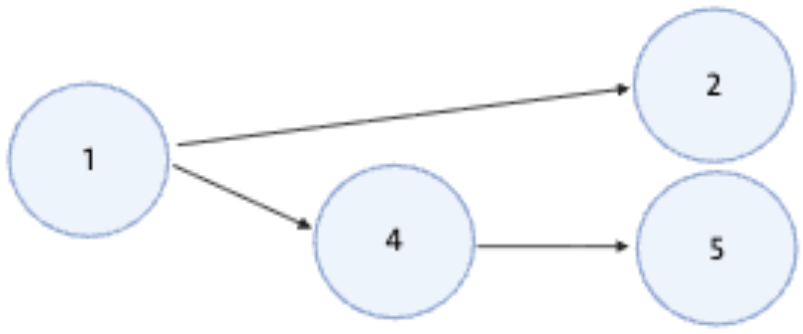  | 8 | R.6 |
| Multi-model clustering comparison | Comparative | 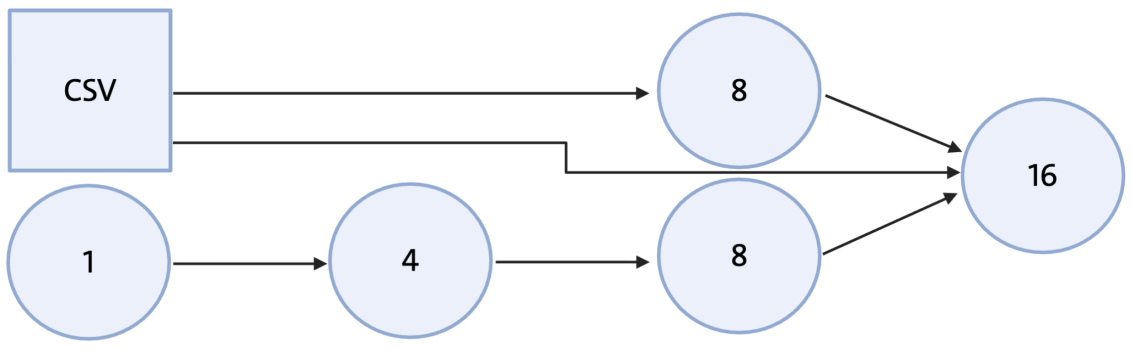  | 9 | R.7 |
| Comparative clustering benchmark | Comparative | 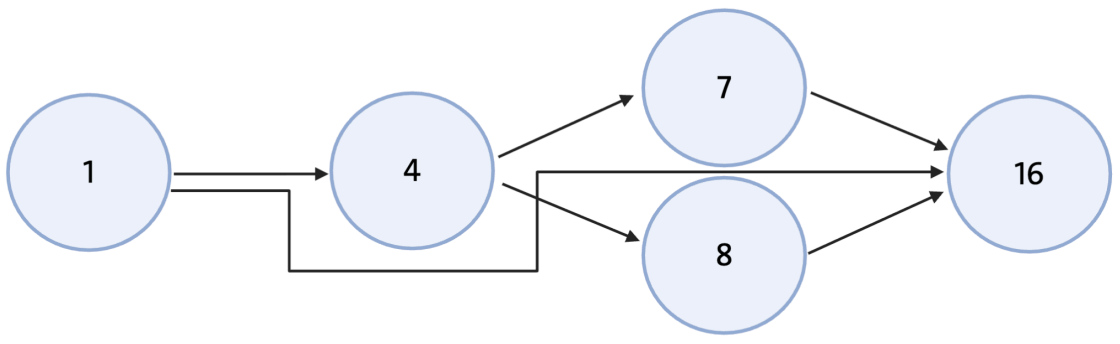  | 10 | R.8 |
| Split-and-merge workflow | Split-and-merge | 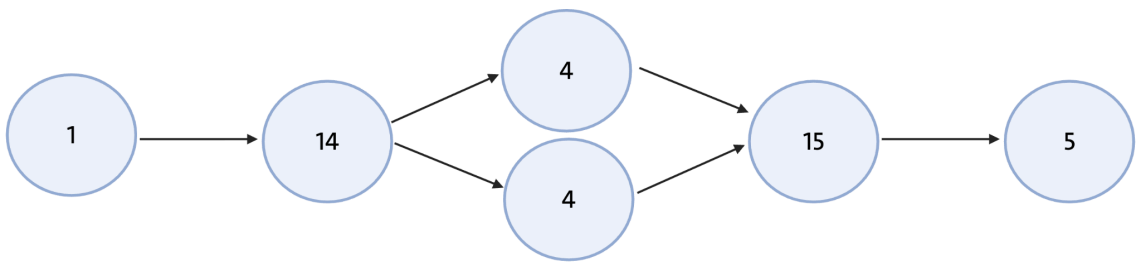  | 11 | R.9 |
| Sample split-and-merge workflow | Split-and-merge | 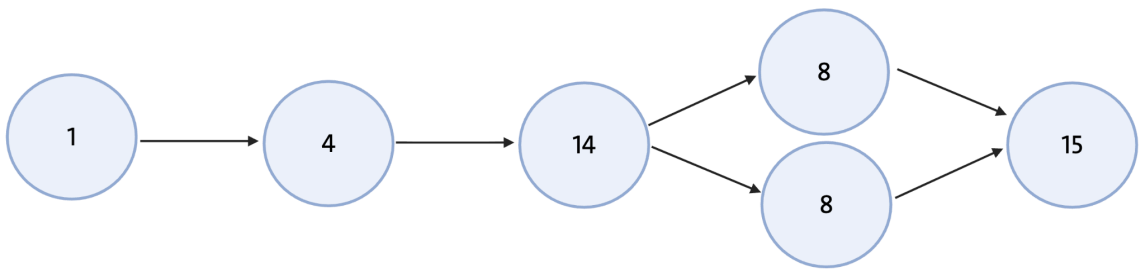  | 12 | R.10 |

In this evaluation design, the success of all 10 workflows is contingent upon their adherence to the following predefined deterministic criteria: semantic validity, correct execution order and parameter binding, and consistent artifact generation across repeated runs. In this configuration, a complete success would be indicated by 12/12 successful workflow executions.

## 5.3 LLM-supported interaction

Evaluation of the LLM-supported interaction indicates that the PoSyMed agent operates robustly within the bounded platform setting (see Fig. 6). Final responses to 120 benchmark questions spanning six thematic topics (see S3) were assessed with an LLM-as-a-Judge score, while internal action sequences were assessed separately via Tool-Calling. The judge score reflects whether the final answer matched the intended benchmark outcome in a complete, comprehensible, and contextually adequate manner. Tool-Calling matches were determined by comparing the tools actually invoked with the expected benchmark sequence. In the strict evaluation, missing expected tools were counted as mismatches. However, in the corrected Tool-Calling evaluation, reasonable exclusions of tools were accepted, as long as the final answer was evaluated as correct.

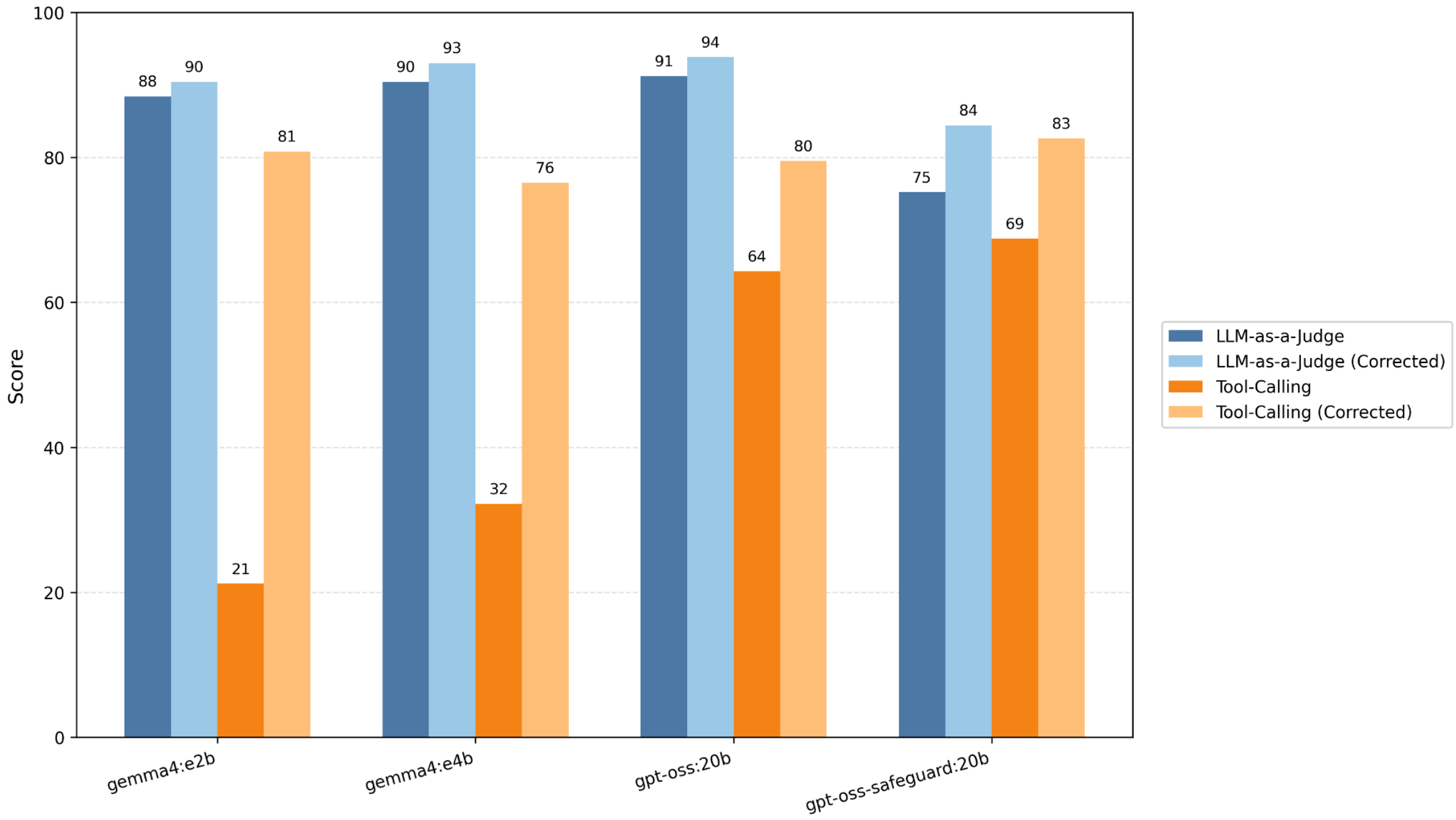

**Figure 6: Comparison of LLM-as-a-Judge and tool-calling performance across evaluated models:** Scores are shown on a normalized 0–100 benchmark-alignment scale. A score of 100 indicates full agreement with the expected outcome under the respective evaluation criterion. LLM-as-a-Judge evaluates the quality of the final answer, i.e., whether the response matches the intended benchmark outcome in a complete, comprehensible, and contextually adequate manner. Tool-Calling evaluates process conformity, or how closely the executed tool sequence matches the expected tool path. Corrected variants reflect a manual review of obvious judge errors and, for Tool-Calling, the acceptance of reasonable tool omissions when the final answer is correct. Across all four models (gemma4:e2b, gemma4:e4b, gpt-oss:20b, and gpt-oss-safeguard:20b), LLM-as-a-Judge scores remain consistently high, whereas raw Tool-Calling scores vary

substantially more between models. This indicates that final-answer quality was often preserved, even when the internal tool path deviated from the benchmarked sequence.

A more nuanced picture emerged when it came to the expected tool calls. Although the final answers were generally rated as correct or largely adequate, the actual tool path executed did not always fully correspond to the sequence defined in the benchmark. In particular, steps involving research, lookup, or file preparation were sometimes condensed without significantly reducing the substantive quality of the answer. For example, in several cases, the model omitted individual intermediate steps such as FETCH_DATA, ANALYZE_DATA, or FETCH_TOOLS, yet still arrived at a factually appropriate answer. These deviations should therefore not be interpreted automatically as a failure of tool use, but rather often reflect a shorter, semantically still sufficient internal action path. Under a relaxed evaluation that required a correct final answer while still treating HUMAN_IN_THE_LOOP and FINALIZE as mandatory, the tool-calling scores increased significantly.

Taken together, these findings support the conclusion that the LLM-based interaction in PoSyMed is functional and scientifically usable within a controlled orchestration framework. However, these results should not be overinterpreted as evidence of superiority over alternative systems. While this evaluation shows that the system is viable within the PoSyMed architecture, it does not constitute a comparative benchmark against other workflow assistants, biomedical copilots, or non-agentic interfaces. The current findings suggest that limited LLM support can be incorporated into a formal, typed, human-supervised biomedical platform without compromising answer quality. We posit that a comprehensive set of benchmark data and protocols for quantitative evaluation of agentic bioinformatics workflow designers is currently absent, though it extends beyond the purview of this publication.

## 6. Discussion and outlook

The results demonstrate that PoSyMed is more than just a workflow system. It is an integrated platform that combines tool provision, controlled execution, provenance-based traceability, and LLM-supported interaction within a unified architecture. Its innovative approach systematically couples formal tool descriptions, server-controlled build and runtime environments, state-based orchestration, and agentic language models. These features link automation to scientific control and support of FAIR principles.

Compared to well-established workflow management systems, such as Nextflow, Snakemake, and Galaxy, PoSyMed fills an important gap. While these systems deterministically and reproducibly describe and execute process chains, they typically require external tools that are technically trustworthy. PoSyMed extends the concept of reproducibility to cover the entire life cycle of scientific analysis. Tools are published in a controlled manner, built reproducibly, validated, and checked for vulnerabilities and malware before being included in analyses as platform artifacts.
Compared to provenance-oriented systems such as AiiDA or REANA, PoSyMed extends this approach to include formal tool typing, semantically controlled workflow composition, and dialogue-based interaction for biomedical applications. We successfully integrated 20 bioinformatics tools that cover various approaches to omics data analysis as well as multiple languages. Similarly, the workflow evaluation revealed that all defined workflow patterns could be specified and executed deterministically during repeated runs. These results imply that the platform not only suggests a conceptual integration of tools and workflows but also can sustain the onboarding of heterogeneous tools and repeatable, DAG-based orchestration.
PoSyMed supports community contributions as a governed process rather than unrestricted deployment. Researchers and developers can independently create and version tools, and the current evaluation indicates that these tools can successfully pass through the platform-managed lifecycle. However, releases to the central instance are intentionally made through the controlled and protected release pipeline rather than by users directly uploading arbitrary container images. We believe that this approach fosters greater security and trust.

We explicitly designed PoSyMed so that it does not use LLMs as autonomous authorities of knowledge, but rather as semantic control layers within formally limited frameworks. Agent decisions are bound to validated tool interfaces, persistent states, and HITL mechanisms. The LLM evaluation validates this design choice. Even when the model did not follow every intermediate step anticipated by the benchmark specification, the quality of the answers remained high across the agent benchmark. The agent demonstrated a pattern of frequently producing the correct response with a reduced number of tool steps compared to the established benchmark. This phenomenon can be attributed to the elevated performance metrics observed in the LLM-As-A-Judge scores, while the strict Tool-Calling scores showed a decline. Therefore, the observed discrepancy is likely attributable to a stringent benchmark rather than to inadequate utilization of the tools. In PoSyMed, therefore, bounded LLM assistance appears to be more than mere rule replay. Within a formally controlled environment, it can produce semantically efficient behavior. This is an encouraging result, as it indicates that strict platform constraints do not preclude flexible and useful interaction. Instead, they may be sufficient to preserve scientific control while allowing the model to compress internal reasoning when explicit decomposition is unnecessary. In biomedical settings, such an architecture may be more robust than open assistance systems lacking formal cross-checking and controlled execution boundaries. At the same time, execution decisions remain with the user, and the LLM provides recommendations, not autonomous action.
Nevertheless, the results should be interpreted with caution. PoSyMed reduces runtime errors and fills an important gap implementing a secure life cycle of scientific analysis but shifts requirements to tool creation. Comparative benchmarks are needed for integration effort, stability, reproducibility, security incidents, user effort, and time savings. Additionally, performance depends significantly on the quality of the formal tool descriptions and curated platform artifacts. Additionally, evaluating LLM components remains methodologically challenging because quality is determined by more than just right-wrong metrics; completeness, traceability, and communicative precision also play a role. Finally, while the controlled build pipeline significantly improves security, it cannot replace complete high-security isolation.

Future research endeavors should prioritize the undertaking of prospective comparative studies, the enhancement of semantic typing and the development of more robust agentic planning and validation mechanisms. An additional important future direction is the integration of federated infrastructures, in which analyses are orchestrated across institutionally separated environments while sensitive data remain local [94,95]. For PoSyMed, this would extend the platform from controlled single-instance execution toward secure cross-site biomedical analysis. Such an extension would be especially relevant for clinical and other governance-sensitive settings, in which data sharing is restricted and analyses must be performed under tightly controlled conditions [94–96].

## 7. Conclusion

This paper introduced PoSyMed, a platform that unifies the orchestration of biomedical software, workflows, and AI-supported interactions within a single, controlled system. Its key contribution is intertwining the requirements for reproducibility, security, traceability, and user accessibility architecturally rather than separately. With formal tool descriptions, reproducible build and runtime processes, state-based workflow control, and HITL supported LLM agents, PoSyMed creates a system designed to facilitate the use of modern bioinformatic methods while maintaining scientific control.

We show that PoSyMed could address key weaknesses in the bioinformatics software landscape, such as fragmentation, high technical barriers to entry, and a lack of operational consistency. Unlike many existing solutions, the platform not only executes workflows but also integrates the controlled provision and validation of the underlying tools. Extending the concept of reproducibility to the entire life cycle of scientific software is what makes this approach particularly relevant.

Concurrently, the work emphasizes that the effective utilization of LLMs in scientific analysis environments is contingent upon the integration of language models within formally controlled, evidence-based, and transparently verifiable processes. PoSyMed exemplifies a viable architectural principle in this regard: automation is not regarded as a substitute for scientific judgment; rather, it is conceived of as structured support for it. It was designed to make bioinformatics tool applications more efficient, more robust, and more reproducible.

Consequently, the study offers a technical platform design and a conceptual contribution to the future design of biomedical research infrastructures. If the approach proves viable in further comparative and real-world evaluation scenarios, PoSyMed could become an important model for a new generation of reproducible, secure, and dialogue-oriented analysis platforms.

## Code availability

The code for PoSyMed is publicly available at the following GitHub group: https://github.com/FedLearnNet (Building-Pipeline, AppWrapper, Backend, Orch-API, Frontend). This repository contains the source code, documentation, and instructions for setting up and running PoSyMed locally. For users who prefer not to set up a local environment, a web-based interface for PoSyMed is available at: https://apps.cosy.bio/posymed. The PoSyMed software is free for academic and non-profit use under the Apache 2.0 License (https://www.apache.org/licenses/LICENSE-2.0.txt). PoSyMed supports Bring Your Own Key (BYOK).

## Technologies used

PoSyMed is built on LangChain4j[5] and the Quarkus[6] Support[7] to allow LLM interactions. The front end (Angular[8]) connects to Quarkus servers, which communicate via WebSocket. In addition to the WebSocket server, the back end offers a REST API and an MCP server[9]. These provide the most important tools for external integration. These features make the front end interchangeable and integrable into other LLM workflows [97]. PoSyMed utilizes gpt-oss:20b [98], a highly capable LLM developed by OpenAI and optimized for helpful responses. It is hosted through a self-hosted Ollama instance [99]. We use Ollama behind a self-hosted OpenWebUI[10] allowing access control.

## Funding

This work was developed as part of the PoSyMed project and is funded by the German Federal Ministry of Research, Technology and Space (BMFTR) under grant number 031L0310A. Also Funded by the European Union under contract no. 101136305. The Hungarian partner is funded by the Hungarian National Research, Development, and Innovation Fund. Views and opinions expressed are however those of the authors only and do not necessarily reflect those of the European Union or the Hungarian National Research, Development and Innovation Fund. Neither the European Union nor the Hungarian National Research, Development and Innovation Fund can be held responsible for them.

## Author contributions statement

Simon S. designed the platform, developed the backend and web interface. Simon S. designed the evaluation. Andreas M., Olga T. and Jan B. supervised the project. Zoe C. developed the tool SpliceDrift. Kester B. supported the testing of the platform. All authors provided critical feedback, participated in discussions, and assisted in interpretation of the results, writing of the manuscript.

## Acknowledgements

We thank Sina Pralle, Asihati Hazaiti, Yanxi Lin, Sesugh Nder, Ranziska Papenhausen and Sven Schmidtke for their contributions to the initial development of platform applications. We also thank Elke Hammer for testing the first prototype of the platform and for providing valuable feedback during the PoSyMed workshop.

[5] https://docs.langchain4j.dev/
[6] https://quarkus.io/
[7] https://docs.quarkiverse.io/quarkus-langchain4j
[8] https://angular.dev/
[9] https://docs.quarkiverse.io/quarkus-mcp-server/dev/index.html
[10] https://github.com/open-webui/open-webui

## Supplementary material 1: Tool-Experiments

### Tool 1: UciFetch
**Typ:** Extractor
**Hyperparameter:** dataset_id
**Input:** None
**Output:** dataset (CSV), report (TEXT)
**Description:** Fetch a UCI dataset by id via ucimlrepo and export a single combined CSV (features + targets).

### Tool 2: SpectralClustering
**Typ:** Algorithmic-Analysis
**Hyperparameter:** n_clusters, affinity, gamma, n_neighbors, assign_labels, n_init, eigen_solver, random_state, standardize, output_embedding
**Input:** features (CSV)
**Output:** labels (CSV), embedding (CSV), report (HTML)
**Description:** Graph-based clustering using scikit-learn SpectralClustering (eigenvectors + k-means).

### Tool 3: SqlSelectExtractor
**Typ:** Self-Extractor
**Hyperparameter:** sql_distribution, host, port, database, username, password, sql_select
**Input:** None
**Output:** dataset (CSV)
**Description:** Secure SQL extractor that connects to a configured database and executes a single read-only SELECT (or WITH…SELECT) query, exporting the result as a CSV table.

### Tool 4: OneHotEncoding
**Typ:** Pre-Processing
**Hyperparameter:** columns, drop_first, bool_to_number, dummy_na, handle_unknown, prefix_sep, keep_original
**Input:** features (CSV)
**Output:** encoded (CSV)
**Description:** CSV-to-CSV one-hot encoding using pandas only; encodes selected (or auto-detected) categorical columns into dummy variables with options for dropping the first category, adding NaN dummies, preserving originals, and converting booleans to 0/1.

### Tool 5: AgglomerativeClustering
**Typ:** Algorithmic-Analysis
**Hyperparameter:** linkage, compute_distances, standardize, metric, use_distance_threshold, n_clusters, plot_title, dendrogram_truncate_mode, dendrogram_p, distance_threshold
Input: features (CSV)
**Output:** dendrogram_png (IMAGE), report (TEXT)
**Description:** Hierarchical (agglomerative) clustering that builds a merge tree using the selected linkage/metric, optionally standardizes features, and produces a publication-ready dendrogram (PNG) plus a lightweight summary report (params, cluster stats, embedded plot).

### Tool 6: unpast

**Typ:** Algorithmic-Analysis
**Hyperparameter:** binarization_method, deep_split, dynamic_tree_cut, ceiling, bidirectional, seed, similarity_cutoff_mode, similarity_cutoff_value, p_value, clustering_method
**Input:** data (TSV)
**Output:** outputFile (CSV)
**Description:** UnPaSt (Unsupervised Patient Stratification) identifies biclusters from an expression matrix by binarizing signals (e.g., via kmeans/Ward/GMM), then applying a clustering strategy (WGCNA or Louvain) with configurable thresholds and reproducibility controls, exporting the detected biclusters as a TSV/CSV table.

### Tool 7: Birch

**Typ:** Algorithmic-Analysis
**Hyperparameter:** threshold, branching_factor, use_global_clustering, n_clusters, compute_labels, standardize
**Input:** features (CSV)
**Output:** subcluster_centers (CSV), report (HTML), labels (CSV)
**Description:** Scalable clustering with scikit-learn BIRCH (CF tree). Builds a clustering-feature tree to form subclusters efficiently, optionally performs a final global clustering step to produce k clusters, and outputs subcluster centers, optional per-sample labels, plus an HTML summary report.

### Tool 8: GaussianMixtureModel

**Typ:** Algorithmic-Analysis
**Hyperparameter:** n_components, covariance_type, tol, reg_covar, max_iter, n_init, init_params, random_state, warm_start, compute_labels, standardize, output_probabilities
**Input:** features (CSV)
**Output:** labels (CSV), responsibilities (CSV), weights (CSV), means (CSV), covariances (CSV), report (HTML)
**Description:** Probabilistic clustering with scikit-learn Gaussian Mixture Models (EM). Fits a mixture of Gaussians and outputs hard labels (argmax), soft assignments (responsibilities), and learned model parameters (weights/means/covariances), with optional standardization and an HTML report.

### Tool 9: Dysregnet

**Typ:** Algorithmic-Analysis
**Hyperparameter:** zscoring, bonferroni_alpha, r2_threshold, normaltest, normaltest_alpha, direction_condition, con_cov, cat_cov, con_col
**Input:** grn (CSV), meta (CSV), expr (CSV)
**Output:** adjacency_heatmap (IMAGE), pvalue_distribution (IMAGE), top_interactions (IMAGE), results (CSV)
**Description:** DysRegNet performs patient-specific, confounder-aware inference of dysregulated regulatory interactions from a given gene regulatory network (TF–target edges), expression matrix, and sample metadata. It supports optional z-scoring, multiple-testing control (Bonferroni alpha), model-fit filtering via $R^2$ threshold, residual normality testing, and optional directionality constraints, and exports both result tables and diagnostic/summary plots.

### Tool 10: Spycone

**Typ:** Algorithmic-Analysis
**Hyperparameter:** data_num_species, clustering_algorithm, reps1, timepts, n_clusters
**Input:** data (CSV), gene_ids (CSV)
**Output:** image (IMAGE)
**Description:** Spycone (SPlicing-aware time-COurse Network Enricher) performs gene-level time-course analysis and clustering over replicated time-series expression profiles, supporting multiple clustering algorithms (e.g., hierarchical, kmeans, dbscan/optics) and generating an image-based summary of the enrichment/cluster results.

### Tool 11: MoSBI

**Typ:** Algorithmic-Analysis
**Hyperparameter:** n_randomizations, margin, metric, n_steps, sn_ratio
**Input:** data (CSV)
**Output:** adjacency_heatmap (IMAGE), histograms (IMAGE), network_by_algorithm (IMAGE), network_plot (IMAGE)
**Description:** MoSBi (Molecular Signature identification using Biclustering) builds similarity-based networks from biclustering/signature results. It estimates similarity cut-offs via randomizations, supports multiple similarity metrics and margin strategies, and outputs diagnostic heatmaps/histograms plus network visualizations.

### Tool 12: clusteval

**Typ:** Evaluation
**Hyperparameter:** clustering_algorithm, evaluation_method, linkage, max_clusters
**Input:** data (CSV)
**Output:** cluster_scores (STRING), silhouette_plot (STRING), scatter_plot (STRING)
**Description:** ClustEval evaluates clustering quality across candidate clusterings using methods such as silhouette, Davies–Bouldin index, or derivative-based heuristics. It supports multiple clustering algorithms (agglomerative/kmeans/dbscan/hdbscan), produces evaluation and diagnostic plots, and summarizes clusterability/score trends.

### Tool 13: LogisticRegression

**Typ:** Analysis
**Hyperparameter:** target_column, standardize, C, max_iter, solver, multi_class, class_weight, random_state, explain_top_k
**Input:** data (CSV)
**Output:** explanation (HTML), report (TEXT), predictions (CSV), coefficients (CSV)
**Description:** Trainable scikit-learn Logistic Regression baseline for classification. Trains on a CSV with a specified target/label column (optional id), optionally standardizes numeric features, generates predictions with per-class probabilities, exports model coefficients (incl. intercept), and produces an explainability HTML report showing the top positive/negative coefficients per class.

### Tool 14: Train-Test-Split

**Typ:** Pre-Processing
**Hyperparameter:** shuffle, test_size, random_state
**Input:** input (CSV)
**Output:** train (CSV), test (CSV)
**Description:** Train-Test-Split

### Tool 15: Branch-Merge-Aggregator

**Typ:** Post-Processing
**Hyperparameter:** mode, key, how
**Input:** left (CSV), right (CSV)
**Output:** output (CSV)
**Description:** Merge two branch outputs into one DataFrame

### Tool 16: Comparative-Clustering-Benchmark

**Typ:** Evaluation
**Hyperparameter:** id_column, truth_label_column, sort_by
**Input:** features (CSV), predictions (CSV), ground_truth (CSV, optional)
**Output:** output (CSV)
**Description:** Compare multiple clustering outputs on the same dataset.

### Tool 17: MofaExploration

**Typ:** Algorithmic-Analysis
**Hyperparameter:** n_top_features, export_weights, export_factors, export_top_features
**Input:** -
**Output:** factors (CSV), weights (CSV), top_features (CSV), report (HTML), weights_plot (IMAGE)
**Description:** Explore a trained MOFA/MOFA+ PBMC10K dataset model and export factors, weights, variance explained, and top features

### Tool 18: Bicon

**Typ:** Algorithmic-Analysis
**Hyperparameter:** l_g_min, l_g_max, max_iter
**Input:** net (TSV), expr (CSV)
**Output:** cluster_map (STRING), network (STRING), result (CSV)
**Description:** Perform network-constrained biclustering on gene expression data to identify patient subgroups and associated subnetworks based on molecular interaction networks

### Tool 19: SCANet

**Typ:** Algorithmic-Analysis
**Hyperparameter:** min_cell, threshold, top_n_genes, cell_annotation, num_rep_cells, n_pcs, n_neighbors, method, lowerbound, upperbound, nbins
**Input:** data (PATH)
**Output:** reduced_data (STRING), cell_annotation (STRING), cluster (STRING), highly_variable_genes (STRING), violin_plot (STRING), processed_data (STRING)
**Description:** Perform single-cell RNA-seq analysis with SCANet, including cell filtering, dimensionality reduction, clustering, representative cell selection, and visualization of quality control metrics and highly variable genes.

**Tool 20: SpliceDrift**
**Typ:** Algorithmic-Analysis
**Hyperparameter:** draws, tune, max_refits, target_accept, min_total
**Input:** junction_counts (TSV)
**Output:** results (CSV), scatter_plot (IMAGE)
**Description:** Bayesian Beta-Binomial regression for age-related splicing drift analysis.

## Datasets

- iris: Iris Species
- uci_dataset17: Breast Cancer Wisconsin
- iris_encoded: Iris Species (One hot encoded)
- uci_dataset17_encoded: Breast Cancer Wisconsin (One hot encoded)
- A.n_genes=500,m=4,std=1,overlap=no.exprs_z.: Random expression dataset
- grn.csv: Gene Regulatory Network edge list
- meta.csv: Sample metadata table
- expr.csv: Gene expression matrix
- spycone_data.csv: Time-series data matrix
- spycone_gene_ids.csv: List of gene identifiers
- random_x: random blobs (n_samples=750, centers=4, n_features=2, cluster_std=0.5)
- gse30219_lung.csv: a Non-Small Cell Lung Cancer dataset from GEO for patients with either adenocarcinoma or squamous cell carcinoma.
- biogrid_net.tsv: pan-cancer dataset with patients that have luminal or basal breast cancer
- pbmc3k.h5ad: data consists in 3k PBMCs from a Healthy Donor and is freely available from 10x Genomics
- junction_counts: synthetic age-related splicing drift dataset

## Results

All experiments can be fully reproduced from the repository X. The corresponding artifacts are stored under tools/. For each run, the generated report R(x).pdf is stored in the root directory, while all additional inputs and outputs are contained in a subfolder named after the run ID (i.e., tools/<run_id>/).

**Table 4 Tool evaluation**: Run-level test report index. Each row corresponds to one tool execution (run) and lists the associated report identifier, the executed tool, and the concrete run configuration (hyperparameters and inputs).

| ID | Report | Tool (Tool-ID) | Hyperparameter | Inputs |
|---|---|---|---|---|
| 1 | R1 | UciFetch (1) | dataset_id: 17 | - |
| 2 | R1 | UciFetch (1) | dataset_id: 13 | - |
| 3 | R2 | SpectralClustering (2) | n_clusters: 3, affinity: rbf, gamma: 1.0, n_neighbors: 10, assign_labels: kmeans, n_init: 10, eigen_solver: arpack, random_state: 42, standardize: 1 | features: iris_encoded |
| 4 | R2 | SpectralClustering (2) | n_clusters: 3, affinity: rbf, gamma: 1.0, n_neighbors: 10, assign_labels: kmeans, n_init: 10, eigen_solver: arpack, random_state: 42, standardize: 1 | features: uci_dataset17_encoded |
| 5 | R3 | SqlSelectExtractor (3) | sql_distribution: postgres, host: host.docker.internal, port: 5432, database: demo, username: demo, password: demo, sql_select: “SELECT * FROM patients LIMIT 10” | - |
| 6 | R4 | OneHotEncoding (4) | columns: “”, drop_first: 0, bool_to_number: 1, dummy_na: 0, handle_unknown: ignore, prefix_sep: “__”, keep_original: 0 | features: iris |
| 7 | R4 | OneHotEncoding (4) | columns: “”, drop_first: 0, bool_to_number: 1, dummy_na: 0, handle_unknown: ignore, prefix_sep: “__”, keep_original: 0 | features: uci_dataset17 |
| 8 | R5 | AgglomerativeClustering (5) | linkage: ward, compute_distances: 1, standardize: 1, metric: euclidean, use_distance_threshold: 1, n_clusters: 2, plot_title: “Agglomerative Clustering Dendrogram”, dendrogram_truncate_mode: none, dendrogram_p: 30, distance_threshold: 0.0 | features: iris |
| 9 | R5 | AgglomerativeClustering (5) | linkage: ward, compute_distances: 1, standardize: 1, metric: euclidean, use_distance_threshold: 1, n_clusters: 2, plot_title: “Agglomerative Clustering Dendrogram”, | features: uci_dataset17 |
| | | | dendrogram_truncate_mode : none, dendrogram_p: 30, distance_threshold: 0.0 | |
| 10 | R6 | UnPaSt (6) | binarization_method: kmeans, deep_split: 2, dynamic_tree_cut: 0.5, ceiling: 3, bidirectional: true, seed: 42, similarity_cutoff_mode: automatic, similarity_cutoff_value: 0.5, p_value: 0.05, clustering_method: WGCNA | data: A.n_genes=500,m=4,std=1, overlap=no.exprs_z |
| 11 | R7 | Birch (7) | threshold: 0.5, branching_factor: 50, use_global_clustering: 1, n_clusters: 3, compute_labels: 1, standardize: 1 | features: iris_encoded |
| 12 | R7 | Birch (7) | threshold: 0.5, branching_factor: 50, use_global_clustering: 1, n_clusters: 3, compute_labels: 1, standardize: 1 | features: uci_dataset17_encoded |
| 13 | R8 | GaussianMixtureModel (8) | n_components: 3, covariance_type: full, tol: 0.001, reg_covar: 0.000001, max_iter: 100, n_init: 1, init_params: kmeans, random_state: 42, warm_start: 0, compute_labels: 1, standardize: 1, output_probabilities: 1 | features: iris_encoded |
| 14 | R8 | GaussianMixtureModel (8) | n_components: 3, covariance_type: full, tol: 0.001, reg_covar: 0.000001, max_iter: 100, n_init: 1, init_params: kmeans, random_state: 42, warm_start: 0, compute_labels: 1, | features: uci_dataset17_encoded |
| | | | standardize: 1, output_probabilities: 1 | |
| 15 | R9 | Dysregnet (9) | zscoring: false, bonferroni_alpha: 0.01, r2_threshold: 0.2, normaltest: false, normaltest_alpha: 0.001, direction_condition: false, con_cov: “birth_days_to”, cat_cov: “race,gender”, con_col: “sample_type” | grn: grn<br>meta: meta<br>expr: expr |
| 16 | R10 | Spycone (10) | data_num_species: 9606, clustering_algorithm: hierarchical, reps1: 5, timepts: 9, n_clusters: 10 | data: spycone_data<br>gene_ids: spycone_gene_ids |
| 17 | R11 | Mosbi (11) | n_randomizations: 5, margin: both, metric: Bray-Curtis, n_steps: 100, sn_ratio: true | data: A.n_genes=500,m=4,std=1, overlap=no.exprs_z |
| 18 | R12 | Clustereval (12) | clustering_algorithm: agglomerative, evaluation_method: silhouette, linkage: ward, max_clusters: 25 | data: random_x |
| 19 | R12 | Clustereval (12) | clustering_algorithm: agglomerative, evaluation_method: silhouette, linkage: ward, max_clusters: 25 | data: iris_encoded |
| 20 | R12 | Clustereval (12) | clustering_algorithm: agglomerative, evaluation_method: silhouette, linkage: ward, max_clusters: 25 | data: uci_dataset17_encoded |
| 21 | R13 | LogisticRegression (13)<br><br>Trained with iris as well via PoSyMed training | target_column: target, standardize: 1, C: 1.0, max_iter: 200, solver: lbfgs, multi_class: auto, class_weight: balanced, random_state: 42, explain_top_k: 15 | data: iris |
| 22 | R14 | train-test-split (14) | shuffle: true, test_size: 0.25, random_state: 42 | input: iris |
| 23 | R14 | train-test-split (14) | shuffle: true, test_size: 0.25, random_state: 42 | input: uci_dataset17 |
| 24 | R15 | Branch-Merge-Aggregator (15) | mode: concat_columns, key:, how: inner | left: iris (split)<br>right: iris (split) |
| 25 | R15 | Branch-Merge-Aggregator (15) | mode: concat_rows, key:, how: outer | left: iris (split)<br>right: iris (split) |
| 26 | R15 | Branch-Merge-Aggregator (15) | mode: merge_on_key, key: variety, how: inner | left: iris (split)<br>right: iris (split) |
| 27 | R16 | Comparative-Clustering-Benchmark (16) | id_column: index, truth_label_column: label, sort_by: silhouette | features: uci_dataset17_encoded<br>predictions: labels (output run 14)<br>ground_truth: labels (output run 14) |
| 28 | R16 | Comparative-Clustering-Benchmark (16) | id_column: index, truth_label_column: label, sort_by: calinski_harabasz | features: uci_dataset17_encoded<br>predictions: labels (output run 12)<br>ground_truth: labels (output run 14) |
| 29 | R17 | MofaExploration (17) | n_top_features: 20, export_weights: 1, export_factors: 1, export_top_features: 1 | - |
| 30 | R17 | MofaExploration (17) | n_top_features: 5, export_weights: 1, export_factors: 1, export_top_features: 1 | - |
| 31 | R18 | Bicon (18) | l_g_min: 10, l_g_max: 15, max_iter: 15 | expr :gse30219_lung.csv,<br>net: biogrid_net.tsv |
| 32 | R19 | Scanet (19) | min_cell: 3, threshold: 0.1, top_n_genes: 500, cell_annotation: leiden, num_rep_cells: 50, n_pcs: 10, n_neighbors: 5, method: | data: pbmc3k.h5ad |
| | | | UMAP, lowerbound: 500, upperbound: 5000, nbins: 10 | |
| 33 | R20 | SpliceDrift (20) | draws: 200, tune: 200, max_refits: 1, target_accept: 0.9, min_total: 5 | junction_counts: junction_counts.tsv |

Table 4 documents the tool-centric evaluation index at the run level. Each row corresponds to a single tool execution and specifies the associated report ID, the tool executed, and the specific run configuration, including hyperparameters and inputs.

## Supplementary material 2: Workflow-Experiments

**Table 5 Workflow evaluation**: Run-level test report index. Each row corresponds to one tool execution (run), lists the associated report identifier and the order in the workflow, the executed tool, and the concrete run configuration (hyperparameters and inputs).

| ID | Order | Report | Tool (Tool-ID) | Hyperparameter | Inputs |
|---|---|---|---|---|---|
| 1 | 1 | R1 | UciFetch (1) | dataset_id: 17 | - |
| 2 | 1 | R1 | UciFetch (1) | dataset_id: 107 | - |
| 3 | 1 | R2 | train-test-split (14) | shuffle: true, test_size: 0.25, random_state: 42 | input: iris |
| 4 | 1 | R2 | train-test-split (14) | shuffle: true, test_size: 0.25, random_state: 42 | input: uci_dataset17 |
| 5 | 1 | R3 | UciFetch (1) | dataset_id: 17 | |
| 5 | 2 | R3 | OneHotEncoding (4) | columns: “”, drop_first: 0, bool_to_number: 1, dummy_na: 0, handle_unknown: ignore, prefix_sep: “__”, keep_original: 0 | input: uci_dataset17 |
| 6 | 1 | R4 | OneHotEncoding (4) | columns: "", drop_first: 0, bool_to_number: 1, dummy_na: 0, handle_unknown: ignore, prefix_sep: "__", keep_original: 0 | input: iris |
| 6 | 2 | R4 | AgglomerativeClustering (5) | linkage: ward, compute_distances: 1, standardize: 1, metric: euclidean, use_distance_threshold: 1, n_clusters: 2, plot_title: "Agglomerative Clustering Dendrogram", dendrogram_truncate_mode: none, dendrogram_p: 30, distance_threshold: 0 | features: iris_encoded |
| 7 | 1 | R5 | UciFetch (1) | dataset_id: 17 | - |
| 7 | 2 | R5 | OneHotEncoding (4) | columns: "", drop_first: 0, bool_to_number: 1, dummy_na: 0, handle_unknown: ignore, prefix_sep: "__", keep_original: 0 | input: uci_dataset17 |
| 7 | 3 | R5 | Birch (7) | threshold: 0.5, branching_factor: 50, use_global_clustering: 1, n_clusters: 3, compute_labels: 1, copy: 1, standardize: 1 | input: encoded_uci_dataset17 |
| 7 | 4 | R5 | GaussianMixtureModel (8) | n_components: 3, covariance_type: full, tol: 0.001, reg_covar: 1e-6, max_iter: 100, n_init: 1, init_params: kmeans, random_state: 42, warm_start: 0, compute_labels: 1, standardize: 1, output_probabilities: 1 | input: encoded_uci_dataset17 |
| 8 | 1 | R6 | UciFetch (1) | dataset_id: 17 | - |
| 8 | 2 | R6 | OneHotEncoding (4) | columns: "", drop_first: 0, bool_to_number: 1, | input: uci_dataset17 |
| | | | | dummy_na: 0, handle_unknown: ignore, prefix_sep: "__", keep_original: 0 | |
| 8 | 3 | R6 | AgglomerativeClustering (5) | linkage: ward, compute_distances: 1, standardize: 1, metric: euclidean, use_distance_threshold: 1, n_clusters: 2, plot_title: “Agglomerative Clustering Dendrogram”, dendrogram_truncate_mode: none, dendrogram_p: 30, distance_threshold: 0.0 | features: uci_dataset17 |
| 8 | 4 | R6 | Birch (7) | threshold: 0.5, branching_factor: 50, use_global_clustering: 1, n_clusters: 3, compute_labels: 1, standardize: 1 | features: uci_dataset17_encoded |
| 9 | 1 | R7 | UciFetch (1) | columns: "", drop_first: 0, bool_to_number: 1, dummy_na: 0, handle_unknown: ignore, prefix_sep: "__", keep_original: 0 | - |
| 9 | 2 | R7 | OneHotEncoding (4) | columns: "", drop_first: 0, bool_to_number: 1, dummy_na: 0, handle_unknown: ignore, prefix_sep: "__", keep_original: 0 | input: uci_dataset17 |
| 9 | 3 | R7 | GaussianMixtureModel (8) | columns: "", drop_first: 0, bool_to_number: 1, dummy_na: 0, handle_unknown: ignore, prefix_sep: "__", keep_original: 0 | input: encoded_uci_dataset17 |
| 9 | 4 | R7 | comparative-clustering | id_column: index, truth_label_column: | features: feature table, predictions: clustering |
| | | | -benchmark (16) | label, sort_by: silhouette | predictions, ground_truth: optional |
| 10 | 1 | R8 | UciFetch (1) | columns: "", drop_first: 0, bool_to_number: 1, dummy_na: 0, handle_unknown: ignore, prefix_sep: "__", keep_original: 0 | - |
| 10 | 2 | R8 | OneHotEncoding (4) | columns: "", drop_first: 0, bool_to_number: 1, dummy_na: 0, handle_unknown: ignore, prefix_sep: "__", keep_original: 0 | input: uci_dataset17 |
| 10 | 3 | R8 | Birch (7) | threshold: 0.5, branching_factor: 50, use_global_clustering: 1, n_clusters: 3, compute_labels: 1, copy: 1, standardize: 1 | input: encoded_uci_dataset17 |
| 10 | 4 | R8 | GaussianMixtureModel (8) | columns: "", drop_first: 0, bool_to_number: 1, dummy_na: 0, handle_unknown: ignore, prefix_sep: "__", keep_original: 0 | input: encoded_uci_dataset17 |
| 10 | 5 | R8 | comparative-clustering-benchmark (16) | id_column: index, truth_label_column: label, sort_by: silhouette | features: feature table, predictions: clustering predictions, ground_truth: optional |
| 11 | 1 | R9 | UciFetch (1) | columns: "", drop_first: 0, bool_to_number: 1, dummy_na: 0, handle_unknown: ignore, prefix_sep: "__", keep_original: 0 | - |
| 11 | 2 | R9 | train-test-split (14) | shuffle: 1, test_size: 0.25, random_state: 42 | input: uci_dataset17 |
| 11 | 3 | R9 | OneHotEncoding (4) | columns: "", drop_first: 0, bool_to_number: 1, dummy_na: 0, handle_unknown: ignore, prefix_sep: "__", keep_original: 0 | input: train_b873 |
| 11 | 4 | R9 | OneHotEncoding (4) | columns: "", drop_first: 0, bool_to_number: 1, dummy_na: 0, handle_unknown: ignore, prefix_sep: "__", keep_original: 0 | input: test_23b8 |
| 11 | 5 | R9 | branch-merge-aggregator (15) | mode: concat_rows, key: "", how: inner | left: encoded_af19, right: encoded_ef25 |
| 11 | 6 | R9 | AgglomerativeClustering (5) | linkage: ward, compute_distances: 1, standardize: 1, metric: euclidean, use_distance_threshold: 1, n_clusters: 2, plot_title: "Agglomerative Clustering Dendrogram", dendrogram_truncate_mode: none, dendrogram_p: 30, distance_threshold: 0 | features: iris_encoded |
| 12 | 1 | R10 | UciFetch (1) | columns: "", drop_first: 0, bool_to_number: 1, dummy_na: 0, handle_unknown: ignore, prefix_sep: "__", keep_original: 0 | - |
| 12 | 2 | R10 | OneHotEncoding (4) | columns: "", drop_first: 0, bool_to_number: 1, dummy_na: 0, handle_unknown: ignore, prefix_sep: "__", keep_original: 0 | input: uci_dataset17 |
| 12 | 3 | R10 | train-test-split (14) | shuffle: 1, test_size: 0.25, random_state: 42 | input: uci_dataset17_encoded |
| 12 | 4 | R10 | GaussianMixtureMode (8) | n_components: 3, covariance_type: full, tol: 0.001, reg_covar: 1e-6, max_iter: 100, n_init: 1, init_params: kmeans, random_state: 42, warm_start: 0, compute_labels: 1, standardize: 1, output_probabilities: 1 | input: test_550e |
| 12 | 5 | R10 | GaussianMixtureMode (8) | n_components: 3, covariance_type: full, tol: 0.001, reg_covar: 1e-6, max_iter: 100, n_init: 1, init_params: kmeans, random_state: 42, warm_start: 0, compute_labels: 1, standardize: 1, output_probabilities: 1 | input: train_2d69 |
| 12 | 6 | R10 | branch-merge-aggregator (15) | mode: concat_columns, key: "", how: inner | left: labels, right: labels |

Table 5 documents the workflow-centric evaluation index at the run level. Each row represents a single execution within a workflow and includes references to the corresponding report ID, the order within the workflow, the tool executed, and the specific run configuration with hyperparameters and inputs.

## Supplementary material 3: LLM evaluation

The evaluation of LLM-supported interaction in PoSyMed examines a different level of performance than the deterministic assessment of tool execution and workflow orchestration. The focus is on whether the agent correctly interprets user requests in the context of uploaded files, selects appropriate internal tools, asks targeted follow-up questions when necessary, and derives a comprehensible response from this information. In the main text, this agent evaluation is described as the third evaluation dimension and is specified through a test set comprising 120 questions on six topics as well as 30 human-in-the-loop (HITL) scenarios.

The test dataset is structured as a platform-specific interaction benchmark. In addition to the question, each instance contains the expected tools (neededTools), a functional target description (expectedAnswer), optional HITL answers (hitlAnswer), and the relevant uploaded files (requiredFiles). Thus, the evaluation focuses not only on the final answer but on the entire interaction path between file reference, tool selection, potential clarification, and answer synthesis.
The six topics we created cover core capabilities, including file-based tool selection, the assessment of preprocessing and analysis readiness, and the planning of next steps with explicit file references.

- **File-aware tool discovery and shortlist selection:** tests whether the agent can identify a suitable tool or a meaningful initial selection based on an uploaded file.
- **File-aware preprocessing and readiness assessment:** evaluates whether the agent correctly assesses the need for preprocessing and a file's basic readiness for analysis.
- **File-centric next-step planning with targeted HITL:** tests whether the agent can plan the next sensible step and, in case of uncertainty, ask a targeted follow-up question.
- **Literature-backed file-aware recommendation:** examines whether recommendations can be justified both in terms of the file and from a plausible literature or research perspective.
- **Multi-file comparison and branch selection:** checks whether the agent can compare multiple uploaded files and derive a preferred analysis path from them.
- **Final-answer quality under explicit file grounding:** evaluates whether the final answer is clear, complete, and explicitly grounded in the file context.

The questions are deliberately formulated in natural user language and repeatedly create situations involving partial uncertainty, in which the agent should not speculate but rather ask for clarification in a controlled manner. This corresponds to the architecture described in the paper, in which LLMs act as bound assistants within a formally controlled framework. The uploaded files play a central role as the primary evidence base. These files are downloaded from the public repository[11].

- breast_cancer_wisconsin.csv: A tabular example dataset used in the benchmark for file-related recommendations and path decisions
- dermatology.csv: A tabular input file used specifically for questions regarding preprocessing and interpretable recommendations
- hepatitis.csv: A tabular dataset that the agent is tasked with using to identify preprocessing needs and potential next analysis steps
- iris.csv: A classic tabular example dataset frequently used in the test set for initial tool recommendations and simple clustering scenarios
- wine.csv: A tabular dataset primarily used for interpretable, user-specific recommendations
- zoo.csv: A tabular file used to evaluate tool selection, preprocessing risks, and the agent's limitations

The agent is expected to explicitly base its recommendations on these artifacts.

The table below contains the complete set of questions and expected actions you requested.

[11] https://archive.ics.uci.edu/datasets

| Question | Needed Tools | Expected Agent Action/Answer | Required Files |
|---|---|---|---|
| **File-aware tool discovery and shortlist selection** | | | |
| Given the uploaded iris.csv, which PoSyMed tool should I inspect first for clustering support? | FETCH_DATA, ANALYZE_DATA, FETCH_TOOLS, FINALIZE | The agent should use the uploaded file context, characterize the dataset at a high level, and recommend a suitable clustering-oriented app without claiming execution. | iris.csv |
| I uploaded breast_cancer_wisconsin.csv. Should I start with a simple baseline workflow or with a more biology-aware path? | FETCH_DATA, ANALYZE_DATA, FINALIZE | The agent should analyze the file context, compare a simple baseline path with a more domain-aware path, and recommend one path with justification. | breast_cancer_wisconsin.csv |
| Using wine.csv, what is the most interpretable first PoSyMed app to discuss with a non-technical audience if I want a dataset-level explanation rather than a target-dependent predictive model? | FETCH_DATA, ANALYZE_DATA, FETCH_TOOLS, FINALIZE | The agent should use the file context to recommend an interpretable first app and explain why it | wine.csv |
| | | suits a non-technical audience. | |
| Given the uploaded zoo.csv, which PoSyMed tool should I inspect first for clustering support? | FETCH_DATA, ANALYZE_DATA, FETCH_TOOLS, FINALIZE | The agent should use the uploaded file context, characterize the dataset at a high level, and recommend a suitable clustering-oriented app without claiming execution. | zoo.csv |
| I uploaded hepatitis.csv. Should I start with a simple baseline workflow or with a more biology-aware path? | FETCH_DATA, ANALYZE_DATA, FINALIZE | The agent should analyze the file context, compare a simple baseline path with a more domain-aware path, and recommend one path with justification. | hepatitis.csv |
| Using dermatology.csv, what is the most interpretable first app to discuss with a non-technical audience? | FETCH_DATA, ANALYZE_DATA, FETCH_TOOLS, FINALIZE | The agent should use the file context to recommend an interpretable first app and explain why it suits a non-technical audience. | dermatology.csv |
| I uploaded zoo.csv. Given that it contains mostly binary features, which PoSyMed app family is best suited to reveal hidden groupings? | FETCH_DATA, ANALYZE_DATA, FETCH_TOOLS, FINALIZE | The agent should fetch file context, analyze the binary feature structure, query the tool catalog, and recommend a clustering-friendly app family. | zoo.csv |
| I uploaded hepatitis.csv. Which PoSyMed tool would help me understand feature importance before building a predictive model? | FETCH_DATA, ANALYZE_DATA, FETCH_TOOLS, FINALIZE | The agent should analyze the file, recognize it as a classification dataset with missing values, and recommend a feature-importance or explainability tool. | hepatitis.csv |
| Using dermatology.csv, should I run dimensionality reduction before or after selecting a PoSyMed clustering app? | FETCH_DATA, ANALYZE_DATA, FINALIZE | The agent should analyze the high-dimensional mixed-type dataset and advise on whether dimensionality reduction is a prerequisite or can be handled by the app itself. | dermatology.csv |
| I uploaded iris.csv. Is there a PoSyMed app that can both cluster the data and explain which features drive each cluster? | FETCH_DATA, ANALYZE_DATA, FETCH_TOOLS, FINALIZE | The agent should analyze the dataset, query the catalog for explainable clustering tools, and recommend one that provides feature-level explanations. | iris.csv |
| I uploaded breast_cancer_wisconsin.csv. If my goal is anomaly detection rather than classification, which PoSyMed tool family should I explore? | FETCH_DATA, ANALYZE_DATA, FETCH_TOOLS, FINALIZE | The agent should analyze the file, note it is typically used for classification but reframe for anomaly detection, and recommend a suitable app family. | breast_cancer_wisconsin.csv |
| Using wine.csv, I want to find natural groupings without using the class label. Which PoSyMed app should I try first? | FETCH_DATA, ANALYZE_DATA, FETCH_TOOLS, FINALIZE | The agent should recognize this as an unsupervised clustering task on numeric features and recommend an appropriate clustering app from the catalog. | wine.csv |
| I uploaded zoo.csv. Can you compare two different PoSyMed app families that could handle this categorical dataset and explain the trade-offs? | FETCH_DATA, ANALYZE_DATA, FETCH_TOOLS, FINALIZE | The agent should analyze the dataset structure, retrieve catalog info on at least two app families, and compare their suitability for categorical data. | zoo.csv |
| I uploaded hepatitis.csv. Given the missing values, which PoSyMed workflow path would you recommend: impute-then-model or a tool that handles missingness natively? | FETCH_DATA, ANALYZE_DATA, FETCH_TOOLS, FINALIZE | The agent should detect missing values in the analysis, compare imputation-based vs native-missingness approaches in the catalog, and recommend a path. | hepatitis.csv |
| Using dermatology.csv, which PoSyMed app would be most useful for a clinician who wants to understand subgroup patterns across skin disease types? | FETCH_DATA, ANALYZE_DATA, FETCH_TOOLS, FINALIZE | The agent should analyze the multi-class clinical dataset, query tools suited for subgroup discovery, and recommend one with clinical interpretability. | dermatology.csv |
| I uploaded iris.csv. Before recommending any app path, ask me whether I care more about explanation or subgroup discovery, and do not choose yet. | FETCH_DATA, ANALYZE_DATA, HUMAN_IN_THE_LOOP, FETCH_TOOLS, FINALIZE | The agent should detect the decision ambiguity, ask one targeted clarification question, then recommend a file-aware app path after the user reply. | iris.csv |
| I uploaded breast_cancer_wisconsin.csv. Before recommending a clinically understandable workflow, ask whether I should favor robustness or exploratory breadth. | FETCH_DATA, ANALYZE_DATA, HUMAN_IN_THE_LOOP, FETCH_TOOLS, FINALIZE | The agent should detect the decision ambiguity, ask one targeted clarification question, then recommend a file-aware app path after the user reply. | breast_cancer_wisconsin.csv |
| I uploaded wine.csv. Before recommending an app path, ask whether this should be treated as a teaching demo or a serious benchmark baseline. | FETCH_DATA, ANALYZE_DATA, HUMAN_IN_THE_LOOP, FETCH_TOOLS, FINALIZE | The agent should detect the decision ambiguity, ask one targeted clarification question, then recommend a file-aware app path after the user reply. | wine.csv |
| I uploaded hepatitis.csv. Before recommending an app path, ask whether I want the simplest workflow or one that better reflects biomedical reasoning. | FETCH_DATA, ANALYZE_DATA, HUMAN_IN_THE_LOOP, FETCH_TOOLS, FINALIZE | The agent should detect the decision ambiguity, ask one targeted clarification question, then recommend a file-aware app path after the user reply. | hepatitis.csv |
| I uploaded dermatology.csv. Before recommending anything, ask whether I want a comparison of several tools or one well-justified first choice. | FETCH_DATA, ANALYZE_DATA, HUMAN_IN_THE_LOOP, FETCH_TOOLS, FINALIZE | The agent should detect the decision ambiguity, ask one targeted clarification question, then recommend a file-aware app path after the user reply. | dermatology.csv |
| **File-aware preprocessing and readiness assessment** | | | |
| I uploaded breast_cancer_wisconsin.csv. Before discussing clustering apps, what preprocessing issue should we discuss first? | FETCH_DATA, ANALYZE_DATA, FINALIZE | The agent should infer the most important likely preprocessing need from the file and propose it as the first discussion point. | breast_cancer_wisconsin.csv |
| For the uploaded wine.csv, would OneHotEncoding likely be relevant before clustering support? | FETCH_DATA, ANALYZE_DATA, FINALIZE | The agent should assess whether categorical encoding is likely relevant and explain the reasoning. | wine.csv |
| I uploaded zoo.csv. Which preprocessing mistake should I avoid before selecting a clustering app? | FETCH_DATA, ANALYZE_DATA, FINALIZE | The agent should identify a plausible preprocessing risk grounded in the file characteristics and explain it clearly. | zoo.csv |
| I uploaded hepatitis.csv. Before discussing clustering apps, what preprocessing issue should we discuss first? | FETCH_DATA, ANALYZE_DATA, FINALIZE | The agent should infer the most important likely preprocessing need from the file and propose it as the first discussion point. | hepatitis.csv |
| For the uploaded dermatology.csv, would OneHotEncoding likely be relevant before clustering support? | FETCH_DATA, ANALYZE_DATA, FINALIZE | The agent should assess whether categorical encoding is likely relevant and explain the reasoning. | dermatology.csv |
| I uploaded iris.csv. Which preprocessing mistake should I avoid before selecting a clustering app? | FETCH_DATA, ANALYZE_DATA, FINALIZE | The agent should identify a plausible preprocessing risk grounded in the file characteristics and explain it clearly. | iris.csv |
| I uploaded wine.csv. Are there any constant or near-constant columns that should be removed before analysis? | FETCH_DATA, ANALYZE_DATA, FINALIZE | The agent should fetch and analyze the file to check for low-variance or constant columns and advise on removal. | wine.csv |
| I uploaded zoo.csv. Since most features are binary, would standardization or normalization still be necessary before clustering? | FETCH_DATA, ANALYZE_DATA, FINALIZE | The agent should analyze the binary-dominated feature set and explain that scaling binary features is unnecessary or even harmful. | zoo.csv |
| I uploaded dermatology.csv. How many missing values are there and what imputation strategy would you recommend? | FETCH_DATA, ANALYZE_DATA, FINALIZE | The agent should quantify missing values and recommend an appropriate imputation strategy given | dermatology.csv |
| | | the mixed feature types. | |
| I uploaded iris.csv. Is the class distribution balanced enough to proceed directly with a classification workflow? | FETCH_DATA, ANALYZE_DATA, FINALIZE | The agent should check class balance in iris and confirm it is balanced (50/50/50), safe to proceed without resampling. | iris.csv |
| I uploaded breast_cancer_wisconsin.csv. Are there highly correlated feature pairs that could cause issues with certain models? | FETCH_DATA, ANALYZE_DATA, FINALIZE | The agent should analyze correlations among the 30 numeric features and flag highly correlated pairs, advising on whether to remove or keep them. | breast_cancer_wisconsin.csv |
| I uploaded hepatitis.csv. Is the target variable suitable for a binary classification workflow or does it need recoding? | FETCH_DATA, ANALYZE_DATA, FINALIZE | The agent should inspect the target column, identify its encoding (DIE/LIVE), and advise if recoding is needed. | hepatitis.csv |
| Using wine.csv, should I apply log transformation to any of the features before running a clustering app? | FETCH_DATA, ANALYZE_DATA, FINALIZE | The agent should analyze feature distributions for skewness and recommend log transformation where appropriate. | wine.csv |
| I uploaded zoo.csv. The legs column is numeric while others are binary. Should I treat it differently in preprocessing? | FETCH_DATA, ANALYZE_DATA, FINALIZE | The agent should recognize the mixed-type issue and recommend appropriate preprocessing for the numeric legs column alongside binary features. | zoo.csv |
| I uploaded dermatology.csv. Which columns contain ordinal data and should they be encoded differently from nominal columns? | FETCH_DATA, ANALYZE_DATA, FINALIZE | The agent should distinguish ordinal from nominal features in the dermatology dataset and advise on appropriate encoding. | dermatology.csv |
| I uploaded zoo.csv. Before proposing preprocessing support, ask whether I should prioritize categorical handling or just start with a simple baseline. | FETCH_DATA, ANALYZE_DATA, HUMAN_IN_THE_LOOP, FETCH_TOOLS, FINALIZE | The agent should ask one decisive clarification when preprocessing priorities are ambiguous, then tailor the preprocessing support path. | zoo.csv |
| I uploaded breast_cancer_wisconsin.csv. Before proposing preprocessing support, ask whether missing-value handling or model selection is the more urgent discussion. | FETCH_DATA, ANALYZE_DATA, HUMAN_IN_THE_LOOP, FETCH_TOOLS, FINALIZE | The agent should ask one decisive clarification when preprocessing priorities are ambiguous, then tailor the preprocessing support path. | breast_cancer_wisconsin.csv |
| I uploaded hepatitis.csv and the file may be messy. Ask me exactly one short clarification before proposing any preprocessing support, and wait for my answer. | FETCH_DATA, ANALYZE_DATA, HUMAN_IN_THE_LOOP, FETCH_TOOLS, FINALIZE | The agent should ask one decisive clarification when preprocessing priorities are ambiguous, then tailor the preprocessing support path. | hepatitis.csv |
| I uploaded dermatology.csv. Before proposing preprocessing support, ask whether I should focus first on normalization assumptions or on explainability. | FETCH_DATA, ANALYZE_DATA, HUMAN_IN_THE_LOOP, FETCH_TOOLS, FINALIZE | The agent should ask one decisive clarification when preprocessing priorities are ambiguous, then tailor the preprocessing support path. | dermatology.csv |
| I uploaded wine.csv and need a lightweight preprocessing path for a classroom demo rather than an exhaustive production workflow. | FETCH_DATA, ANALYZE_DATA, FINALIZE | The agent should propose a lightweight preprocessing path grounded in the file and tailored to a classroom demo. | wine.csv |
| **File-centric next-step planning with targeted HITL** | | | |
| Analyze the uploaded wine.csv and tell me whether it better supports a simple clustering discussion or a more specialized biomedical discussion. | FETCH_DATA, ANALYZE_DATA, FINALIZE | The agent should analyze the file and compare two plausible discussion paths, then recommend one. | wine.csv |
| Using the uploaded zoo.csv, what should be the next support step after file inspection? | FETCH_DATA, ANALYZE_DATA, FINALIZE | The agent should inspect the file context and recommend one concrete | zoo.csv |
| | | next support step. | |
| I uploaded hepatitis.csv. Does this file look ready for direct app selection, or should PoSyMed inspect more before recommending tools? | FETCH_DATA, ANALYZE_DATA, FINALIZE | The agent should assess whether the file is ready for direct app guidance or whether more intermediate inspection is needed. | hepatitis.csv |
| Analyze the uploaded dermatology.csv and tell me whether it better supports a simple clustering discussion or a more specialized biomedical discussion. | FETCH_DATA, ANALYZE_DATA, FINALIZE | The agent should analyze the file and compare two plausible discussion paths, then recommend one. | dermatology.csv |
| Using the uploaded iris.csv, what should be the next support step after file inspection? | FETCH_DATA, ANALYZE_DATA, FINALIZE | The agent should inspect the file context and recommend one concrete next support step. | iris.csv |
| I uploaded breast_cancer_wisconsin.csv. Does this file look ready for direct app selection, or should PoSyMed inspect more before recommending tools? | FETCH_DATA, ANALYZE_DATA, FINALIZE | The agent should assess whether the file is ready for direct app guidance or whether more intermediate inspection is needed. | breast_cancer_wisconsin.csv |
| I uploaded iris.csv. After inspecting the file, should the next step be feature selection or directly choosing a clustering app? | FETCH_DATA, ANALYZE_DATA, FINALIZE | The agent should analyze the low-dimensional clean dataset and recommend skipping feature selection in favor of direct app selection. | iris.csv |
| I uploaded breast_cancer_wisconsin.csv. What should be the priority: handling feature correlations or selecting a baseline model? | FETCH_DATA, ANALYZE_DATA, FINALIZE | The agent should weigh the impact of correlated features and recommend a practical next step. | breast_cancer_wisconsin.csv |
| Using wine.csv, is the dataset ready for a direct comparison of two PoSyMed clustering apps, or do I need a preparation step first? | FETCH_DATA, ANALYZE_DATA, FETCH_TOOLS, FINALIZE | The agent should assess readiness and recommend whether preparation is needed before running clustering app comparison. | wine.csv |
| I uploaded zoo.csv. Given the binary nature of most features, what is the logical next step before selecting a tool? | FETCH_DATA, ANALYZE_DATA, FINALIZE | The agent should recognize the binary features and recommend an appropriate next step such as distance | zoo.csv |
| | | metric selection. | |
| I uploaded hepatitis.csv. Should I address the missing data first, or can I select a PoSyMed app that handles missingness? | FETCH_DATA, ANALYZE_DATA, FETCH_TOOLS, FINALIZE | The agent should analyze missingness patterns and check the catalog for tools that handle missing data natively. | hepatitis.csv |
| Using dermatology.csv, after the initial file inspection, should I merge similar diagnostic categories before proceeding? | FETCH_DATA, ANALYZE_DATA, FINALIZE | The agent should analyze the multi-class structure and advise on whether merging categories makes sense for the user's goal. | dermatology.csv |
| I uploaded iris.csv. After analyzing the file, plan a two-step workflow: what should step one and step two be? | FETCH_DATA, ANALYZE_DATA, FETCH_TOOLS, FINALIZE | The agent should propose a concrete two-step plan grounded in the file analysis and available tools. | iris.csv |
| I uploaded breast_cancer_wisconsin.csv. Is this dataset complex enough to justify a multi-step workflow, or can I go straight to a single app? | FETCH_DATA, ANALYZE_DATA, FINALIZE | The agent should evaluate dataset complexity and recommend single-step or multi-step | breast_cancer_wisconsin.csv |
| | | based on the analysis. | |
| Using wine.csv, the file looks clean. Should the next step focus on exploratory visualization or jump directly to model selection? | FETCH_DATA, ANALYZE_DATA, FINALIZE | The agent should consider the dataset properties and recommend whether exploration or direct modeling is the better next step. | wine.csv |
| I uploaded iris.csv and want you to inspect the file first, but before you recommend the next step ask whether I care more about teaching value or methodological depth. | FETCH_DATA, ANALYZE_DATA, HUMAN_IN_THE_LOOP, FETCH_TOOLS, FINALIZE | The agent should analyze the uploaded file, detect the decisive ambiguity, ask one clarification, and then recommend the next step. | iris.csv |
| I uploaded breast_cancer_wisconsin.csv. Before recommending the next step, ask whether the audience is clinical or technical. | FETCH_DATA, ANALYZE_DATA, HUMAN_IN_THE_LOOP, FETCH_TOOLS, FINALIZE | The agent should analyze the uploaded file, detect the decisive ambiguity, ask one clarification, and then recommend the next step. | breast_cancer_wisconsin.csv |
| I uploaded wine.csv. Before recommending the next step, ask whether I want a baseline comparison or one recommended app. | FETCH_DATA, ANALYZE_DATA, HUMAN_IN_THE_LOOP, FETCH_TOOLS, FINALIZE | The agent should analyze the uploaded file, detect the decisive ambiguity, ask one clarification, and then recommend the next step. | wine.csv |
| I uploaded hepatitis.csv. Before recommending the next step, ask whether I care more about explanation or exploratory signal hunting. | FETCH_DATA, ANALYZE_DATA, HUMAN_IN_THE_LOOP, FETCH_TOOLS, FINALIZE | The agent should analyze the uploaded file, detect the decisive ambiguity, ask one clarification, and then recommend the next step. | hepatitis.csv |
| I uploaded dermatology.csv. Before recommending the next step, ask whether I want to emphasize data inspection or tool discovery. | FETCH_DATA, ANALYZE_DATA, HUMAN_IN_THE_LOOP, FETCH_TOOLS, FINALIZE | The agent should analyze the uploaded file, detect the decisive ambiguity, ask one clarification, and then recommend the next step. | dermatology.csv |
| **Literature-backed file-aware recommendation** | | | |
| I uploaded zoo.csv. Can you use the file context to formulate a literature-backed recommendation for the next workflow step? | FETCH_DATA, ANALYZE_DATA, RESEARCH_PAPERS, FINALIZE | The agent should use the uploaded file to shape a focused literature-backed recommendation. | zoo.csv |
| Given the uploaded hepatitis.csv, based on the file characteristics, what literature question would best support the next workflow decision? | FETCH_DATA, ANALYZE_DATA, FINALIZE | The agent should derive a focused literature query from the file context and the implied workflow decision. | hepatitis.csv |
| I uploaded dermatology.csv. Using the file context and a brief literature check, should literature search come before or after app shortlist discussion? | FETCH_DATA, ANALY_DATA, RESEARCH_PAPERS, FINALIZE | The agent should decide whether literature is warranted at this point and justify the ordering. | dermatology.csv |
| I uploaded iris.csv. Can you use the file context to formulate a literature-backed recommendation for the next workflow step? | FETCH_DATA, ANALYZE_DATA, RESEARCH_PAPERS, FINALIZE | The agent should use the uploaded file to shape a focused literature-backed recommendation. | iris.csv |
| Given the uploaded breast_cancer_wisconsin.csv, based on the file characteristics, what literature question would best support the next workflow decision? | FETCH_DATA, ANALYZE_DATA, FINALIZE | The agent should derive a focused literature query from the file context and the implied | breast_cancer_wisconsin.csv |
| | | workflow decision. | |
| I uploaded wine.csv. Using the file context and a brief literature check, should literature search come before or after app shortlist discussion? | FETCH_DATA, ANALYZE_DATA, RESEARCH_PAPERS, FINALIZE | The agent should decide whether literature is warranted at this point and justify the ordering. | wine.csv |
| I uploaded iris.csv. Is there published work suggesting that certain clustering algorithms work particularly well for small, clean numeric datasets like this? | FETCH_DATA, ANALYZE_DATA, RESEARCH_PAPERS, FINALIZE | The agent should analyze the dataset, search literature for clustering benchmarks on small numeric datasets, and tie findings to a PoSyMed recommendation. | iris.csv |
| I uploaded breast_cancer_wisconsin.csv. Based on the literature, which feature selection approach is most validated for this type of biomedical tabular data? | FETCH_DATA, ANALYZE_DATA, RESEARCH_PAPERS, FINALIZE | The agent should analyze the file, search for literature on feature selection for breast cancer datasets, and recommend a validated approach. | breast_cancer_wisconsin.csv |
| Using wine.csv, does the literature support using MOFA or similar multi-omics tools on simple food science datasets? | FETCH_DATA, ANALYZE_DATA, RESEARCH_PAPERS, FINALIZE | The agent should assess whether multi-omics tools are appropriate for this dataset based on literature and the file's characteristics. | wine.csv |
| I uploaded zoo.csv. What does the literature say about handling binary-dominant datasets for taxonomic classification? | FETCH_DATA, ANALYZE_DATA, RESEARCH_PAPERS, FINALIZE | The agent should analyze the binary features, search for relevant literature on binary feature classification, and provide grounded advice. | zoo.csv |
| Given hepatitis.csv, is there evidence in the literature that certain imputation strategies work better for clinical datasets with this pattern of missingness? | FETCH_DATA, ANALYZE_DATA, RESEARCH_PAPERS, FINALIZE | The agent should analyze the missing data pattern, find literature on imputation for clinical data, and recommend an evidence-based strategy. | hepatitis.csv |
| I uploaded dermatology.csv. Does recent literature recommend any specific approach for multi-class skin disease classification from tabular features? | FETCH_DATA, ANALYZE_DATA, RESEARCH_PAPERS, FINALIZE | The agent should analyze the multi-class clinical dataset, find relevant literature, and recommend an approach validated for dermatological classification. | dermatology.csv |
| I uploaded wine.csv. From a literature perspective, is PCA a good preprocessing step before clustering on this kind of chemical analysis data? | FETCH_DATA, ANALYZE_DATA, RESEARCH_PAPERS, FINALIZE | The agent should analyze the dataset, check literature on PCA for chemical datasets, and give an evidence-based recommendation. | wine.csv |
| Given the uploaded hepatitis.csv, what does the literature suggest about the minimum sample size needed for reliable classification with this many features? | FETCH_DATA, ANALYZE_DATA, RESEARCH_PAPERS, FINALIZE | The agent should analyze the feature-to-sample ratio, find literature on sample size requirements, and advise on feasibility. | hepatitis.csv |
| I uploaded breast_cancer_wisconsin.csv. Is there literature evidence that ensemble methods outperform single models on this type of medical dataset? | FETCH_DATA, ANALYZE_DATA, RESEARCH_PAPERS, FINALIZE | The agent should analyze the dataset, search literature comparing ensemble vs single models for breast cancer data, and recommend accordingly. | breast_cancer_wisconsin.csv |
| I uploaded breast_cancer_wisconsin.csv. Before you search papers or recommend anything, ask whether I care more about clinical explainability or methodological novelty. | FETCH_DATA, ANALYZE_DATA, HUMAN_IN_THE_LOOP, RESEARCH_PAPERS, FINALIZE | The agent should ask one high-value clarification that changes the literature framing, then provide a file-aware literature-backed recommendation. | breast_cancer_wisconsin.csv |
| I uploaded iris.csv. Before you frame the literature angle, ask whether this is a classroom demo or a methods benchmark. | FETCH_DATA, ANALYZE_DATA, HUMAN_IN_THE_LOOP, RESEARCH_PAPERS, FINALIZE | The agent should ask one high-value clarification that changes the literature framing, then provide a file-aware literature-backed recommendation. | iris.csv |
| I uploaded hepatitis.csv and want a literature-backed answer, but before you search papers ask whether I want disease-specific evidence or generic workflow evidence. | FETCH_DATA, ANALYZE_DATA, HUMAN_IN_THE_LOOP, RESEARCH_PAPERS, FINALIZE | The agent should ask one high-value clarification that changes the literature framing, then provide a file-aware literature-backed recommendation. | hepatitis.csv |
| I uploaded dermatology.csv. Before you frame the literature recommendation, ask whether I prioritize interpretability or robustness. | FETCH_DATA, ANALYZE_DATA, HUMAN_IN_THE_LOOP, RESEARCH_PAPERS, FINALIZE | The agent should ask one high-value clarification that changes the literature framing, then provide a file-aware literature-backed recommendation. | dermatology.csv |
| I uploaded wine.csv. Before you frame the literature need, ask whether I am comparing app families or just justifying one chosen baseline. | FETCH_DATA, ANALYZE_DATA, HUMAN_IN_THE_LOOP, RESEARCH_PAPERS, FINALIZE | The agent should ask one high-value clarification that changes the literature framing, then provide a file-aware literature-backed recommendation. | wine.csv |
| **Multi-file comparison and branch selection** | | | |
| I uploaded both iris.csv and wine.csv. Which file should PoSyMed use as the primary basis for the next recommendation? | FETCH_DATA, ANALYZE_DATA, FINALIZE | The agent should compare the uploaded files and select the stronger primary basis for the next recommendation. | iris.csv, wine.csv |
| With both breast_cancer_wisconsin.csv and hepatitis.csv uploaded, should the workflow branch into two possible app paths or stay on one path? | FETCH_DATA, ANALYZE_DATA, FINALIZE | The agent should compare the files and decide whether a single path or branching guidance is more appropriate. | breast_cancer_wisconsin.csv, hepatitis.csv |
| I uploaded dermatology.csv and zoo.csv. Which file is safer for a first demonstration and which file is better for a richer discussion? | FETCH_DATA, ANALYZE_DATA, FINALIZE | The agent should assign the two uploaded files to safer versus richer discussion roles with justification. | dermatology.csv, zoo.csv |
| I uploaded both iris.csv and breast_cancer_wisconsin.csv. Which file should PoSyMed use as the primary basis for the next recommendation? | FETCH_DATA, ANALYZE_DATA, FINALIZE | The agent should compare the uploaded files and select the stronger primary basis for the next recommendation. | iris.csv, breast_cancer_wisconsin.csv |
| With both wine.csv and hepatitis.csv uploaded, should the workflow branch into two possible app paths or stay on one path? | FETCH_DATA, ANALYZE_DATA, FINALIZE | The agent should compare the files and decide whether a single path or branching guidance is more appropriate. | wine.csv, hepatitis.csv |
| I uploaded dermatology.csv and breast_cancer_wisconsin.csv. Which file is safer for a first demonstration and which file is better for a richer discussion? | FETCH_DATA, ANALYZE_DATA, FINALIZE | The agent should assign the two uploaded files to safer versus richer discussion roles with justification. | dermatology.csv, breast_cancer_wisconsin.csv |
| I uploaded iris.csv and hepatitis.csv. Which dataset is more suitable for a first-time user to explore PoSyMed's clustering capabilities? | FETCH_DATA, ANALYZE_DATA, FINALIZE | The agent should compare both datasets on complexity, missing values, and suitability for beginners, then recommend one. | iris.csv, hepatitis.csv |
| I uploaded wine.csv and zoo.csv. One is fully numeric, the other mostly binary. Which one would better demonstrate a typical PoSyMed workflow? | FETCH_DATA, ANALYZE_DATA, FINALIZE | The agent should contrast the two datasets and recommend which better showcases a typical workflow based on feature types. | wine.csv, zoo.csv |
| I uploaded dermatology.csv and hepatitis.csv. Both are clinical datasets. Which one has fewer data quality issues and is safer to model first? | FETCH_DATA, ANALYZE_DATA, FINALIZE | The agent should compare data quality metrics (missing values, feature types) and recommend the cleaner dataset. | dermatology.csv, hepatitis.csv |
| I uploaded iris.csv and dermatology.csv. If I only have time for one analysis, which dataset would yield more actionable insights with a single PoSyMed app? | FETCH_DATA, ANALYZE_DATA, FINALIZE | The agent should compare the potential insight value of each dataset given a single-app constraint. | iris.csv, dermatology.csv |
| I uploaded breast_cancer_wisconsin.csv and wine.csv. Should they share the same preprocessing pipeline or do they need separate handling? | FETCH_DATA, ANALYZE_DATA, FINALIZE | The agent should compare feature characteristics and recommend whether a shared or separate preprocessing pipeline is appropriate. | breast_cancer_wisconsin.csv, wine.csv |
| I uploaded zoo.csv and iris.csv. Which dataset would be a better candidate for a PoSyMed subgroup discovery workflow? | FETCH_DATA, ANALYZE_DATA, FETCH_TOOLS, FINALIZE | The agent should analyze both for subgroup discovery suitability, considering feature types and class structure. | zoo.csv, iris.csv |
| I uploaded hepatitis.csv and wine.csv. Can both datasets be analyzed with the same PoSyMed app, or do they need different tools? | FETCH_DATA, ANALYZE_DATA, FETCH_TOOLS, FINALIZE | The agent should compare both datasets, check tool catalog compatibility, and advise if one tool handles both or if separate tools are needed. | hepatitis.csv, wine.csv |
| I uploaded breast_cancer_wisconsin.csv and zoo.csv. Which dataset is more appropriate for demonstrating feature importance analysis in PoSyMed? | FETCH_DATA, ANALYZE_DATA, FINALIZE | The agent should compare which dataset better demonstrates feature importance given their different feature structures. | breast_cancer_wisconsin.csv, zoo.csv |
| I uploaded dermatology.csv and wine.csv. If my goal is unsupervised exploration, which dataset provides a richer analysis opportunity? | FETCH_DATA, ANALYZE_DATA, FINALIZE | The agent should compare both for unsupervised analysis potential considering dimensionality, feature types, and natural structure. | dermatology.csv, wine.csv |
| I uploaded iris.csv and breast_cancer_wisconsin.csv, but you should first ask whether I want the easier demo or the more clinically relevant demo. | FETCH_DATA, ANALYZE_DATA, HUMAN_IN_THE_LOOP, FETCH_TOOLS, FINALIZE | The agent should compare the uploaded files, ask one clarification that changes the primary-file choice, then recommend the next workflow path. | iris.csv, breast_cancer_wisconsin.csv |
| I uploaded wine.csv and hepatitis.csv, but the choice depends on whether I value cleaner structure or stronger biomedical framing. | FETCH_DATA, ANALYZE_DATA, HUMAN_IN_THE_LOOP, FETCH_TOOLS, FINALIZE | The agent should compare the uploaded files, ask one clarification that changes the primary-file choice, then recommend the next workflow path. | wine.csv, hepatitis.csv |
| I uploaded dermatology.csv and zoo.csv, but you need to ask whether interpretability or categorical richness should dominate the recommendation. | FETCH_DATA, ANALYZE_DATA, HUMAN_IN_THE_LOOP, FETCH_TOOLS, FINALIZE | The agent should compare the uploaded files, ask one clarification that changes the primary-file choice, then recommend the next workflow path. | dermatology.csv, zoo.csv |
| I uploaded iris.csv and wine.csv, but the next step depends on whether I want a teaching path or a benchmark path. | FETCH_DATA, ANALYZE_DATA, HUMAN_IN_THE_LOOP, FETCH_TOOLS, FINALIZE | The agent should compare the uploaded files, ask one clarification that changes the primary-file choice, then recommend the next workflow path. | iris.csv, wine.csv |
| I uploaded breast_cancer_wisconsin.csv and dermatology.csv, but you should ask which audience I have in mind before choosing the primary file. | FETCH_DATA, ANALYZE_DATA, HUMAN_IN_THE_LOOP, FETCH_TOOLS, FINALIZE | The agent should compare the uploaded files, ask one clarification that changes the primary-file choice, then recommend the next workflow path. | breast_cancer_wisconsin.csv, dermatology.csv |
| **Final-answer quality under explicit file grounding** | | | |
| I uploaded hepatitis.csv. Give me a final recommendation that is grounded in the file, concise, and easy to judge for completeness. | FETCH_DATA, ANALYZE_DATA, FINALIZE | The agent should provide a concise final recommendation grounded in the uploaded file and structured for easy evaluation. | hepatitis.csv |
| Using the uploaded dermatology.csv, provide a final answer that includes a recommendation, one alternative, and one explicit uncertainty. | FETCH_DATA, ANALYZE_DATA, FINALIZE | The agent should provide a balanced final answer with recommendation, alternative, and uncertainty, all grounded in the file context. | dermatology.csv |
| I uploaded iris.csv. Explain the next step and also make clear what the agent cannot do. | FETCH_DATA, ANALYZE_DATA, FINALIZE | The agent should state the next step, explain the boundary of its capabilities, and remain grounded in the file context. | iris.csv |
| I uploaded breast_cancer_wisconsin.csv. Give me a final recommendation that is grounded in the file, concise, and easy to judge for completeness. | FETCH_DATA, ANALYZE_DATA, FINALIZE | The agent should provide a concise final recommendation grounded in the uploaded file and structured for easy evaluation. | breast_cancer_wisconsin.csv |
| Using the uploaded wine.csv, provide a final answer that includes a recommendation, one alternative, and one explicit uncertainty. | FETCH_DATA, ANALYZE_DATA, FINALIZE | The agent should provide a balanced final answer with recommendation, alternative, | wine.csv |
| | | and uncertainty, all grounded in the file context. | |
| I uploaded zoo.csv. Explain the next step and also make clear what the agent cannot do. | FETCH_DATA, ANALYZE_DATA, FINALIZE | The agent should state the next step, explain the boundary of its capabilities, and remain grounded in the file context. | zoo.csv |
| I uploaded wine.csv. Summarize the key dataset characteristics and give a concrete first action I should take in PoSyMed. | FETCH_DATA, ANALYZE_DATA, FINALIZE | The agent should provide a grounded summary of wine.csv characteristics and recommend a specific, actionable first step. | wine.csv |
| I uploaded zoo.csv. Give me a final answer that clearly states what the agent can do for me with this file and what I would need to do manually. | FETCH_DATA, ANALYZE_DATA, FINALIZE | The agent should clearly separate automated capabilities from manual steps the user needs to take. | zoo.csv |
| I uploaded hepatitis.csv. Provide a recommendation that acknowledges the data quality issues and explains how they affect the recommended workflow. | FETCH_DATA, ANALYZE_DATA, FINALIZE | The agent should ground the answer in specific data quality findings and explain their impact on the workflow choice. | hepatitis.csv |
| I uploaded iris.csv. Give a final recommendation that a biology student with no ML background could follow step by step. | FETCH_DATA, ANALYZE_DATA, FINALIZE | The agent should provide a beginner-friendly, step-by-step recommendation grounded in the iris dataset properties. | iris.csv |
| I uploaded breast_cancer_wisconsin.csv. Provide a final answer that includes both a recommended PoSyMed workflow and the specific limitations of that approach for this dataset. | FETCH_DATA, ANALYZE_DATA, FETCH_TOOLS, FINALIZE | The agent should recommend a workflow and explicitly state its limitations given the dataset properties. | breast_cancer_wisconsin.csv |
| I uploaded dermatology.csv. Give a final recommendation and explicitly cite which file characteristics informed your choice. | FETCH_DATA, ANALYZE_DATA, FINALIZE | The agent should tie the recommendation directly to observed file characteristics like feature count, types, and missing values. | dermatology.csv |
| I uploaded iris.csv. Produce a final answer that compares two possible approaches and argues for one, grounded in the dataset profile. | FETCH_DATA, ANALYZE_DATA, FINALIZE | The agent should present two viable approaches, compare them based on the file analysis, and recommend one with justification. | iris.csv |
| I uploaded hepatitis.csv. Give me a one-paragraph final answer that covers: dataset readiness, recommended tool, and expected next steps. | FETCH_DATA, ANALYZE_DATA, FINALIZE | The agent should deliver a concise but complete answer covering all three requested aspects. | hepatitis.csv |
| I uploaded breast_cancer_wisconsin.csv. Give a final answer and explicitly state which of the 30 features are most relevant to your recommendation. | FETCH_DATA, ANALYZE_DATA, FINALIZE | The agent should reference specific features from the dataset analysis in its recommendation. | breast_cancer_wisconsin.csv |
| I uploaded breast_cancer_wisconsin.csv and want a final recommendation, but you should first ask whether the audience is clinical or technical. | FETCH_DATA, ANALYZE_DATA, HUMAN_IN_THE_LOOP, FETCH_TOOLS, FINALIZE | The agent should ask one audience or preference clarification before generating the final answer, then produce a grounded and comprehensible | breast_cancer_wisconsin.csv |
| | | recommendation. | |
| I uploaded iris.csv and want a compact final answer, but you should first ask whether I value teaching clarity or stronger benchmarking logic. | FETCH_DATA, ANALYZE_DATA, HUMAN_IN_THE_LOOP, FETCH_TOOLS, FINALIZE | The agent should ask one audience or preference clarification before generating the final answer, then produce a grounded and comprehensible recommendation. | iris.csv |
| I uploaded hepatitis.csv and want a final answer, but you should first ask whether I need one tool recommendation or a comparative shortlist. | FETCH_DATA, ANALYZE_DATA, HUMAN_IN_THE_LOOP, FETCH_TOOLS, FINALIZE | The agent should ask one audience or preference clarification before generating the final answer, then produce a grounded and comprehensible recommendation. | hepatitis.csv |
| I uploaded dermatology.csv and want the answer tuned to my audience, so ask whether this is for clinicians or data scientists. | FETCH_DATA, ANALYZE_DATA, HUMAN_IN_THE_LOOP, FETCH_TOOLS, FINALIZE | The agent should ask one audience or preference clarification before generating the final answer, then produce a grounded and comprehensi | dermatology.csv |
| | | ble recommendation. | |
| I uploaded wine.csv and want a final answer, but you should first ask whether interpretability or breadth matters more. | FETCH_DATA, ANALYZE_DATA, HUMAN_IN_THE_LOOP, FETCH_TOOLS, FINALIZE | The agent should ask one audience or preference clarification before generating the final answer, then produce a grounded and comprehensible recommendation. | wine.csv |

## Supplementary material 4: Implementation details

The platform's architecture is based on two clearly separated APIs with distinct responsibilities. The central backend API acts as the stable long-term control and access layer for the platform. Its core functions include managing tool definitions, versioning, metadata, runs and run states, and enforcing a state-based model of the analysis lifecycle. Analyses go through defined phases from initialization to execution to completion, with each transition being controlled and persisted by the backend API. This yields a traceable execution history in which analyses are represented not as short-lived compute jobs, but as persistent system states that remain referable over time. This principle aligns with provenance-oriented process models in which execution and data dependencies are recorded explicitly [1]. Actual execution is handled by a separate orchestration API that functions as a specialized, technically constrained service. Its role is limited to coordinating containerized execution, managing infrastructure-level state transitions, and controlling access to resources. This separation encapsulates domain and business logic within the backend API, while limiting the orchestration API to operational tasks. This reduces complexity, facilitates independent validation of both APIs, and lowers security risk by isolating direct access to the container runtime.

The PoSyMed platform adopts key zero-trust principles to allow the analysis of sensitive data. Each analysis job runs in an isolated Docker container. Role-based access control, authentication, and authorization are implemented through OAuth2/OpenID Connect. In this architecture, the combination of controlled images, role-based access control, and separate runtime instances establishes a security-oriented operating model, even without additional virtualization layers. This operating model adopts key zero-trust principles such as explicit authentication, least-privilege access, and isolated execution contexts. Authentication and authorization are implemented through OAuth2/OpenID Connect. The architecture follows data-visiting platform principles in which analyses are executed within controlled infrastructure boundaries rather than moving sensitive data into uncontrolled environments [2,3].

Each tool is a containerized application that communicates with the platform through HTTP and WebSocket (WS) interfaces. These channels support controlled interaction with the PoSyMed platform, including initialization, status monitoring, result transfer, and structured error handling. Each tool may exist in multiple versions, enabling iterative development and the traceable reproduction of prior results.

A tool's formal specification is stored centrally in the platform's tool database and locally as a YAML file within the respective application. This dual anchoring supports consistency between technical implementation and platform-level orchestration. It also enables validation by the platform, the tool runtime, and human reviewers. Tools are therefore not treated as opaque black boxes but as verifiable units with both technical and functional contracts.

Files can be loaded into a tool container via files already present in the container, files uploaded by the Orch-API, or, for small payloads, data transmitted directly over the WebSocket channel.

## Supplementary material 5: Controlled building pipeline

To ensure that only reproducible, inspectable and security-checked tool versions become executable within PoSyMed, each tool release is processed through a dedicated, server-controlled building pipeline. Every pipeline run is executed in an isolated container context and follows a fixed sequence of operations that begins with backend-defined configuration retrieval and ends with publication of a validated image together with traceable release metadata.

**Fetch Config**
In Fetch Config, the pipeline requests the complete run configuration from the backend. This step defines the repository source and target image identity for the build and initializes the server-side metadata required for the remaining pipeline stages.

**Fetch Repository**
In Fetch Repository, the target workspace is created to ensure a clean build context. The linked source repository is then cloned into this fresh directory.

**Fetch Files**
In Fetch Files, additional model weights files can be injected into the repository workspace when required. These files are transferred as a ZIP archive, unpacked into the configured model directory, and written through path-safe extraction, enabling controlled delivery of external assets such as model weights or runtime files.

**Build Image**
In Build Image, the pipeline constructs the Docker image from the prepared repository using docker build with BuildKit enabled. The image name is derived deterministically from platform metadata.

**Test**
In Test, the newly built image is launched in an isolated container runtime with platform-defined environment variables, including *TEST_MODE=true*. The pipeline then executes the tool entry point inside the container.

**Trivy Scan**
In Trivy Scan, the built container image is analyzed with Trivy using JSON-formatted reporting restricted to HIGH and CRITICAL vulnerabilities. The resulting scan is stored as a structured vulnerability summary together with the raw report content.

**Malware Scan**
In Malware Scan, the full repository workspace is recursively screened with ClamAV. The scan covers the checked-out repository together with any injected files, and any detected finding causes immediate failure of the release process.

**Push Image**
In Push Image, the validated image is published to the configured registry. The pipeline prepares a dedicated Docker buildx builder and context and performs a multi-platform release for the configured target platforms, followed by cleanup of the temporary buildx resources.

**Push Summary**
In Push Summary, the pipeline resolves the exact Git commit hash of the built repository, derives repository file links at that commit, records the published image name, and aggregates the vulnerability and malware scan results into a final release summary. This summary is uploaded to the backend, providing a traceable record of the source state, validation outcome and published executable artifact associated with the released tool version.

## Supplementary material 6: Platform demonstration

The Tool-Store is the central entry point for discovering platform-integrated analysis components. It exposes tool cards together with high-level metadata and filter options, thereby supporting structured navigation across heterogeneous tools, workflows, and trained models within one controlled interface (see Fig. 7).

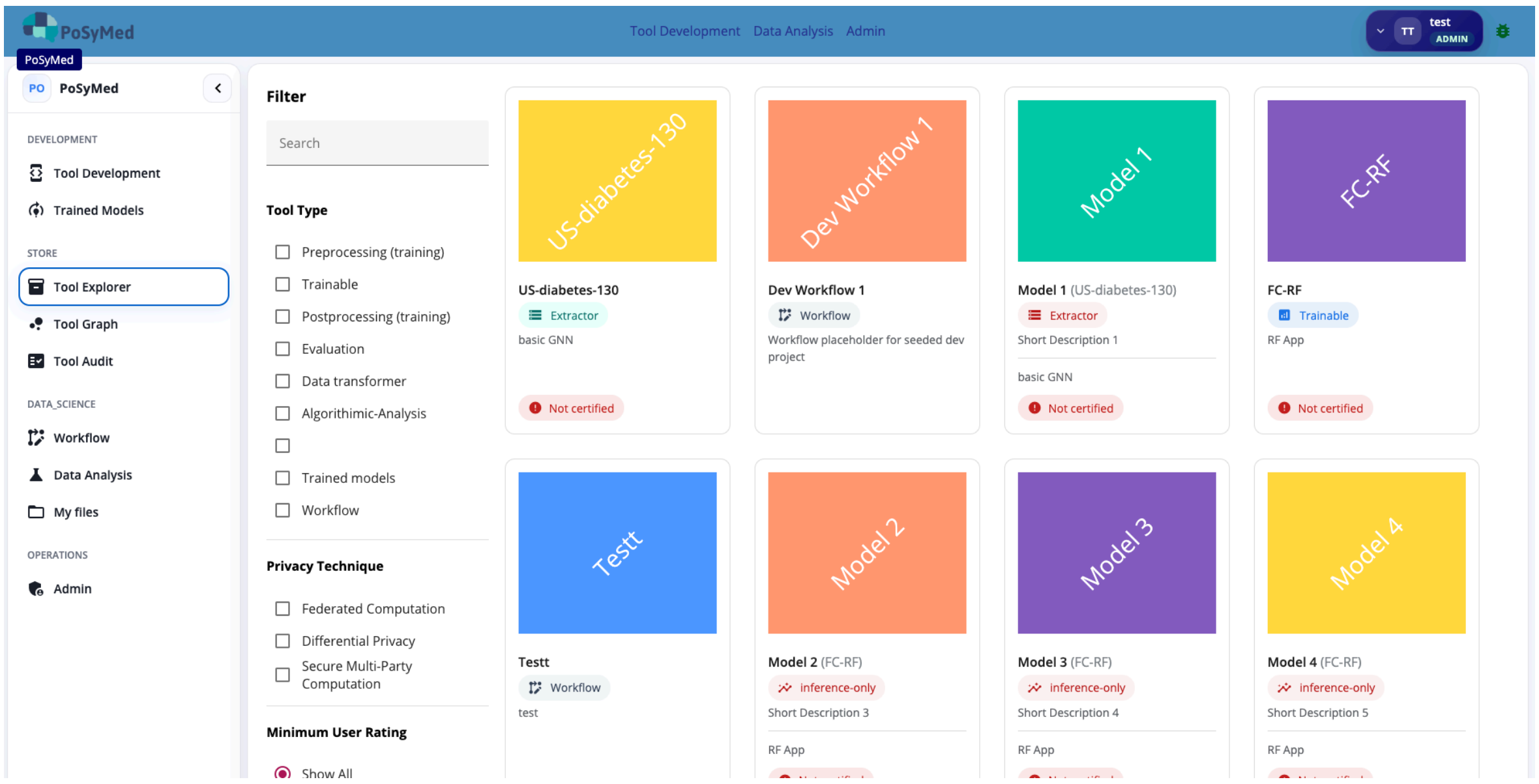

**Figure 7: Store overview page.** The PoSyMed Tool-Store interface provides a searchable and filterable overview of available platform artifacts. Items are presented as cards with concise metadata, including tool category and publication-related status indicators, enabling users to browse analysis components, workflows, and trained models within a unified interface

Once a tool has been selected, PoSyMed exposes its detailed metadata in a dedicated detail view. This page makes the formal specification of the application visible to the user and thereby supports transparency about the tool's role, description, configuration, and release context (see Fig. 8).

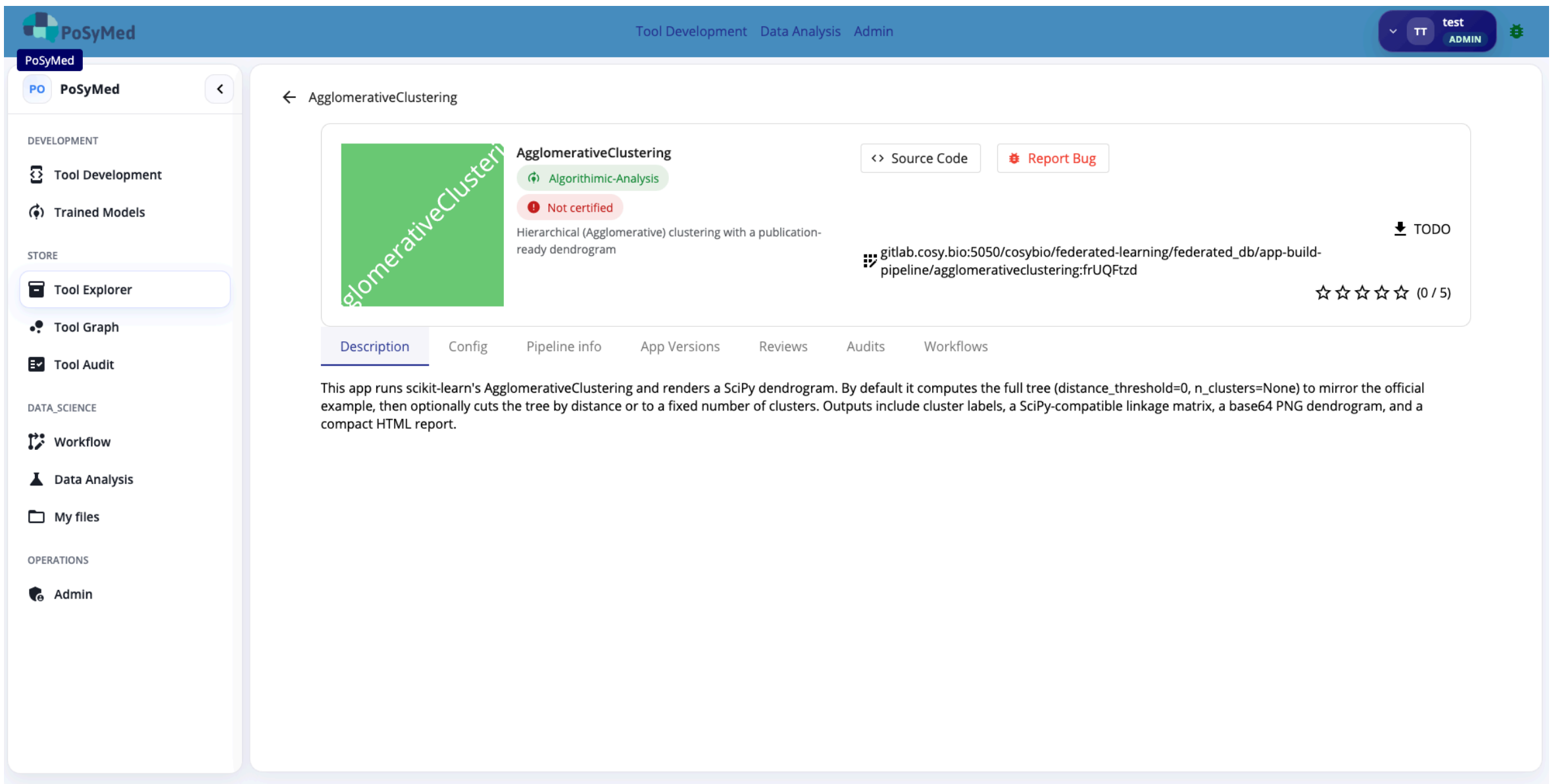

**Figure 8: App Detail Page.** The application detail page exposes the selected tool's core metadata, including its title, type, short description, repository linkage, and configuration-related tabs. This view makes the formally specified tool identity visible

within the platform and serves as the main entry point for inspection of configuration, versions, audits, and workflow usage.

Workflow construction in PoSyMed is based on typed, platform-managed DAGs in which nodes refer to concrete tool definitions and edges represent valid input-output mappings. The workflow editor therefore makes the structural organization of multi-step analyses explicit before execution (see Fig. 9).

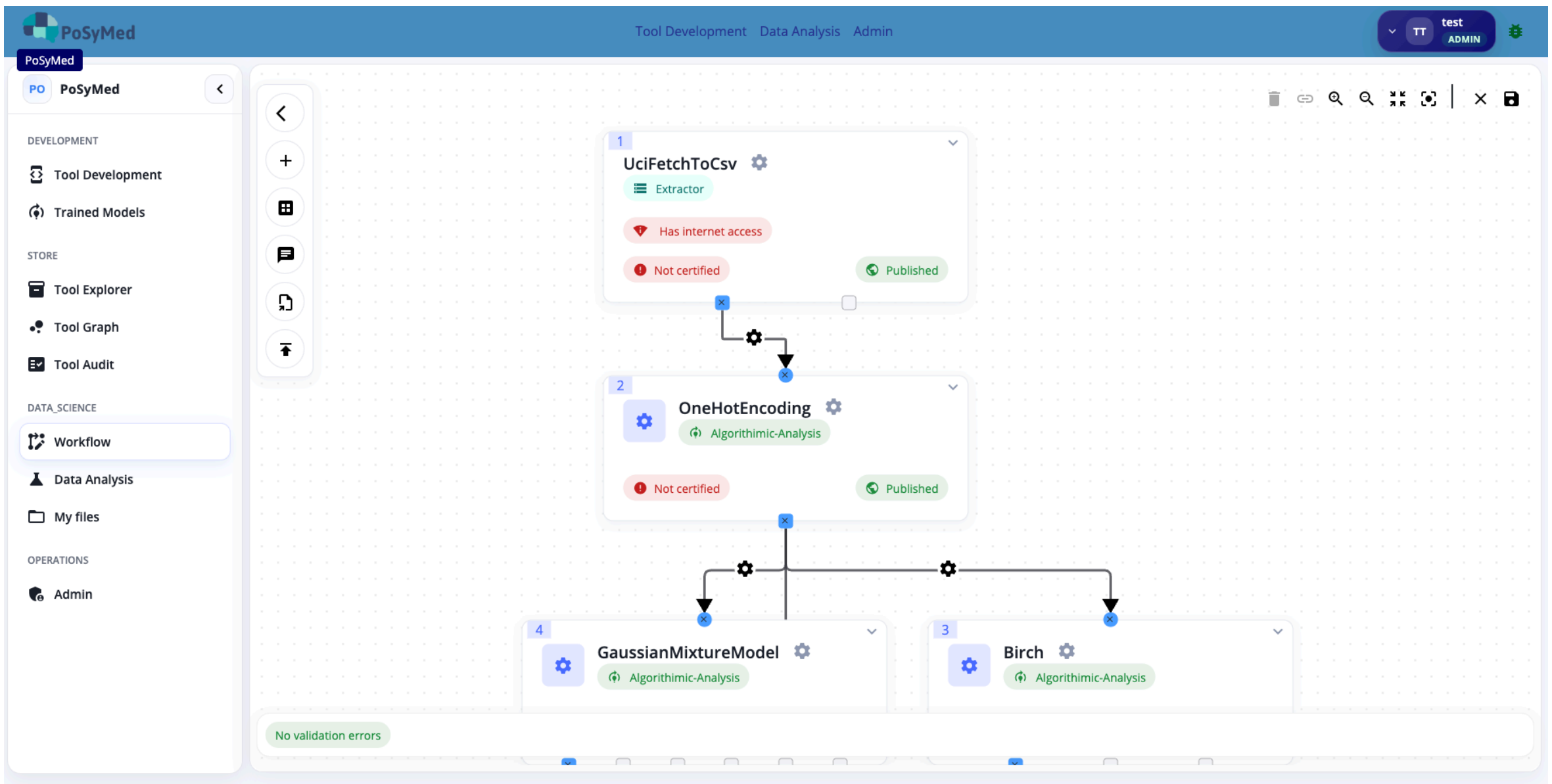

**Figure 9: Workflow Detail Page.** The workflow interface displays a platform-managed directed acyclic graph in which individual nodes correspond to concrete tools and edges represent valid downstream data flow. This view illustrates how extraction, preprocessing, and analysis tools can be composed into explicit, inspectable workflow structures prior to execution.

In addition to descriptive metadata, PoSyMed also exposes build and validation results for each tool version. This links the user-facing application page directly to the controlled build pipeline described in the manuscript and makes publication status auditable from within the interface (see Fig. 10).

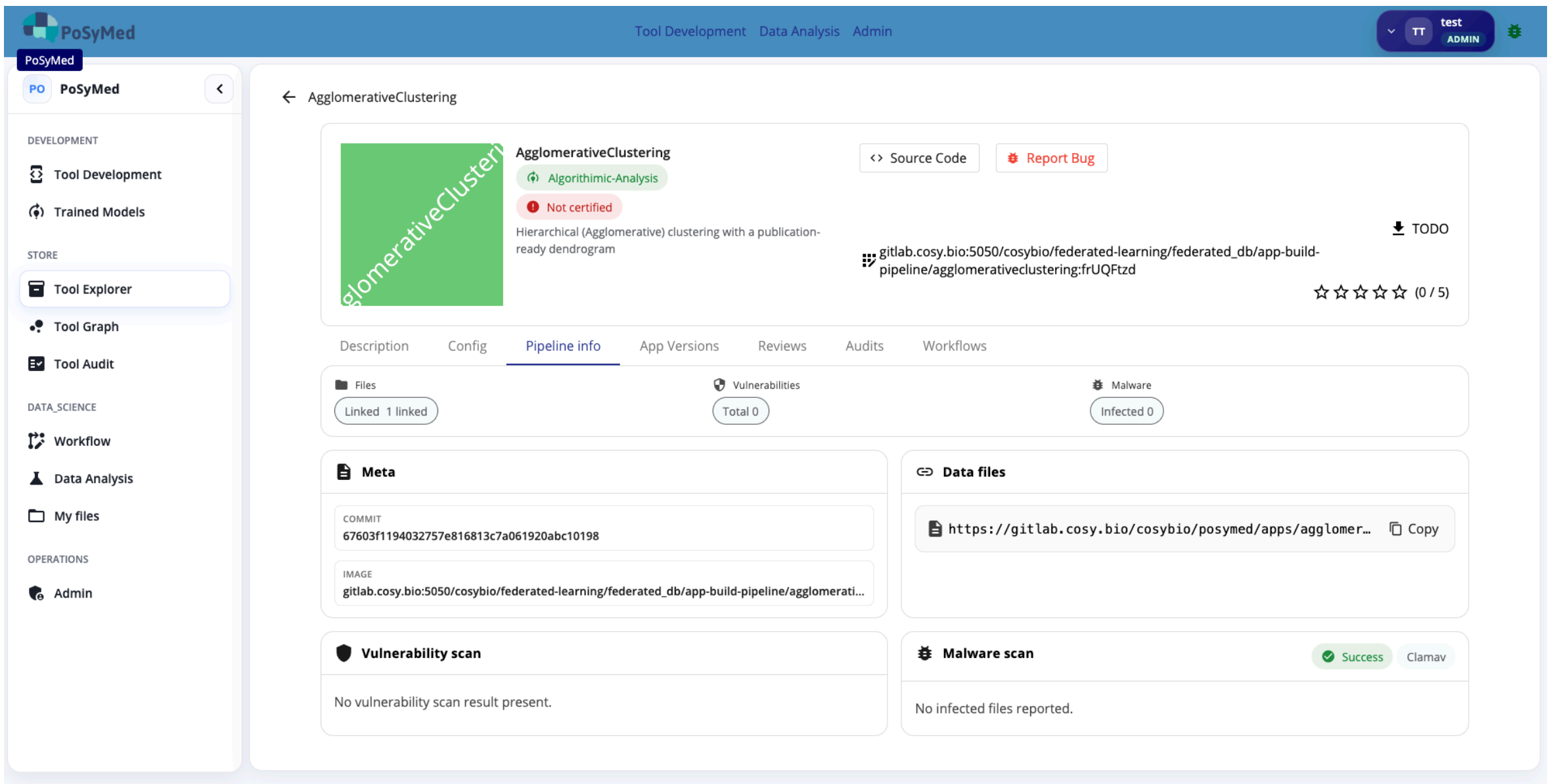

**Figure 10: App detail view with pipeline results.** The pipeline information tab links the selected tool to its centrally managed build process and exposes build-related metadata such as linked files, commit references, container image information, and security checks. This view illustrates how PoSyMed connects tool publication to a controlled validation pipeline rather than treating tools as opaque uploaded artifacts.

Tool development in PoSyMed is guided by a formal application configuration that defines hyperparameters, inputs, outputs, and local development settings. The corresponding development interface supports editing and inspection of these specification elements in a structured, platform-aware form (see Fig. 11).

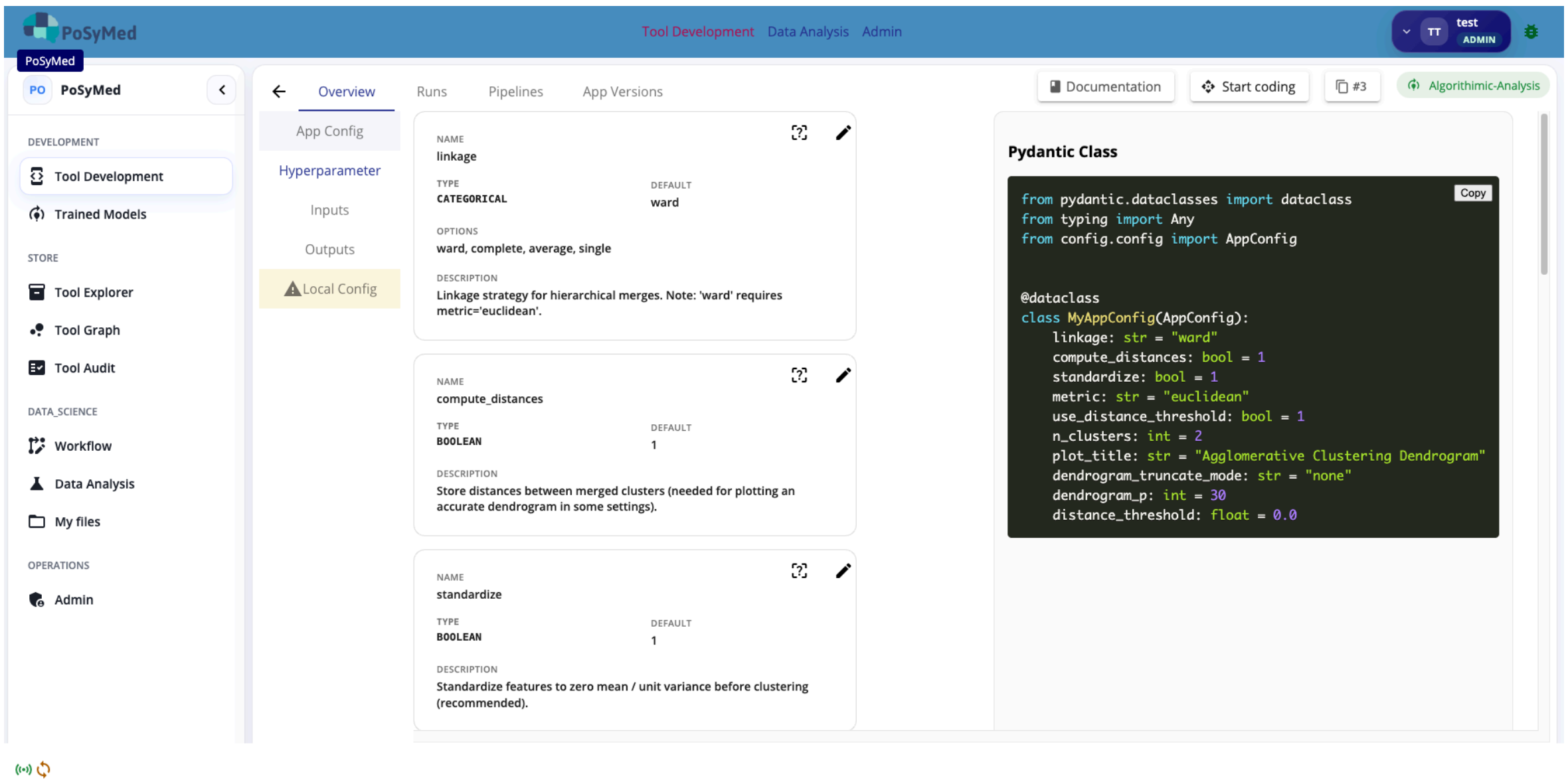

**Figure 11: App development interface.** The tool development view exposes the application configuration used to define the formal contract of a PoSyMed tool. Hyperparameters, inputs, outputs, and local configuration are presented alongside

code-oriented documentation, illustrating how tool behavior is specified through explicit, machine-readable metadata.

Because PoSyMed supports interactive development, the platform can surface runtime output directly while a tool is being configured or tested. This shortens the feedback loop between application specification and observed runtime behavior (see Fig. 12).

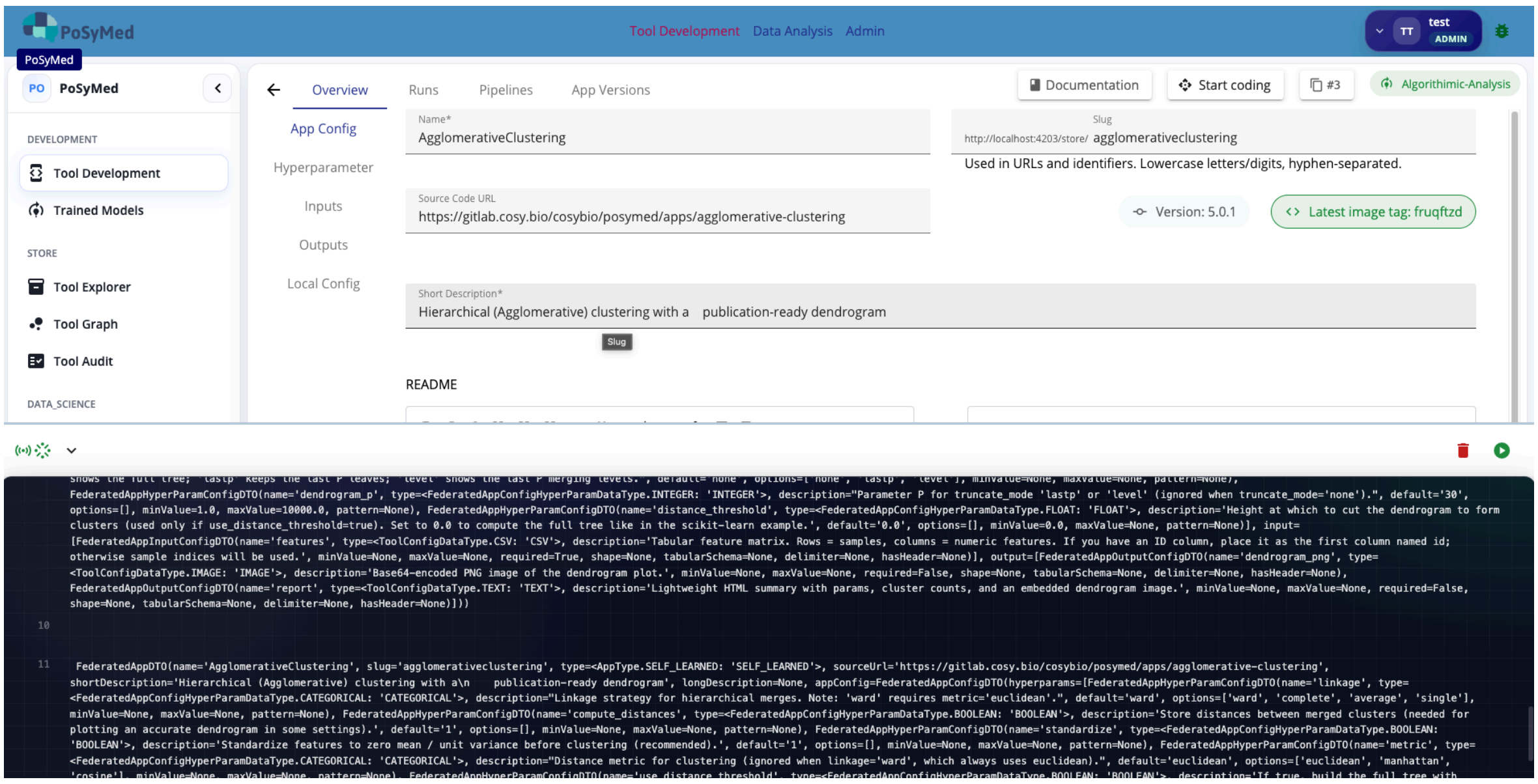

**Figure 12: App development with live console output.** During development, PoSyMed exposes live runtime output directly in the user interface. The combined configuration and console view illustrates how local tool execution, configuration editing, and streamed runtime logs are integrated into one interactive development environment.

Before publication, tools can be started in a controlled local test mode. This allows developers to provide concrete hyperparameters and input files while preserving the same typed configuration logic used by the platform during later execution (see Fig. 13).

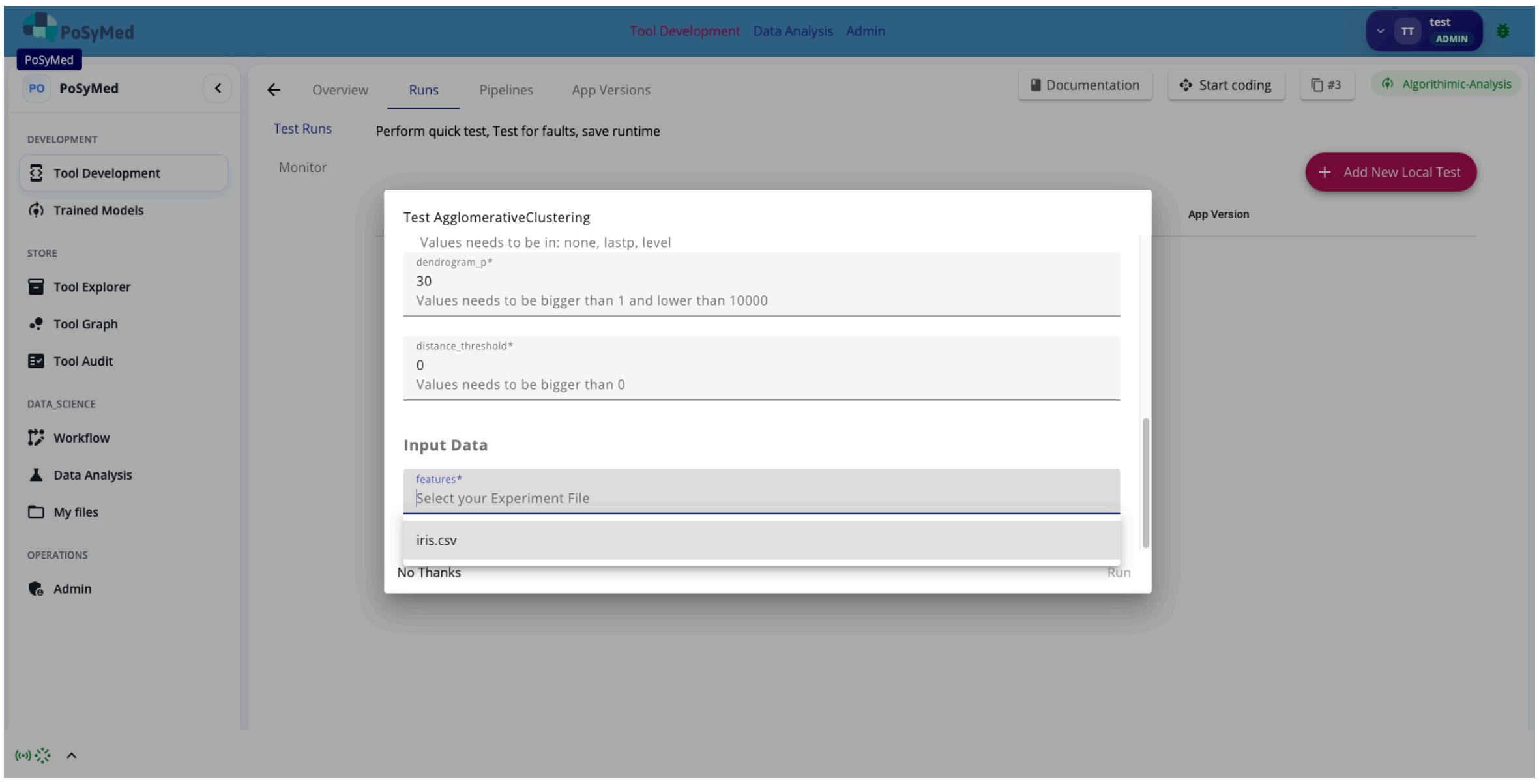

**Figure 13: App development test initialization.** The test-run dialog allows a developer to start a controlled local test by supplying hyperparameter values and selecting input data files. This interface demonstrates how PoSyMed validates test configuration against the tool's declared schema before execution begins.

After a test has been executed, the resulting outputs can be inspected directly within the platform. This includes both status information and returned artifacts, thereby linking test execution to the same traceability principles that govern production runs (see Fig. 14).

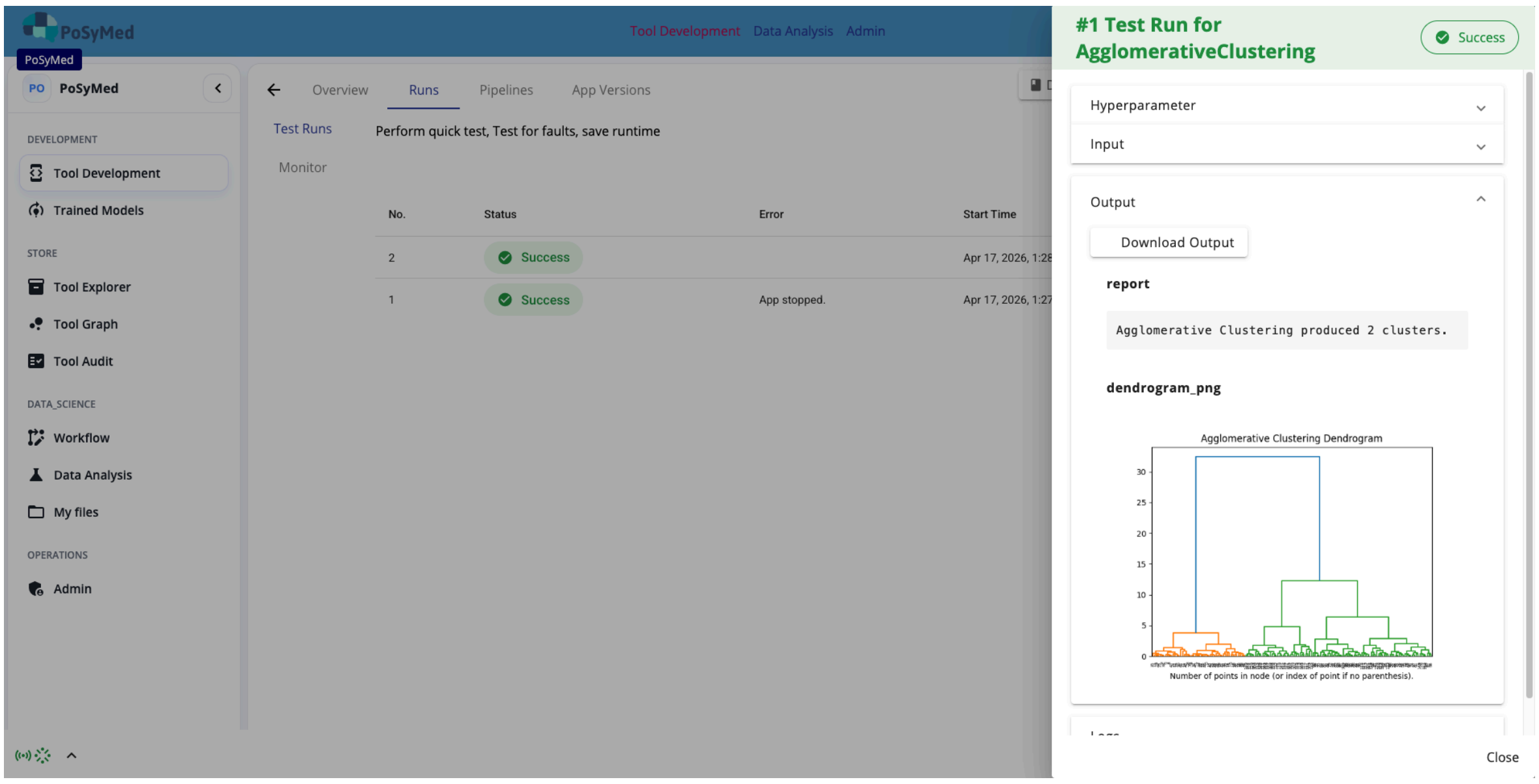

**Figure 14: App development test results.** The test results view summarizes completed local test runs together with their status and returned artifacts. In this example, the generated report and dendrogram output can be inspected and downloaded, illustrating how PoSyMed records and exposes test outcomes within the same platform environment used for later analyses.

The controlled build pipeline is represented as an explicit multi-step process rather than a hidden background operation. Exposing this process in the interface makes the publication state of a tool version transparent and supports inspection of intermediate build steps and logs (see Fig. 15).

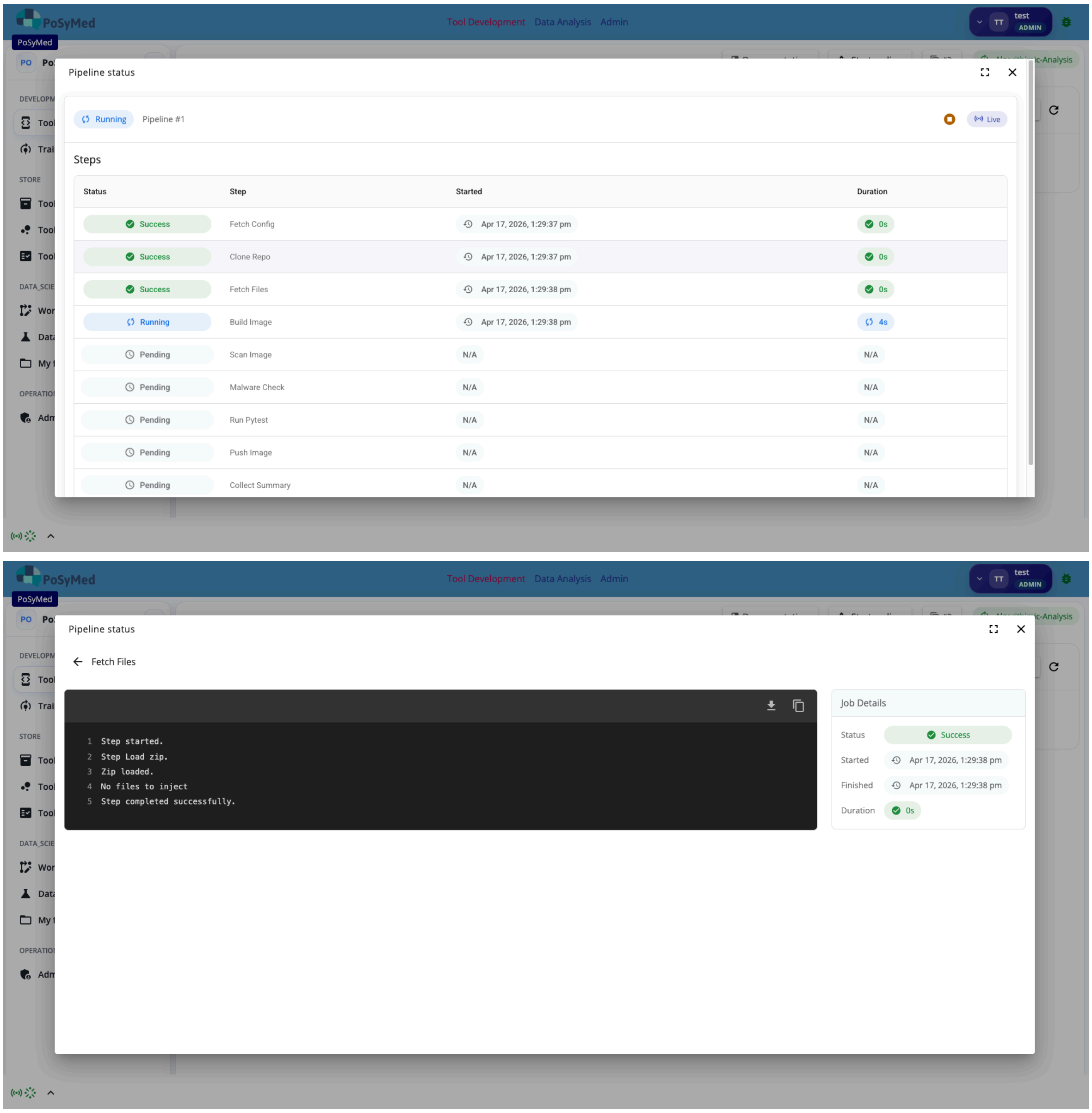

**Figure 15: App development pipeline view.** The pipeline status interface exposes the sequential build and validation steps performed before a tool becomes executable on the platform. Shown steps include configuration loading, repository cloning, file fetching, image creation, and downstream validation and publication stages, together with step-wise job details and runtime logs.

At execution time, PoSyMed treats an analysis as a persistent run state linked to concrete inputs and outputs. The experiment interface therefore combines uploaded files, runtime logs, and output artifacts within a single view (see Fig. 16).

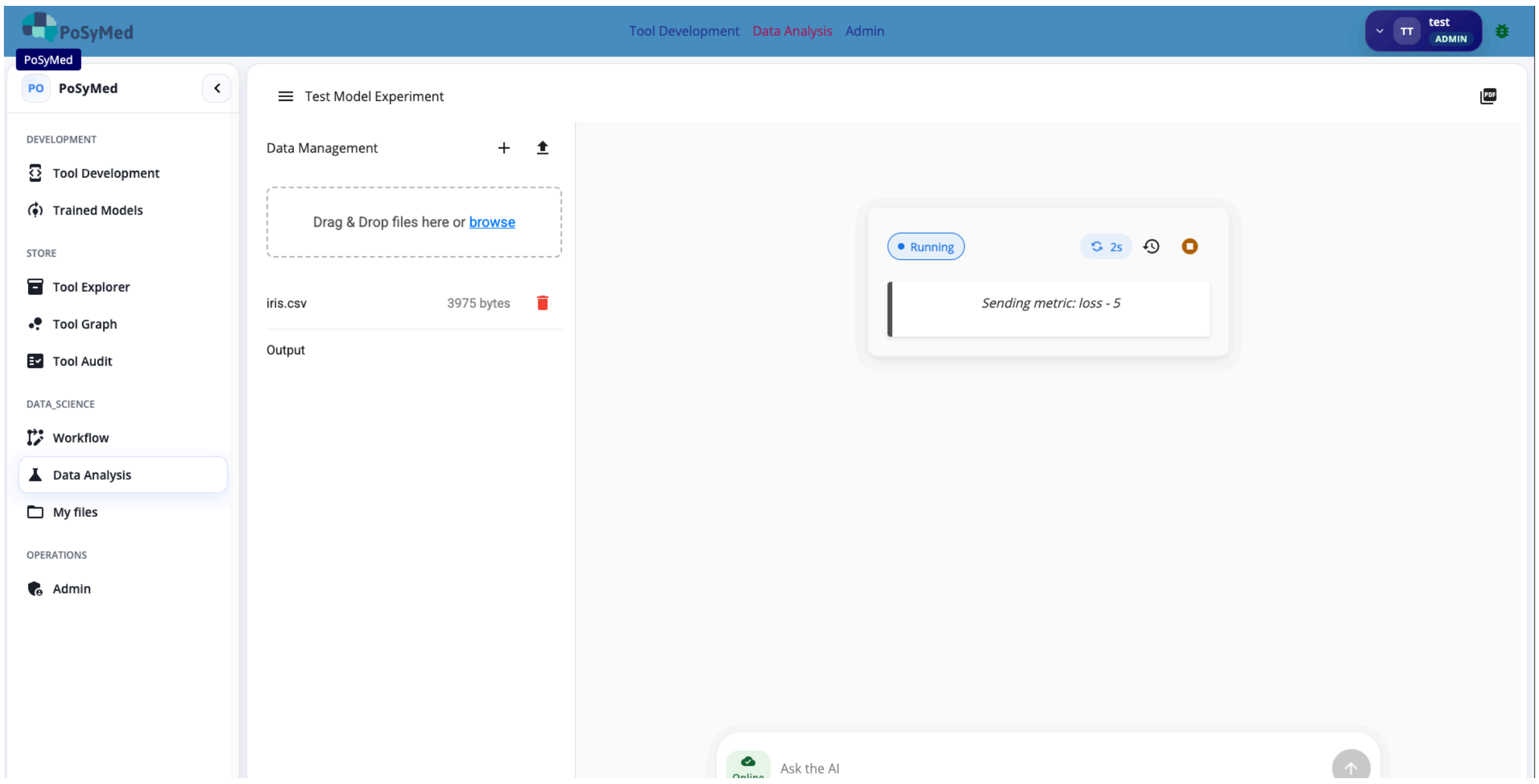

**Figure 16: Experiment start with live runtime log.** The experiment interface shows a newly started analysis run with uploaded input data on the left and live runtime telemetry on the right. This view illustrates how PoSyMed exposes ongoing execution state directly to the user while preserving the association between a run, its inputs, and streamed status messages.

After completion, output artifacts are made available for inspection within the same experimental context. PoSyMed thus links execution not only to status transitions, but also to directly viewable results that can be opened, analyzed, or reused (see Fig. 17).

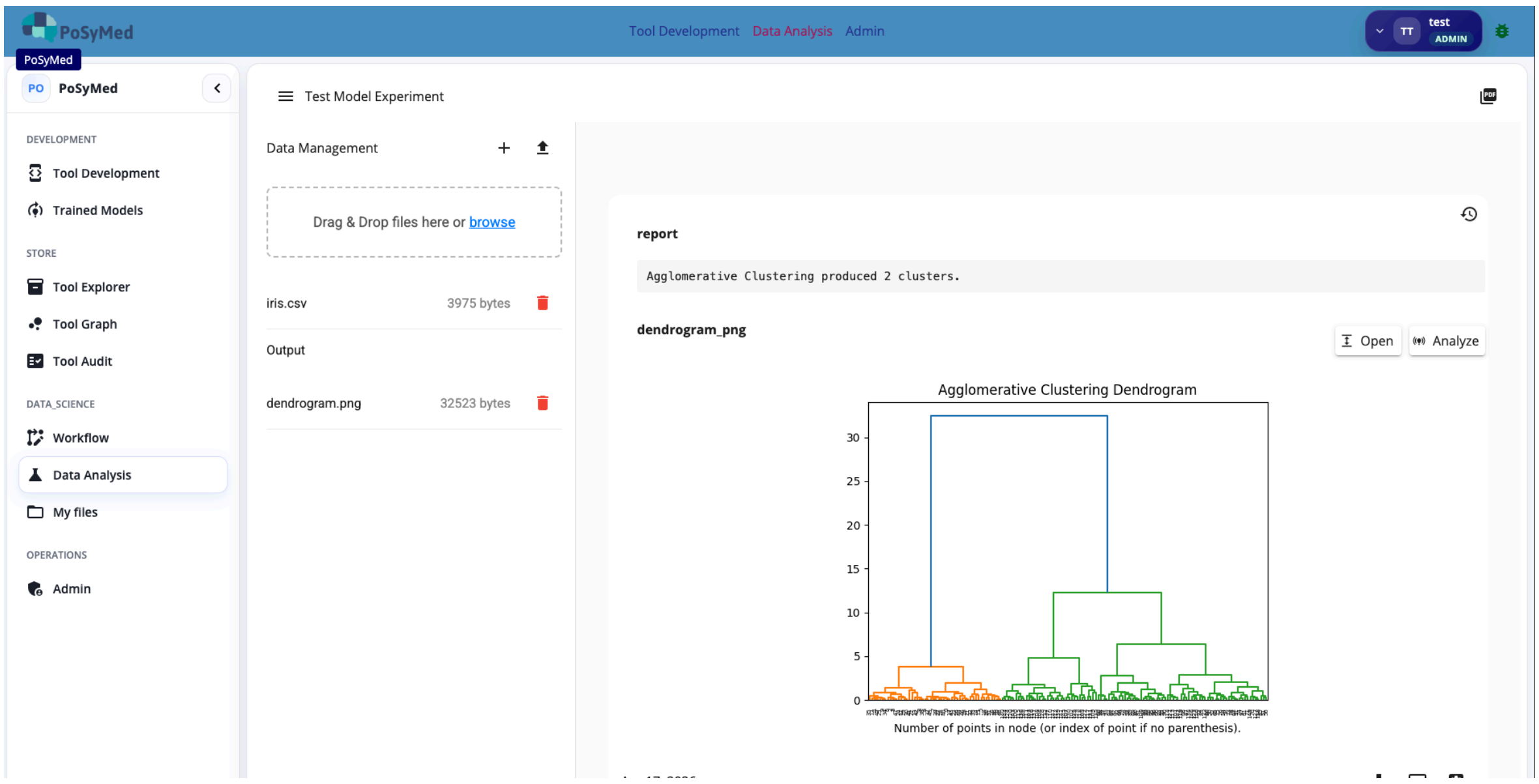

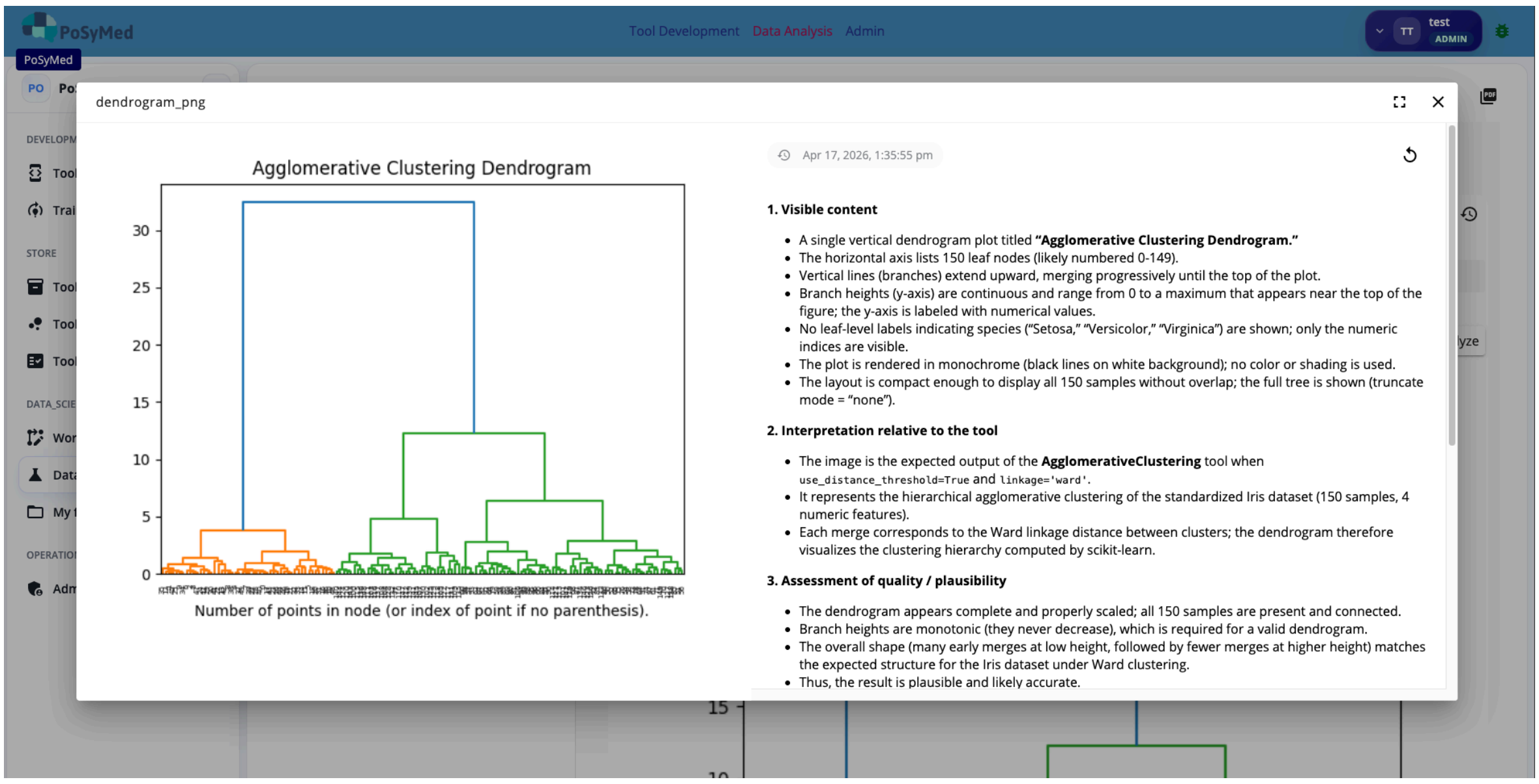

**Figure 17: Experiment result view.** The completed experiment page displays returned artifacts and rendered outputs within the same analysis context in which the run was started. In the example shown, the tool report and generated dendrogram are available for direct inspection, illustrating PoSyMed's coupling of persistent run state with explicit result artifacts.

Input handling in PoSyMed is governed by tool-specific schemas and validation logic. As a result, hyperparameters and selected inputs are checked before execution, preventing structurally invalid runs from starting (see Fig. 18).

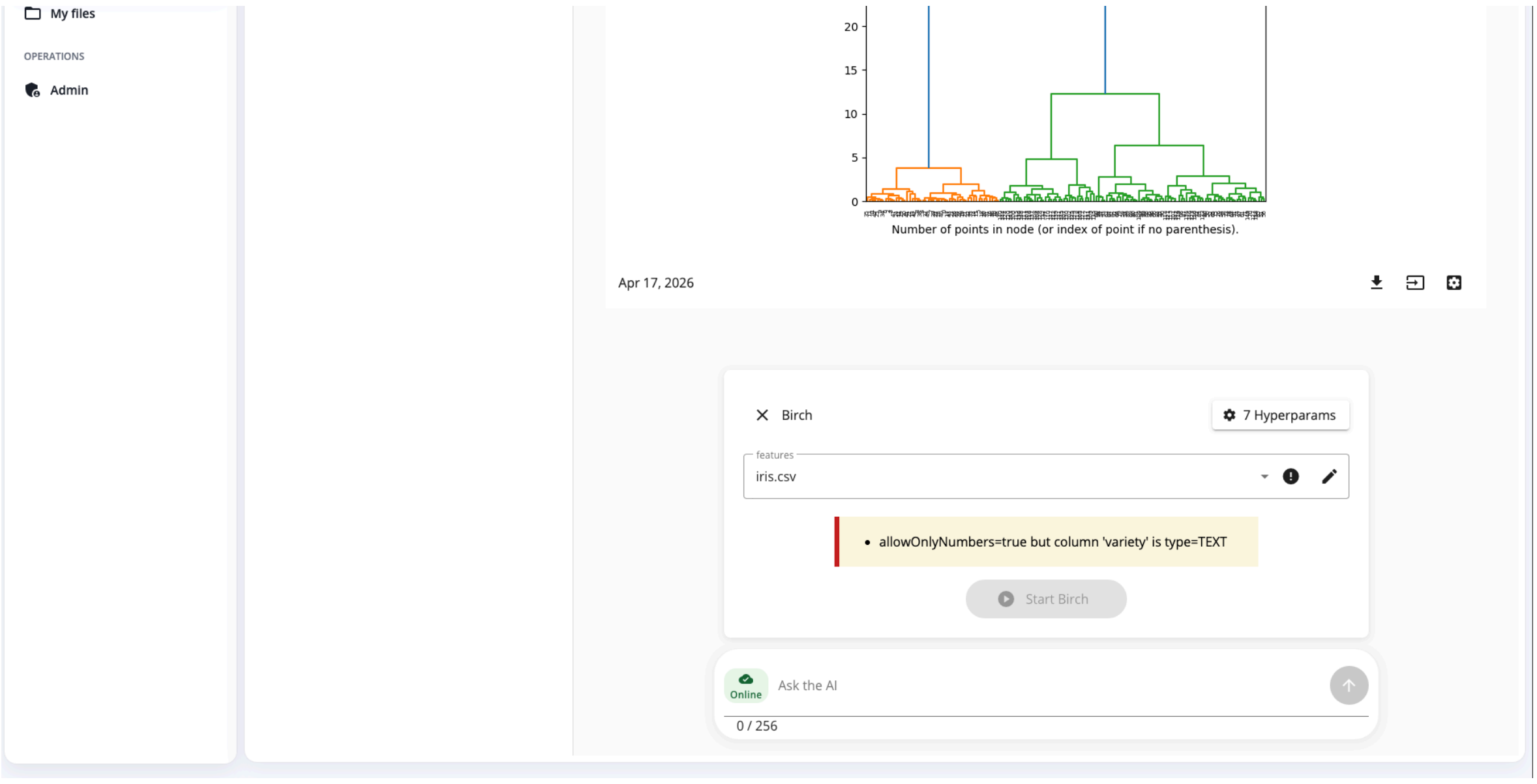

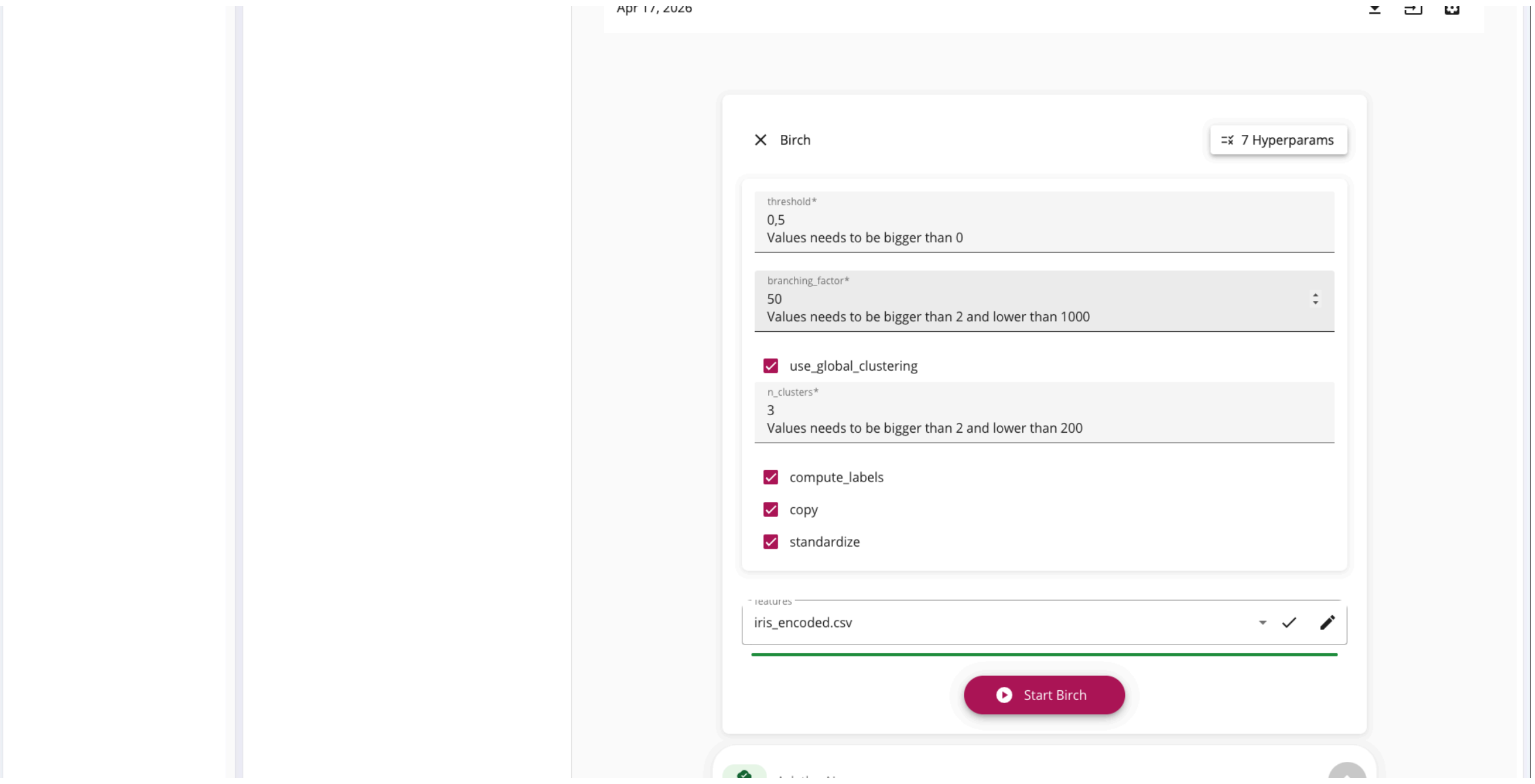

**Figure 18: Experiment input validation.** The experiment configuration panel shows how hyperparameters and selected inputs are validated against the tool's declared configuration before execution. Constraint messages are surfaced directly in the interface, demonstrating how PoSyMed prevents invalid analyses through schema-bound input checking at run time.

Returned outputs are not restricted to a single fixed format. Instead, PoSyMed can expose tabular, graphical, and file-based result representations, depending on the tool's declared outputs and the generated artifact type (see Fig. 19).

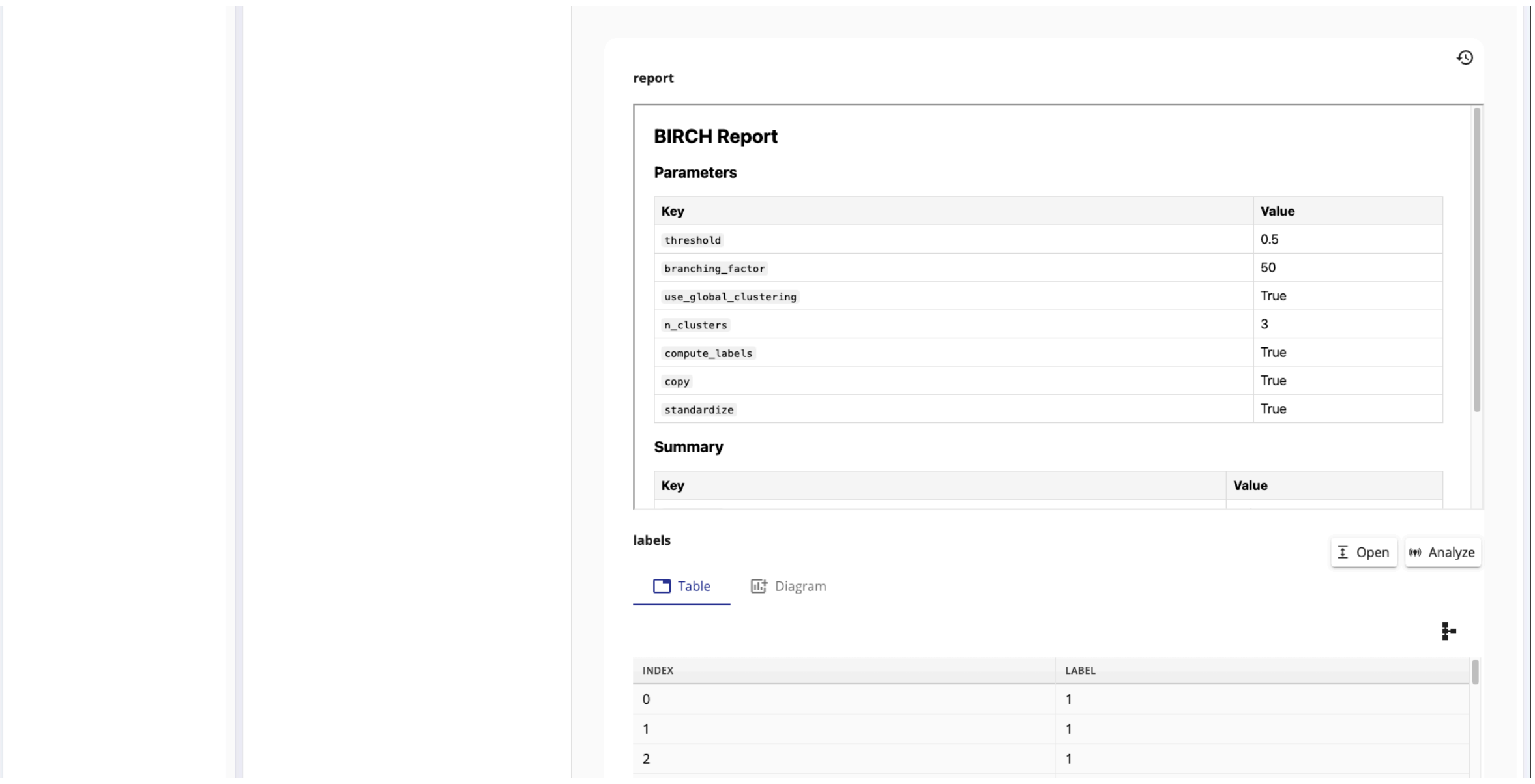

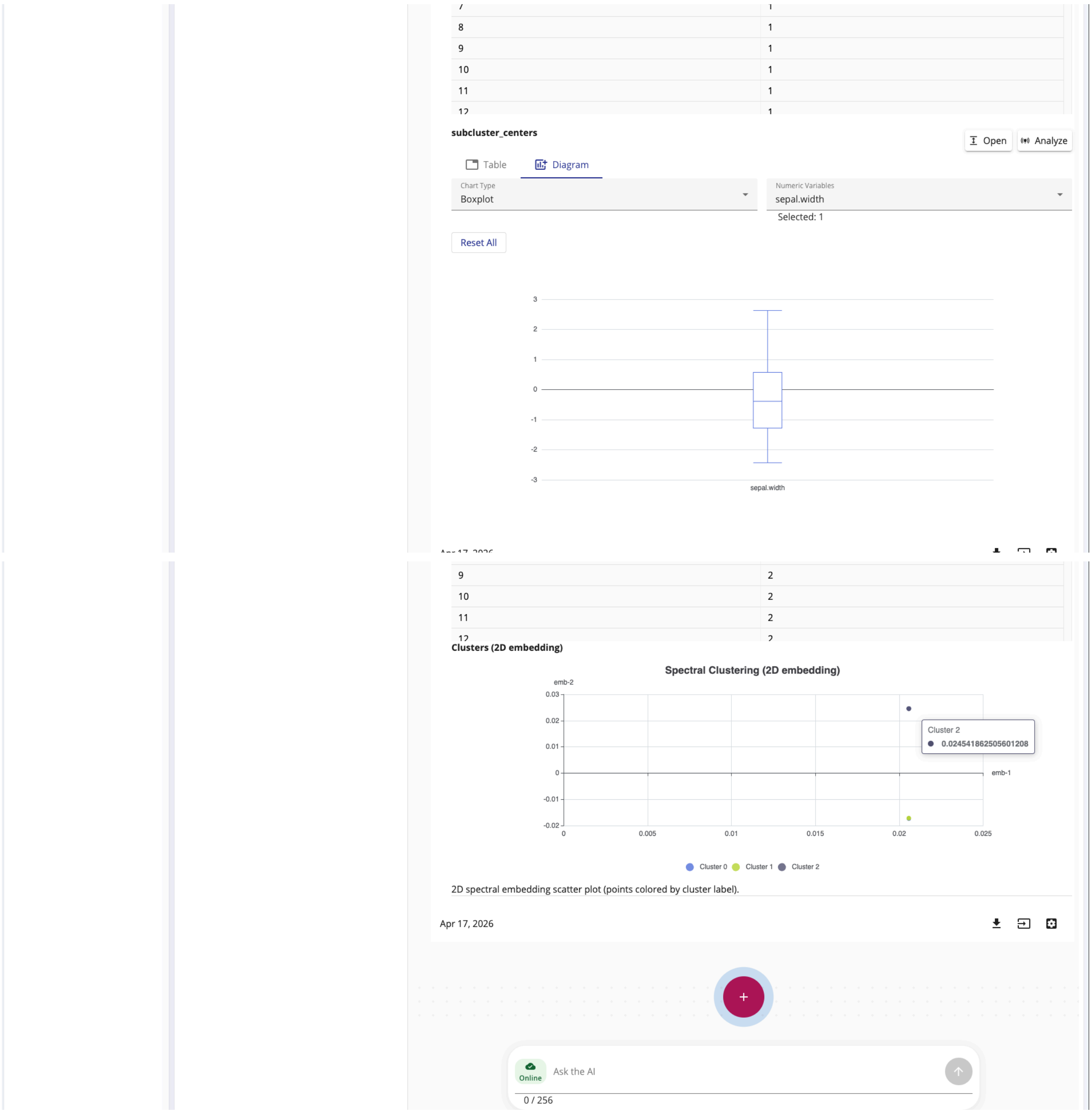

**Figure 19: Experiment output variation.** The result interface supports multiple output representations, including tables, diagrams, charts, and downloadable files, depending on the tool output type. The example illustrates how PoSyMed can render heterogeneous result artifacts within a single analysis environment while preserving their association with the originating run.

LLM support in PoSyMed is embedded into the platform as a bounded interface layer rather than as an unrestricted assistant. The interface therefore exposes structured suggestions, tool-oriented next steps, and explicit uncertainty statements grounded in the available platform context (see Fig. 20).

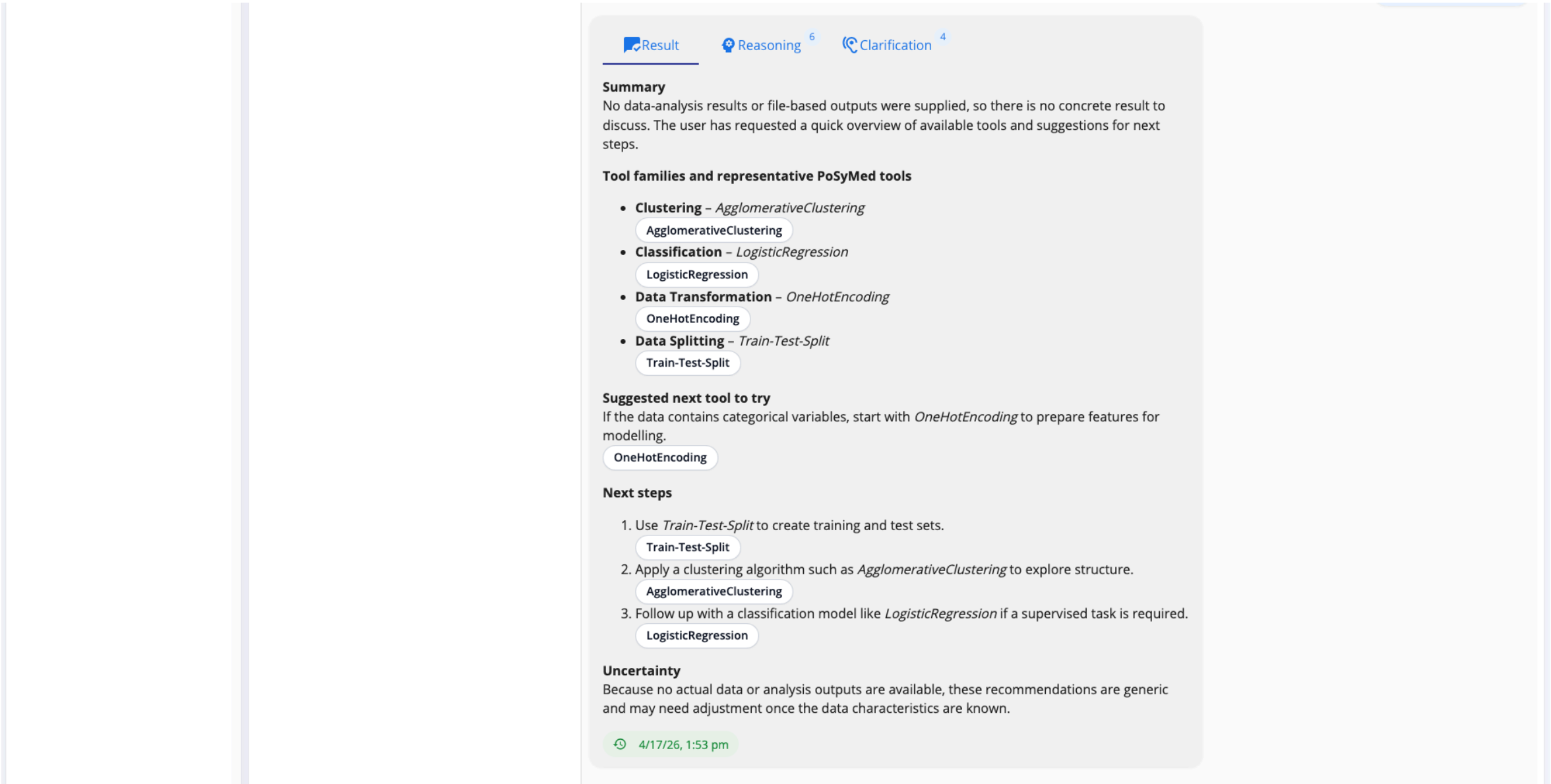

**Figure 20: LLM integration in the analysis interface.** The integrated LLM interface provides a structured response within the analysis context, including summarized findings, representative tool suggestions, proposed next steps, and explicit uncertainty statements. This illustrates the manuscript's bounded human-in-the-loop design, in which language-model support is tied to platform state and tool-aware reasoning rather than unconstrained free-form generation.

Behind the user-facing analysis layer, PoSyMed separates long-term platform logic from infrastructure-level execution control. The orchestration administration view makes this lower-level operational layer inspectable by exposing engine status, running containers, persisted runs, and related execution metadata (see Fig. 21).

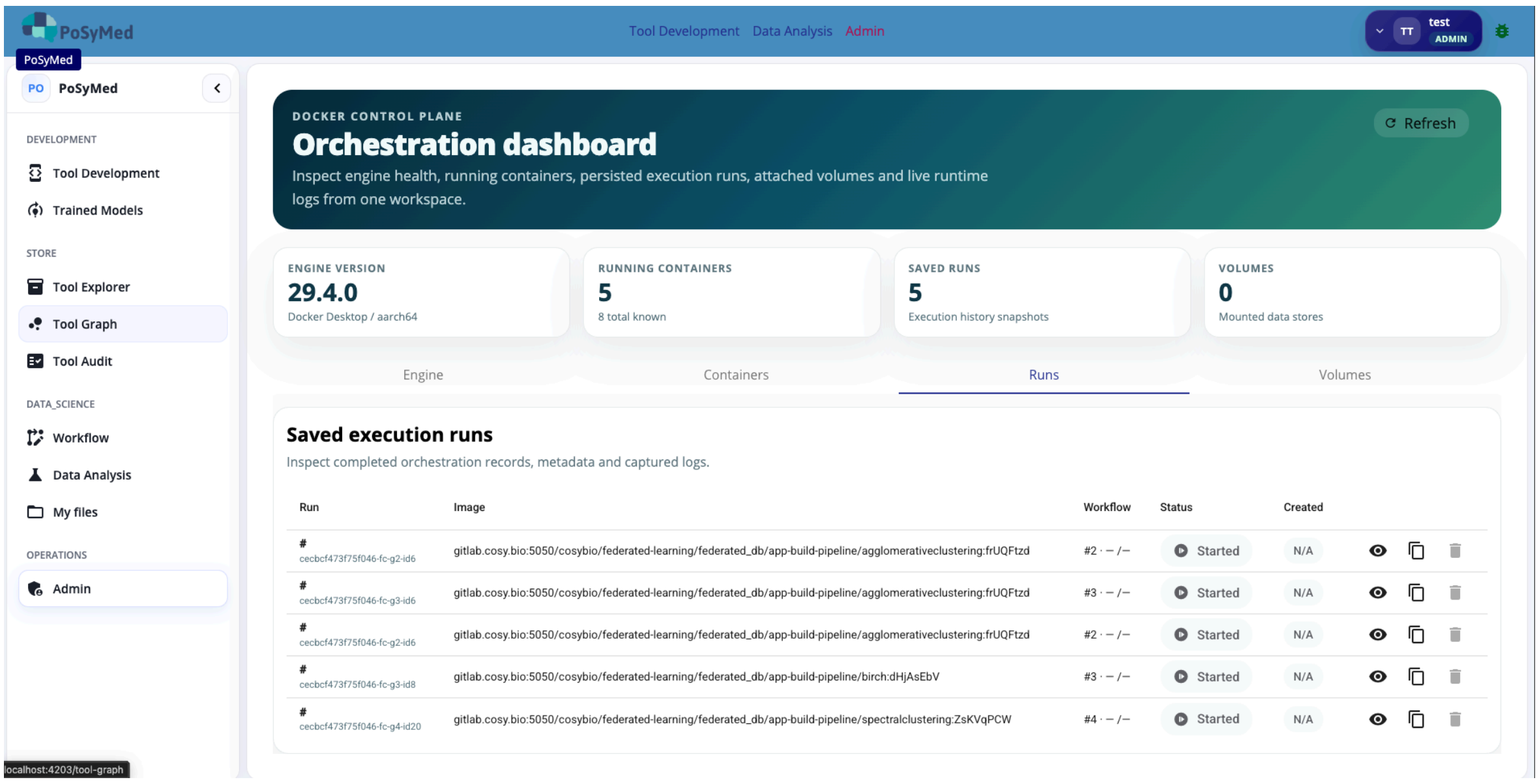

**Figure 21: Orchestration administration dashboard.** The orchestration administration interface exposes infrastructure-level execution state, including engine information, running containers, saved execution runs, and attached resources. This view illustrates the manuscript's separation between the central backend API and a distinct orchestration layer responsible for controlled container execution and operational runtime management.